\documentclass[
  journal=pasa,
  manuscript=research-paper,
  year=2023,
  volume=,
]{cup-journal}

\usepackage{amsmath}
\usepackage{amssymb}
\usepackage{gensymb}
\usepackage[nopatch]{microtype}
\usepackage{booktabs}
\usepackage{verbatim}
\usepackage[hidelinks]{hyperref}
\hypersetup{colorlinks=true,linkcolor=red,citecolor=orange,filecolor=cyan,urlcolor=magenta}
\usepackage{orcidlink}

\newcommand{\swift}{\textit{Swift}}
\newcommand{\hst}{\textit{HST}}
\newcommand{\jwst}{\textit{JWST}}
\newcommand{\spitzer}{\textit{Spitzer}}
\newcommand{\rst}{\textit{Roman}}
\newcommand{\euclid}{\textit{Euclid}}
\newcommand{\halpha}{$\mathrm{H}\alpha$}
\newcommand{\hbeta}{$\mathrm{H}\beta$}
\newcommand{\abump}{$A_\mathrm{bump}$}
\newcommand{\abumpnorm}{$A_\mathrm{bump}/E(B-V)_{\mathrm{gas}}$}
\newcommand{\kbump}{$k_\mathrm{bump}$}
\newcommand{\rpah}{$R_\mathrm{PAH}$}
\newcommand{\EBVgas}{$E(B-V)_{\mathrm{gas}}$}
\newcommand{\EBVstar}{$E(B-V)_{\mathrm{star}}$}
\newcommand{\EBVratio}{$E(B-V)_{\mathrm{star}}/ E(B-V)_{\mathrm{gas}}$}
\newcommand{\EBVratioavg}{$\langle E(B-V)_{\mathrm{star}}\rangle/\langle E(B-V)_{\mathrm{gas}}\rangle$}
\newcommand\arcmin{\mbox{$^\prime$}}%
\newcommand\arcsec{\mbox{$^{\prime\prime}$}}%
\newcommand{\sigmaSFR}{$\Sigma_{SFR}$}
\newcommand{\magphys}{\texttt{MAGPHYS}}

\title{Constraining the link between the 2175\AA\ dust absorption feature and PAHs in Nearby Star-Forming Galaxies using \swift/UVOT and \jwst/MIRI}

\author{A. J. Battisti\orcidlink{0000-0003-4569-2285}}
\affiliation{Research School of Astronomy and Astrophysics, Australian National University, Cotter Road, Weston Creek, ACT 2611, Australia}
\alsoaffiliation{International Centre for Radio Astronomy Research (ICRAR), University of Western Australia, M468, 35 Stirling Highway, Crawley, WA 6009, Australia}
\alsoaffiliation{ARC Centre of Excellence for All Sky Astrophysics in 3 Dimensions (ASTRO 3D), Australia}
\email[A. J. Battisti]{andrew.battisti@anu.edu.au}

\author{I. Shivaei\orcidlink{0000-0003-4702-7561}}
\affiliation{Centro de Astrobiolog\'{i}a (CAB), CSIC-INTA, Ctra. de Ajalvir km 4, Torrej\'{o}n de Ardoz, E-28850, Madrid, Spain}

\author{H.-J. Park\orcidlink{0000-0002-9809-6631}}
\affiliation{Research School of Astronomy and Astrophysics, Australian National University, Cotter Road, Weston Creek, ACT 2611, Australia}
\alsoaffiliation{ARC Centre of Excellence for All Sky Astrophysics in 3 Dimensions (ASTRO 3D), Australia}

\author{M. Decleir\orcidlink{0000-0001-9462-5543}
}
\affiliation{European Space Agency (ESA), ESA Office, Space Telescope Science Institute, 3700 San Martin Drive, Baltimore, MD 21218, USA}
\alsoaffiliation{ESA Research Fellow}

\author{D. Calzetti\orcidlink{0000-0002-5189-8004}}
\affiliation{Department of Astronomy, University of Massachusetts-Amherst, Amherst, MA 01003, USA}

\author{J. Mathew\orcidlink{0000-0003-0459-5964}}
\affiliation{Advanced Instrumentation and Technology Centre, Research School of Astronomy and Astrophysics, Australian National University, Cotter Road, Weston Creek, ACT 2611, Australia}

\author{E. Wisnioski\orcidlink{0000-0003-1657-7878}}
\affiliation{Research School of Astronomy and Astrophysics, Australian National University, Cotter Road, Weston Creek, ACT 2611, Australia}
\alsoaffiliation{ARC Centre of Excellence for All Sky Astrophysics in 3 Dimensions (ASTRO 3D), Australia}

\author{Elisabete da Cunha\orcidlink{0000-0001-9759-4797}}
\affiliation{International Centre for Radio Astronomy Research (ICRAR), University of Western Australia, M468, 35 Stirling Highway, Crawley, WA 6009, Australia}
\alsoaffiliation{ARC Centre of Excellence for All Sky Astrophysics in 3 Dimensions (ASTRO 3D), Australia}


\begin{document}

\begin{abstract}
The 2175\AA\ bump is a prominent absorption feature at ultraviolet (UV) wavelengths in dust extinction and attenuation curves. Understanding the relative strength of this feature is important for making accurate dust corrections at both low- and high-redshift. This feature is postulated to arise from polycyclic aromatic hydrocarbon (PAH) dust grains; however, the carrier has not been definitively established. We present results on the correlation between the 2175\AA\ feature and PAH abundances in a spatially-resolved manner for 15 local galaxies in the PHANGS-\jwst\ survey that have NUV and mid-IR imaging data from \swift/UVOT and \jwst/MIRI, respectively. We find a moderate positive correlation between the 2175\AA\ feature strength and PAH abundance (spearman coefficient, $0.3 \lesssim \rho \lesssim 0.5$), albeit with large intrinsic scatter. However, most of this trend can be attributed to a stronger negative correlation of both quantities with SFR surface density and specific-SFR (proxies of ionising radiation; $\rho\sim-0.6$). 
The latter trends are consistent with previous findings that both the 2175\AA\ carrier and PAHs are small grains that are easily destroyed by UV photons,
although the proxy for PAH abundance (based on photometry) could also be influenced by dust heating.  
When controlling for SFR surface density, we find weaker correlations between the 2175\AA\ feature and PAH abundances ($\rho \lesssim 0.3$), disfavouring a direct link. However, analyses based on spectroscopic (instead of photometric) measurements of the 2175\AA\ feature and PAH features are required to verify our findings.
No significant trends with gas-phase metallicity or galactocentric radii are found for the 2175\AA\ feature and PAHs, however the metallicity range of our sample is limited ($8.40 < 12+\log[\mathrm{O/H}] < 8.65$). We provide prescriptions for the strength of the 2175\AA\ feature and PAHs in local massive (metal-rich) galaxies with SFR surface density and specific-SFR, however the former should be used with caution due to the fact that bump strengths measured from \swift/UVOT are expected to be underestimated.  
\end{abstract}

\section{Introduction}
Interstellar dust grains within galaxies can significantly alter their observed spectral energy distributions (SEDs) by acting to obscure light from ultraviolet (UV) to near-infrared (near-IR) wavelengths and reemitting this energy at infrared (IR) wavelengths \citep{galliano18}. Accurately characterizing the wavelength dependence of dust attenuation curves and their variation \citep{salim&narayanan20} is important because the assumed shape of dust curves has a strong impact on derived physical properties of galaxies from SED modeling \citep{conroy13} and can also affect the accuracy of photometric distance (photo-$z$) estimates \citep{salvato19, battisti19}. The latter is particularly important for upcoming precision dark energy studies with the \euclid\ \citep{euclid24} and \rst\ \citep{wangYun22} space telescopes, where galaxies are used to trace the large-scale structure of the Universe. 

A common characteristic of dust extinction and attenuation\footnote{We define `extinction' as the absorption and scattering of light out of the line of sight by dust, which has no dependence on geometry. `Attenuation' is the combination of extinction, scattering of light into the line of sight by dust, and geometrical effects due to the star-dust geometry (for review see \citealt{salim&narayanan20}).} curves is that the strongest extinction/attenuation typically occurs at bluer wavelengths and it decreases towards redder wavelengths in a gradual, continuous manner, except for the possible presence of a broad absorption feature centered at 2175\AA\ (referred to as the 2175\AA\ feature or bump) and weaker features at longer wavelengths \citep{draine03}: optical \citep[e.g.,][]{fitzpatrick19}, near-IR \citep[e.g.,][]{decleir22}, and mid-IR \citep[e.g.,][]{gordon21, hensley&draine21}. 
The overall shape of a dust extinction curve is primarily a consequence of the grain size distribution in the interstellar medium \citep[ISM; e.g.,][]{weingartner&draine01, hensley&draine23}. 
The origin of the 2175\AA\ feature remains unclear but is postulated to arise from small carbonaceous dust grains \citep[e.g.,][]{draine03, bradley05, papoular&papoular09}, with polycyclic aromatic hydrocarbons (PAHs) being the favoured carrier \citep[e.g.,][]{li&draine01, lin23}, although silicate carriers have also been proposed \citep[e.g.,][]{bradley05}. Understanding when the 2175\AA\ feature is present in dust attenuation curves of galaxies is important, particularly when limited rest-frame UV filters are available, because it can introduce variation of $\sim$25\% in the value of dust corrections at UV wavelengths \citep[affecting UV-based SFRs;][]{kennicutt&evans12} and also introduce a bias of $\sim$10\% on photo-$z$ estimates \citep[e.g.,][]{battisti19}.

The 2175\AA\ feature is observed in extinction curves measured using individual stars for many sightlines of the Milky Way \citep[MW; e.g.,][]{cardelli89, fitzpatrick99, valencic04, gordon09, fitzpatrick19}, the Large Magellanic Cloud \citep[LMC; e.g.,][]{gordon03}, and the Andromeda galaxy \citep[M31; e.g.,][]{bianchi96,clayton15}. The feature is usually weak or absent in the Small Magellanic Cloud \citep[SMC; e.g.,][]{gordon03}, although exceptions to this have recently been reported \citep{gordon24}. 
Beyond the local group, it has been observed in a small subset of extinction curves for quasars absorption systems \citep[e.g.,][]{ma17} and gamma-ray bursts \citep[GRBs; e.g.,][]{zafar11, zafar12}. However, it is often absent in a majority of quasar hosts \citep[e.g.,][]{hopkins04, gallerani10}, quasar absorption systems \citep[e.g.,][]{york06}, and GRBs \citep[e.g.,][]{schady12, zafar18b}. However, it is worth noting that these distant sources are likely probing more extreme environments than the typical ISM in star forming galaxies.
For extinction curves, the absence of the feature can be directly attributed to an absence of the carrier along the line of sight.

Measuring the 2175\AA\ feature in \textit{attenuation} curves is more complicated. Theoretical studies have shown that the apparent strength of the 2175\AA\ feature in attenuation curves, relative to the intrinsic extinction curve, can be considerably reduced (but not \textit{removed}) due to the additional geometric and scattering effects at play \citep[e.g.,][]{gordon97, witt&gordon00, seon&draine16}. Observational studies have found large variation in the strength of the 2175\AA\ feature. \citet{calzetti94} found that starburst galaxies (strongly star-forming galaxies; SFGs) lack the feature in their attenuation curves, whereas studies of local ``normal" SFGs can have a weak feature \citep[relative to the MW; e.g.,][]{conroy10, wild11, battisti17b}, although large individual variation is evident \citep[e.g.,][]{salim18, belles23}. Large variation is also evident at higher redshifts, with some studies favoring the inclusion of a weak feature \citep[e.g.,][]{noll09a, buat11b, buat12, kriek&conroy13, scoville15, shivaei20a, battisti20, kashino21, shivaei22b, witstok23, markov24} and others suggesting a feature does not need to be included \citep[e.g.,][]{reddy15, zeimann15, salmon16}. These results indicate that there is no consensus regarding the importance of the 2175\AA\ feature in galaxy attenuation curves. 

Previous studies have attempted to link the 2175\AA\ feature to PAHs by comparing the strength of the 2175\AA\ absorption relative to the abundance of PAHs. Ideally, the most straightforward manner to test this would be to compare features in extinction. However, the extreme difference in the degree of extinction 
between UV wavelengths \citep[requiring stars with $A_V\lesssim6$; e.g.,][]{clayton03} and mid-IR wavelengths \citep[requiring stars with $A_V\gtrsim10$; e.g.,][]{hensley&draine20} makes direct comparison difficult with current facilities, although it is possible that this situation may change with new observations from \jwst. Instead, PAH abundance can also be inferred from PAH emission in the mid-IR \citep[e.g.,][]{draine&li07, draine21}. 
The strongest PAH emission features occur at rest-frame 3.3, 6.2, 7.7, 8.6, 11.3, 12.7, and 16.4$\mu$m \citep{tielens08, li20}.
If PAHs are the carrier, we might expect a correlation to exist between the 2175\AA\ absorption and PAH abundance based on emission. However, care needs to be taken to account for the fact that extinction/attenuation measures path-length(s) between stars and the observer whereas emission will arise from the entire path-length along the line-of-sight. 
\cite{massa22} used a carefully selected sample of MW stars with UV and IR spectroscopy and found a strong correlation between the 2175\AA\ feature and the 8.6 and 11.3$\mu$m PAH emission features, a moderate correlation with the 7.7$\mu$m feature, and no correlation with the 6.2$\mu$m feature. Recently, \cite{gordon24} performed a comparable analysis for the SMC and LMC and also found a moderate correlation between the 2175\AA\ bump area and the abundance of PAHs (via emission). Studies attempting to look at the link between the 2175\AA\ feature and PAH emission in external galaxies have had mixed findings; \cite{decleir19} combined \swift/UVOT and \spitzer/IRAC data for NGC~0628 and found no significant trend between the bump strength and a proxy for PAH abundance for regions within the galaxy. 
\cite{shivaei22b} combined MUSE and \spitzer/MIPS data for integrated $z\sim2$ galaxies and found a moderate trend between these quantities. However, in both studies the accuracy of the PAH measurements are limited by the \spitzer\ data (in terms of mid-IR filter coverage and sensitivity). With the availability of \jwst/MIRI data for nearby galaxies, which has more mid-IR filters and better sensitivity than \spitzer, it is now possible to revisit this comparison from a new perspective. 

In this work, we use a sample of 15 nearby galaxies with excellent UV, optical, and IR data for characterising the presence of the 2175\AA\ feature and the PAH abundance in a spatially-resolved manner, and to examine their relationship to each other and to the properties of the ISM. This paper is organised as follows: Section~\ref{data} describes the observational data and sample, Section~\ref{method} describes our methodology for deriving the various quantities used in our analysis, Section~\ref{results} shows our results, Section~\ref{discussion} is our discussion, and Section~\ref{conclusion} summarises our main conclusions. 

\section{Data and Sample}\label{data}
Our sample consists of 15 (out of 19) PHANGS-JWST galaxies that have UV data from \swift/UVOT (Section~\ref{swift_data}). These galaxies represent a subset of the PHANGS-\jwst\ \citep{lee23} sample, which coincide exactly with the PHANGS-MUSE \citep{emsellem22} sample (Section~\ref{phangs_data}). Therefore all galaxies in this sample have both mid-IR data from \jwst/MIRI and optical integral field spectroscopic (IFS) data from VLT/MUSE. We also use \spitzer/IRAC data for each galaxy from the \spitzer\ Survey of Stellar Structure in Galaxies \citep[S4G;][]{sheth10} (except NGC2835, see Section~\ref{spitzer_data}).
The UVOT and MIRI data provide constraints on the 2175\AA\ absorption feature and PAH features, respectively.
The MUSE data provide constraints on various ISM properties and the IRAC data are used for correcting the MIRI data for stellar continuum and to derive stellar mass maps. 
A gallery of the data for each galaxy is shown in Figure~\ref{fig:sample_demo}. 
A list of the galaxies and their properties is shown in Table~\ref{tab:properties}. All photometry and emission line measurements are corrected for foreground MW dust extinction by using the Galactic dust reddening maps\footnote{\url{https://irsa.ipac.caltech.edu/applications/DUST/}} and assuming the average MW extinction curve from \cite{fitzpatrick19}.

\begin{figure*}
\includegraphics[width=0.99\textwidth]{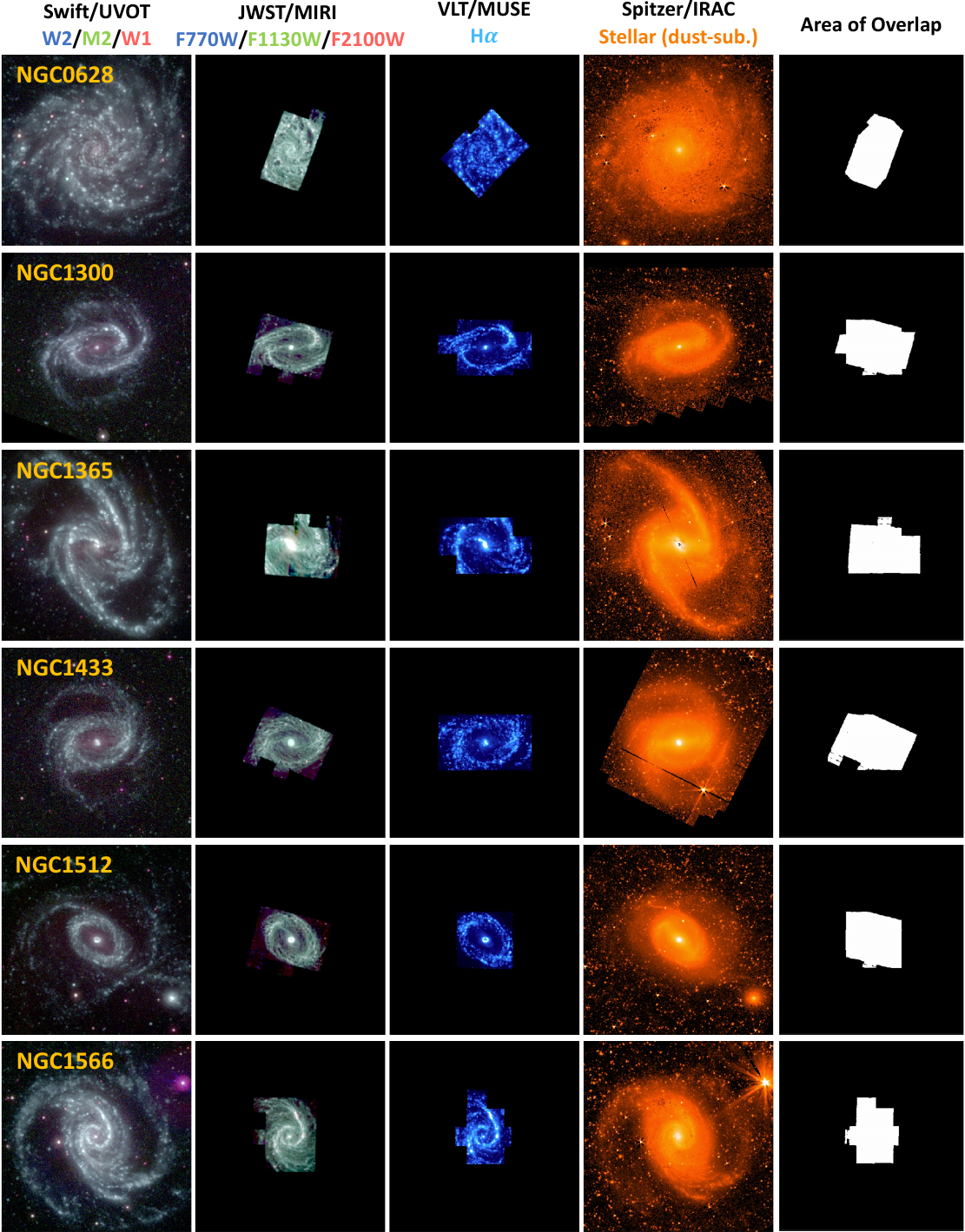}
\caption{Gallery of data used in our study. For each galaxy we show a 10\arcmin$\times$10\arcmin\ postage stamp of the Swift/UVOT RGB composite, \jwst/MIRI RGB composite, VLT/MUSE \halpha, and \spitzer/IRAC 3.6~$\mu$m (dust-corrected), and the area of mutual overlap (limited by MIRI and MUSE data). All images are log-scale. Our main analysis is restricted to the region of overlap between the datasets. (figure continues on the next page)
 \label{fig:sample_demo}}
\end{figure*}

\begin{figure*}\ContinuedFloat 
\includegraphics[width=0.98\textwidth]{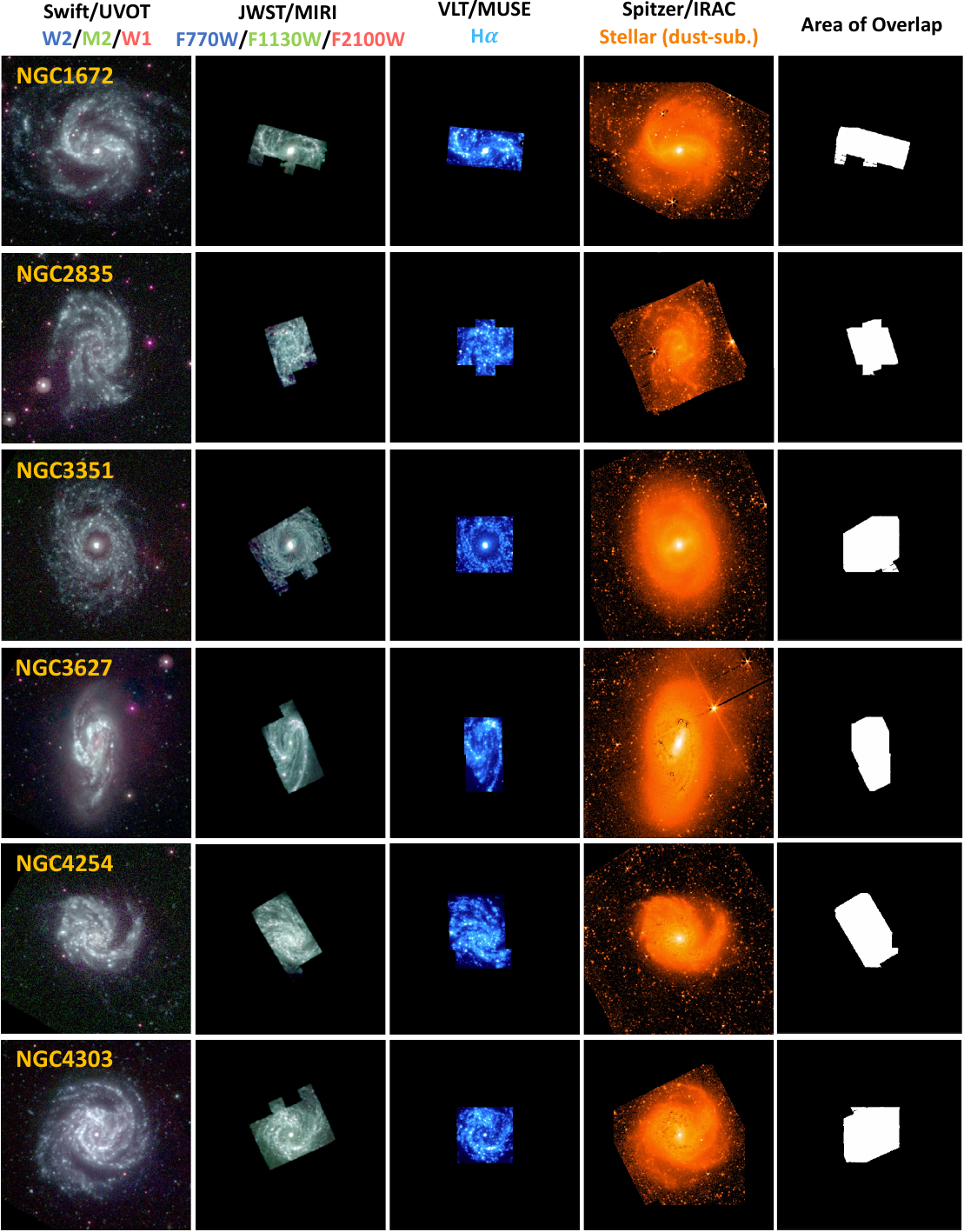}
\caption{(continued figure) 
 \label{fig:sample_demo_2}}
\end{figure*}

\begin{figure*}\ContinuedFloat 
\includegraphics[width=0.98\textwidth]{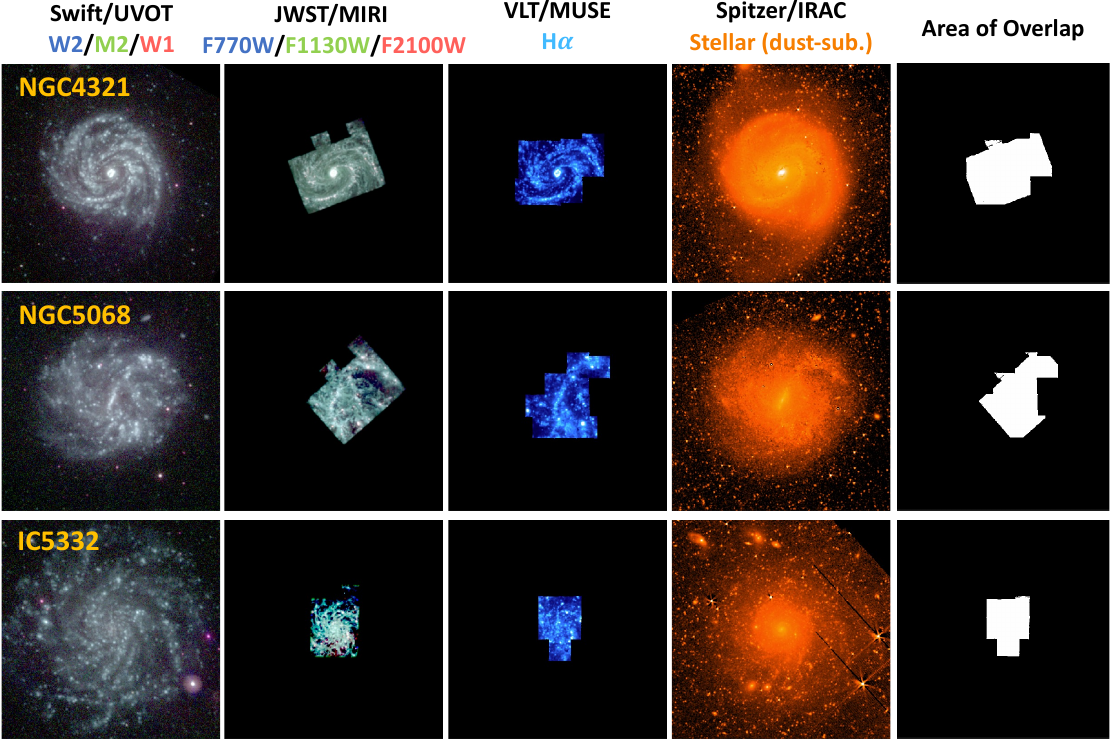}
\caption{(continued figure) 
 \label{fig:sample_demo_3}}
\end{figure*}

\setlength{\tabcolsep}{5pt} 
\begin{table*}
\begin{threeparttable}
\caption{Galaxy sample with \swift/UVOT, VLT/MUSE, and \jwst/MIRI data}
\label{tab:properties}
\begin{tabular}{lcccccccc}
\toprule
\headrow Name & Dist. & log($M_\star$) & log(SFR) & $i$ & $\sigma_i$  & $E(B-V)_\mathrm{MW}$ & $\langle E(B-V)_{\mathrm{gas}}\rangle$ & Hubble Type/Morphology \\
 & [Mpc] & [$M_\odot$] & [$M_\odot$ yr$^{-1}$] & [deg] & [deg] &  [mag] &  [mag] & \\
\midrule
NGC 0628 & 9.8 & 10.34 & 0.24 & 8.9 & 12.2 &  0.0607 & 0.18 & SA(s)c \\
NGC 1300 & 19.0 & 10.62 & 0.07 & 31.8 & 6.0 & 0.0260 & 0.23 & SB(rs)bc \\
NGC 1365 & 19.6 & 10.99 & 1.23 & 55.4 & 6.0 & 0.0176 & 0.21 & SB(s)b \\
NGC 1433 & 18.6 & 10.87 & 0.05 & 28.6 & 6.0 & 0.0078 & 0.15 & (R')SB(r)ab \\
NGC 1512 & 18.8 & 10.71 & 0.11 & 42.5 & 6.0 & 0.0091 & 0.15 & SB(r)a \\
NGC 1566 & 17.7 & 10.78 & 0.66 & 29.5 & 10.7 & 0.0078 & 0.21 & SAB(s)bc \\
NGC 1672 & 19.4 & 10.73 & 0.88 & 42.6 & 6.0 & 0.0200 & 0.21 & SB(s)b \\
NGC 2835 & 12.2 & 10.00 & 0.09 & 41.3 & 5.3 & 0.0030 & 0.11 & SB(rs)c \\
NGC 3351 & 10.0 & 10.36 & 0.12 & 45.1 & 6.0 & 0.0238 & 0.21 & SB(r)b \\
NGC 3627 & 11.3 & 10.83 & 0.58 & 57.3 & 1.0 & 0.0287 & 0.37 & SAB(s)b \\
NGC 4254 & 13.1 & 10.42 & 0.49 & 34.4 & 0.9 & 0.0334 & 0.30 & SA(s)c \\
NGC 4303 & 17.0 & 10.52 & 0.73 & 23.5 & 9.2 & 0.0052 & 0.27 & SAB(rs)bc \\
NGC 4321 & 15.2 & 10.75 & 0.55 & 38.5 & 2.4 & 0.0052 & 0.31 & SAB(s)bc \\
NGC 5068 &  5.2 & 9.40 & -0.56 & 35.7 & 10.9 & 0.0022 & 0.09 & SAB(rs)cd \\
IC 5332  & 9.0 &  9.67 & -0.39 & 26.9 & 6.0 & 0.0144 & 0.05 & SA(s)d \\

\bottomrule
\end{tabular}
\begin{tablenotes}[hang]
\item[] Values in the first six columns are from \cite{emsellem22} (based on \citealt{leroy21} and \citealt{lang20}). $E(B-V)_\mathrm{MW}$ are from NASA/IPAC IRSA Galactic dust reddening maps. $\langle E(B-V)_{\mathrm{gas}}\rangle$ are the median gas reddening (derived from the Balmer decrement; see Section~\ref{method_EBV_gas}) in star forming regions (see Section~\ref{selection_cuts} criteria (1)-(3)). Hubble Type/Morphology are from the NASA/IPAC Extragalactic Database.
\end{tablenotes}
\end{threeparttable}
\end{table*}

\subsection{\swift/UVOT}\label{swift_data}
The \textit{Neil Gehrels Swift Observatory} \citep{gehrels04} can observe gamma-ray, X-ray, UV, and optical wavebands, but our focus will be to use UV data from its UVOT instrument \citep{roming05}. UVOT has a field of view of 17\arcmin$\times$17\arcmin, a spatial resolution of 2.5\arcsec, and provides UV observations in three filters, UVW2 ($\lambda_\mathrm{eff}=0.1991\mu$m; $\mathrm{FWHM}=0.0657\mu$m), UVM2 ($\lambda_\mathrm{eff}=0.2221\mu$m; $\mathrm{FWHM}=0.0498\mu$m), and UVW1 ($\lambda_\mathrm{eff}=0.2486\mu$m; $\mathrm{FWHM}=0.0693\mu$m) \citep{poole08, decleir19}. 
These three filters are ideally suited to study the 2175\AA\ feature in the local universe because the two `wide' filters (UVW2 and UVW1) lie off the feature and the medium filter (UVM2) lies on top of the feature. The UVOT filters have been used for the purpose of measuring the 2175\AA\ feature by numerous studies \citep[e.g.,][]{hoversten11, hagen17, decleir19, decleir19PhDT, ferreras21, wangY22, zhou23, belles23}. 

We retrieve all \swift/UVOT data from the NASA High Energy Astrophysics Science Archive Research Center (HEASARC) service\footnote{\url{https://heasarc.gsfc.nasa.gov/docs/archive.html}}. \swift\ has a large focus on transient science (e.g., gamma ray bursts, supernovae) and many of these galaxies were observed during supernovae events. We exclude all observations of galaxies that coincide within three months after a supernovae event. 

Data were reduced and mosaiced using the publicly available \texttt{DRESSCode}\footnote{\url{https://spacetelescope.github.io/DRESSCode/}} (Data Reduction
of Extended Swift Sources Code – Decleir et al., in prep.). \texttt{DRESSCode} is an automated pipeline that executes the different steps of the data reduction to all UVOT images. The code uses several tasks from the HEASoft software\footnote{\url{https://heasarc.gsfc.nasa.gov/docs/software/heasoft/}}, and has been optimized for extended sources. The first version of this pipeline is described in detail in \cite{decleir19}, where it was used to reduce UVOT images of NGC~628. An updated version of the pipeline is explained and demonstrated in detail in chapter 2 of \cite{decleir19PhDT}. Since then, additional updates have been made, mostly to enhance the efficiency and flexibility of the pipeline. The latest version of the code will be described in a dedicated paper by Decleir et al. (in prep.). Here, we summarize the different steps of the current version of the pipeline as it was used to reduce the images of our sample.

Raw data and calibration files were retrieved from the HEASARC Archive. First, the \texttt{DRESSCode} converts the raw data into ``sky" images, adding World Coordinate System (WCS) coordinates to the images. Then, aspect corrections are calculated and applied to the images to enhance the accuracy of the astrometry. Subsequently, the pipeline performs flux corrections to account for: 1) coincidence loss, 2) large-scale sensitivity variations, and 3) time-dependent sensitivity loss. Once these corrections have been applied to all individual frames, they are co-added (summed) per UVOT filter. Finally, the combined images are converted to flux density units (Jy) using the appropriate calibration factors, and an (inverse) aperture correction is applied to account for the fact that these calibration factors were determined for apertures with a 5\arcsec\ radius. We refer the reader to \cite{decleir19} and \cite{decleir19PhDT} for more details.

We manually crop the final reduced images to remove the outer edges with low exposure time where the noise is significantly higher. This provides us with more reliable sky regions to estimate the background level for subtraction and error estimation. A list of the total exposure times for the UVOT data for each galaxy are shown in Table~\ref{tab:swift_exposure}.

\begin{table}
\begin{threeparttable}
\caption{\swift/UVOT Exposure Times}
\label{tab:swift_exposure}
\begin{tabular}{lrrr}
\toprule
\headrow Name & $t_\mathrm{exp}$(UVW2) & $t_\mathrm{exp}$(UVM2) & $t_\mathrm{exp}$(UVW1) \\
 & [s]\hspace{3mm} & [s]\hspace{3mm} & [s]\hspace{3mm} \\
\midrule
NGC 0628 &  17035 &  23162 &  14464 \\ 
NGC 1300 &   8752 &  13307 &   8495 \\ 
NGC 1365 &  20331 &  59554 &  23046 \\ 
NGC 1433 &   4542 &   4044 &   3466 \\ 
NGC 1512 &  18719 &  18375 &  16901 \\ 
NGC 1566 &  14216 &  10645 &  10547 \\ 
NGC 1672 &   9335 &   8317 &   7570 \\ 
NGC 2835 &   4906 &   4457 &   4083 \\ 
NGC 3351 &   5219 &   4569 &   2875 \\ 
NGC 3627 &  13112 &  28393 &   9694 \\ 
NGC 4254 &   6016 &   2132 &   6006 \\ 
NGC 4303 &   7654 &  69934 &   4825 \\ 
NGC 4321 &   8654 &   8983 &   5287 \\ 
NGC 5068 &   7346 &   5663 &   6162 \\ 
IC 5332  &   4175 &   4019 &   3621 \\ 

\bottomrule
\end{tabular}
\begin{tablenotes}[hang]
\item[]Each exposure time corresponds to the maximum co-added depth, which typically covers the entire galaxy (FoV=17\arcmin$\times$17\arcmin).
\end{tablenotes}
\end{threeparttable}
\end{table}

\subsection{PHANGS-\jwst\ and PHANGS-MUSE}\label{phangs_data}
The Physics at High Angular resolution in Nearby GalaxieS (PHANGS)–\jwst\ survey \citep{lee23} is a Cycle 1 \jwst\ Large Treasury Program (GO 2107) to obtain NIRCam and MIRI imaging of 19 nearby galaxies from the PHANGS-MUSE survey (detailed below). For this project we use only the MIRI data from F770W ($\lambda_\mathrm{eff}=7.528\mu$m), F1130W ($\lambda_\mathrm{eff}=11.298\mu$m), and F2100W ($\lambda_\mathrm{eff}=20.563\mu$m). These three filters are ideally suited to study the strong PAH features at 7.7 and 11.3~$\mu$m because the F770W and F1130W filters lie on these features and the F2100W filter provides the baseline for the warm dust continuum emission \citep[e.g.,][]{chastenet23a, sutter24}. We use the publicly available reduced MIRI data from the PHANGS team\footnote{\url{https://sites.google.com/view/phangs/home/data}}, which is stored by the Canadian Advanced Network for Astronomical Research (CANFAR)\footnote{\url{https://www.canfar.net/storage/vault/list/phangs/RELEASES/PHANGS-JWST}}.

PHANGS-MUSE is a Large Program using the VLT/MUSE that obtained optical IFS data, spanning 470-935~nm ($R\sim1800-3600$), for 19 nearby galaxies. We also use the publicly available MUSE line maps produced by the PHANGS team\footnote{\url{https://www.canfar.net/storage/vault/list/phangs/RELEASES/PHANGS-MUSE}}, which are described in \cite{emsellem22}. These provide the following emission line maps: \hbeta, [OIII]$\lambda$4959, [OIII]$\lambda$5007, [NII]$\lambda$6548, \halpha, [NII]$\lambda$6584, [SII]$\lambda$6717, and [SII]$\lambda$6731. 

 Our analysis will focus on the region of overlap between the PHANGS-\jwst\ and PHANGS-MUSE data, where the former was designed to maximise overlap with the latter while also allowing flexibility for scheduling (see Figure~1 from \citealt{lee23}). The area of overlap is shown in Figure~\ref{fig:sample_demo}, and typically covers a $\sim$3-4\arcmin\ wide region in the centre of each galaxy.

\subsection{\spitzer/IRAC}\label{spitzer_data}
The reduced \spitzer/IRAC ch1 ($\lambda_\mathrm{eff}=3.550\mu$m) and ch2 ($\lambda_\mathrm{eff}=4.493\mu$m) mosaics were produced by the S4G survey \citep{sheth10} and retrieved from the NASA/IPAC Extragalactic Database\footnote{\url{https://ned.ipac.caltech.edu/}}, with the exception of NGC2835 which was taken as part of pid 14033 (PI: J.C. Mu\~noz-Mateos). Reduced mosaics for NGC2835 were provided by M. Querejeta (priv comm.) and were observed and reduced using a similar strategy to S4G \citep{querejeta21}. 

\subsection{Foreground Milky Way Star Masks}
We construct masks of foreground Milky Way (MW) stars based on the \swift\ UVW1 images. These are constructed using the \texttt{PTS-7/8} software \citep{verstocken20}, the Python Toolkit for SKIRT, the radiative transfer code (\citealt{camps&baes15} and \citealt{camps&baes20}). Initially, the software retrieves the source catalogue from the 2MASS all-sky catalogue of point sources (\citealt{cutri03}). It then subtracts the background level of a small patch surrounding each point. \texttt{PTS} identifies a local peak as a foreground star and creates masks around it if it satisfies two conditions: (1) the local peak within the small patch is three times brighter than the background and (2) its coordinate matches that of a 2MASS point source (see \citealt{clark18}, \citealt{decleir19}, and \citealt{decleir19PhDT}).

\subsection{Image Resampling}\label{image_resampling}
\begin{figure*}[hbt!]
\centering
\includegraphics[width=0.95\linewidth]{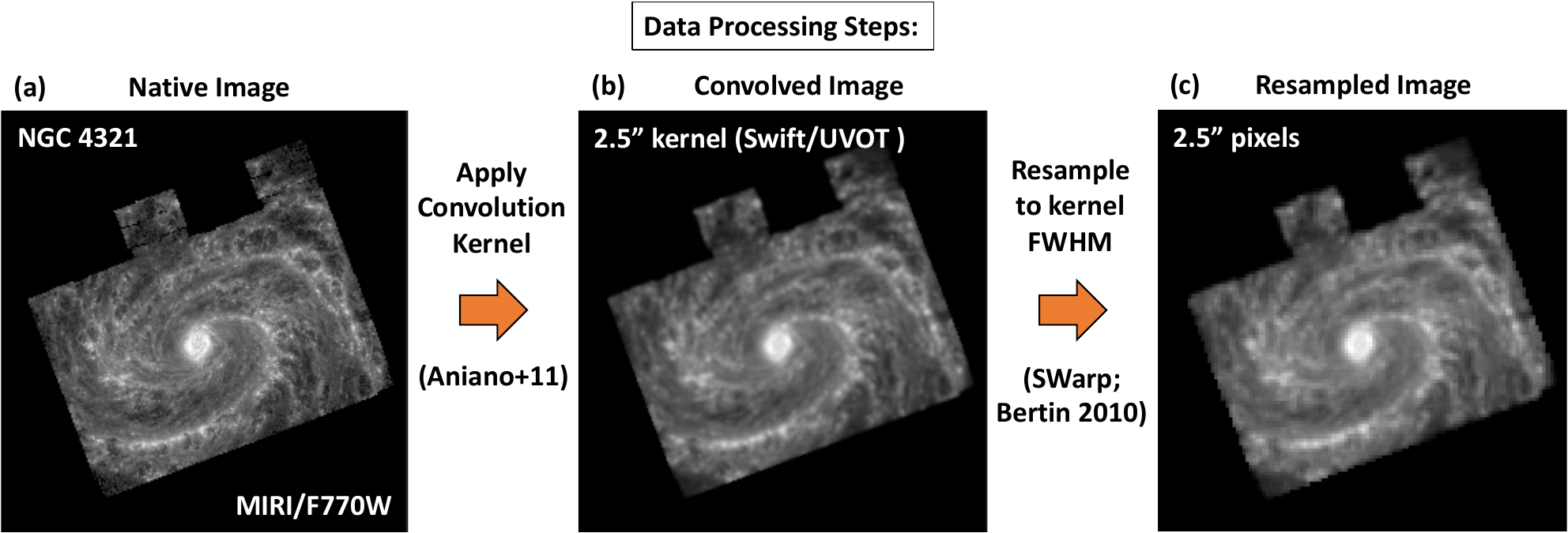}
\caption{The data processing workflow to enable consistent photometric and spectroscopic comparison. (\textbf{(a)} $\Rightarrow$ \textbf{(b)}) We start with a fully-reduced and calibrated image or line map at its native resolution and convolve it with a 2.5\arcsec\ Gaussian kernel (Swift/UVOT PSF), which is the largest PSF among the data, using the techniques and kernels available from \citet{aniano11}. (\textbf{(b)} $\Rightarrow$ \textbf{(c)}) Next, the convolved images are resampled to a pixel size of 2.5\arcsec\ using the \texttt{SWarp} software \citep{bertin10}. The panels show this process for MIRI/F770W data of NGC~4321. All data from Swift, \jwst, VLT, and \spitzer\ were convolved and resampled to the same 2.5\arcsec\ grid.} 
\label{fig_resampling}
\end{figure*}

In order to make self-consistent comparisons between the different datasets, the data are convolved and resampled to the Swift/UVOT point-spread function (PSF) because it has the lowest resolution among the datasets. This is done in two steps for each image and line map (i.e., MUSE data). First the data are convolved to match the Swift/UVOT resolution, which is approximated as a 2.5\arcsec\ Gaussian kernel\footnote{\url{https://swift.gsfc.nasa.gov/analysis/uvot_digest/psf.html}}, using the techniques and kernels available from \citet{aniano11}. Second, the data are resampled to 2.5\arcsec\ pixels, using the \texttt{SWarp} software \citep{bertin10} and adopting \texttt{RESAMPLING\_TYPE=LANCZOS3}. Using a resampled pixel size equal to the convolved PSF ensures that the regions can be considered independent. A visual representation of this workflow is shown in Figure~\ref{fig_resampling}. For reference, the physical size of 2.5\arcsec\ ranges from 60--240~pc for the distance of our sample. 

\section{Methodology}\label{method}
\begin{figure*}[hbt!]
\centering
\includegraphics[width=0.95\linewidth]{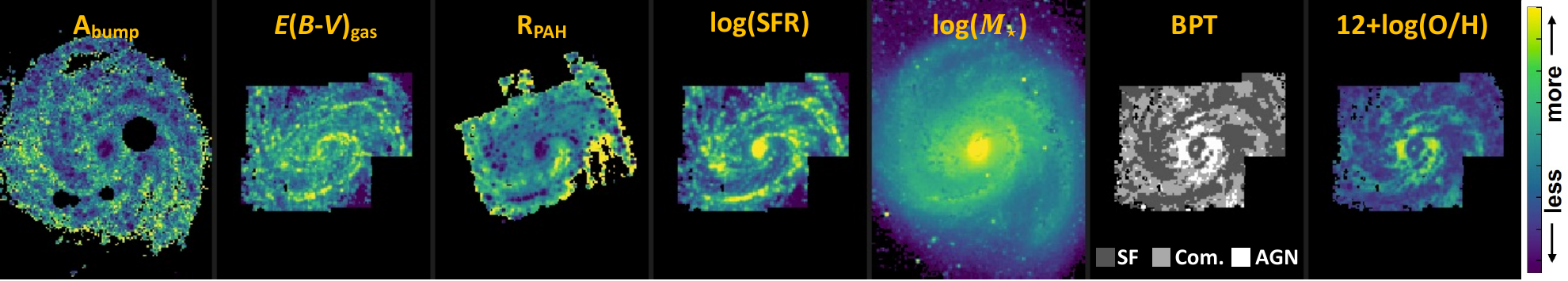}
\caption{Example of derived property maps for NGC~4321. From left to right, the 2175\AA\ strength \abump, ionised gas reddening \EBVgas, PAH abundance \rpah, log(SFR), log($M_\star$), BPT classifications, and gas-phase metallicity (using Scal). The methods used to derive each property are described in Section~\ref{method}. The holes of missing data in \abump\ correspond to regions masked due to foreground MW stars.} 
\label{fig_property_example}
\end{figure*}

In this section we describe the methods for quantifying the various physical properties of interest in this study. Examples of the property maps are shown in Figure~\ref{fig_property_example}.

\subsection{2175\AA\ Absorption Feature Strength - \abump}\label{method:abump}
\begin{figure}[hbt!]
\centering
\includegraphics[width=0.95\linewidth]{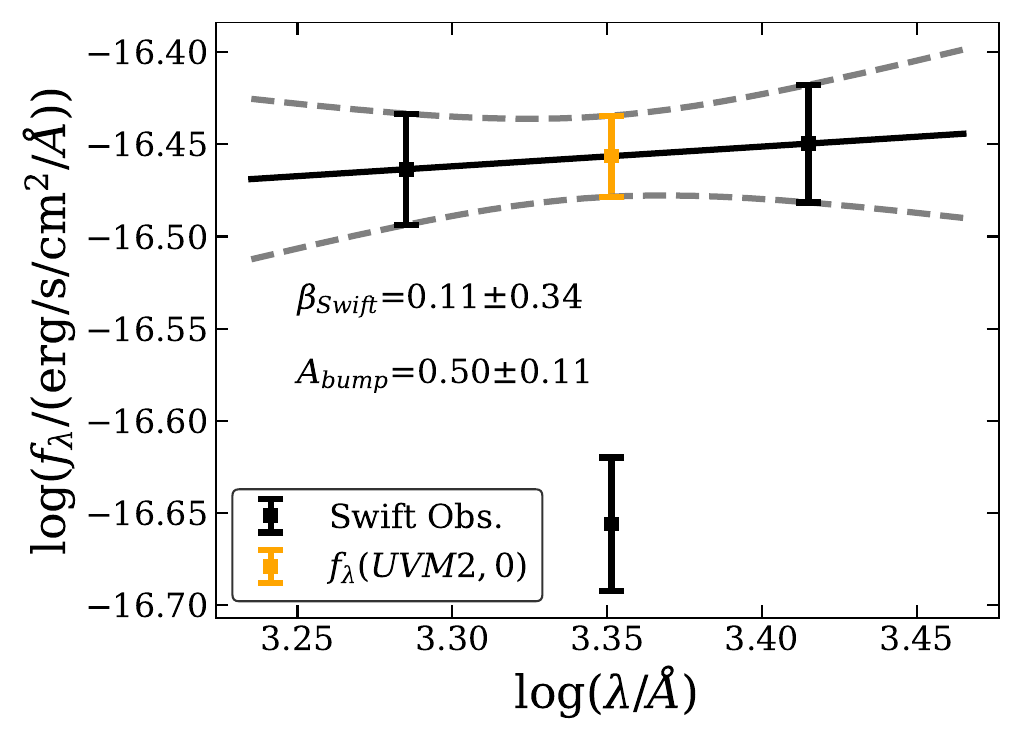}
\caption{Example of how the 2175\AA\ feature strength, \abump, is derived for each region based on the Swift data (black squares). We fit the UV continuum slope, $\beta_{\rm{Swift}}$ (black line), from the two off-feature Swift filters (UVW2 and UVW1) and determine the expected flux density at UVM2 (orange square). This value is then compared to the observed flux density at UVM2 (middle observation). The 1$\sigma$ uncertainty on the UV slope and normalisation is shown by dashed gray lines and accounted for in the uncertainty of \abump.}
\label{fig_Abump_example}
\end{figure}

The 2175\AA\ feature is quantified using a simple analytic combination of the three Swift/UVOT filters in three steps. First, the UV continuum slope, $\beta$, is measured from the two (off-feature) wide filters under the assumption that the UV flux density follows a power-law shape over the range $1250\le\lambda\le2600$~\AA\ \citep[e.g.,][]{calzetti94},
\begin{equation}
\begin{aligned}\label{eq:beta}
\log f_\lambda(\lambda)= C + \beta \times \log \lambda  \,,
\end{aligned}
\end{equation}
where $f_\lambda(\lambda)$ is the flux density, in units of erg~s$^{-1}$~cm$^{-2}$~\AA$^{-1}$, $\lambda$ is the wavelength in \AA, and $C$ is a constant normalisation term. 
Using the Swift filters for the UV slope, $\beta_{\rm{Swift}}$,
\begin{equation}
\begin{aligned}\label{eq:beta_swift}
\beta_{\rm{Swift}} = \frac{\log[f_\lambda(\mathrm{UVW2})/f_\lambda(\mathrm{UVW1})]}{\log[\lambda_{\mathrm{UVW2}}/\lambda_{\mathrm{UVW1}}]} \,.
\end{aligned}
\end{equation}
Second, we estimate the expected flux density for the UVM2 filter assuming this UV continuum slope, and corresponding to the expected value in the absence of a 2175\AA\ feature, 
\begin{equation}
\begin{aligned}\label{eq:UVM2int}
\log f_{\lambda}(\mathrm{UVM2},0) = C + \beta \times \log \lambda_{\rm{eff,UVM2}}  \,.
\end{aligned}
\end{equation}
Third, the 2175\AA\ feature strength, \abump, is estimated from the ratio of the observed flux density in the UVM2 filter relative to the value derived from the UV slope, 
\begin{equation}
\begin{aligned}\label{eq:Abump}
A_{\rm{bump}} = -2.5 \log (f_{\lambda}(\mathrm{UVM2})/f_{\lambda}(\mathrm{UVM2},0))  \,,
\end{aligned}
\end{equation}
where \abump\ has units of magnitudes. An example of these steps for an example pixel in our sample is shown in Figure~\ref{fig_Abump_example}. This formalism is effectively the same as that used in \cite{zhou23} for estimating the bump strength from the three \swift\ filters (see their Section~3.2).
The uncertainty on \abump\ is a combination of the uncertainty from the UV-slope and normalisation and the uncertainty of the observation at UVM2 (see Figure~\ref{fig_Abump_example}).

The apparent strength of the 2175\AA\ feature (i.e., \abump) will scale with the total amount of dust attenuation, with dustier regions exhibiting larger absorption features. Therefore, to characterise the intrinsic bump strength, it is necessary to normalise \abump\ with respect to the total attenuation in a given band (e.g., $A_V$ is commonly used) or with respect to the reddening $E(B-V)$. For the purpose of this paper we normalise with respect to the reddening to get the intrinsic bump strength, \kbump,
\begin{equation}
\begin{aligned}\label{eq:kbump}
k_{\rm{bump}} = A_{\rm{bump}}/E(B-V)  \,.
\end{aligned}
\end{equation}
The method for characterising the reddening is described in the next section. For reference, the MW extinction curve has an average $k_{\rm{bump}}\sim 3.3$ \citep{fitzpatrick99, salim&narayanan20}.

Finally, we note that the values of \abump\ (and \kbump) based on the UVOT filters will typically be an \textit{underestimate} of the true values that would be inferred based on a spectroscopic method. This occurs for two reasons: (1) the width of the UVM2 filter will suppress the peak amplitude of the feature and (2) the UVW2 and UVW1 filters have tails that extend into the 2175\AA\ feature such that they do not provide a completely clean baseline for the UV continuum. 
However, the UVW2 and UVW1 filters also have extended red tails (see Figure~\ref{fig_filters_vs_bump}) such that the measured UV continuum baseline can be higher than the true value for very red SEDs (e.g., old stellar populations). This second effect is not seen to significantly impact the measurements of the continuum baseline in our sample, based on SED modelling (e.g., see moderately reddened SED fit for an example region of NGC~4321 in Figure~\ref{fig_NGC1300_NGC4321_AV_compare}), and therefore we do not correct for this. This is likely because our selection criteria (Section~\ref{selection_cuts}) tend to restrict our analysis to star-forming regions with younger average stellar populations.
We detail the reliability of the \swift/UVOT filters to measure the UV slopes and 2175\AA\ feature in \ref{app:swift_reliability}. 

In summary, the true bump amplitude is expected to scale linearly with the value inferred from the UVOT filters such that the correlations observed and presented in this work should be robust but that the exact parameters of our fits should be treated with caution.

\subsection{Ionised gas Reddening - \EBVgas}\label{method_EBV_gas}
The 2175\AA\ absorption feature is measured relative to the stellar continuum, hence its strength is expected to be linked to the reddening on the stellar continuum, \EBVstar. However, accurately measuring the reddening on the stellar continuum is non-trivial due to the degeneracy in SED colour with stellar population age and typically requires full SED coverage from UV to IR to break this degeneracy via an energy balance assumption or through spectral modeling of the stellar continuum and absorption features. Performing such modeling is computationally expensive and also prone to uncertainty at the scales of our resampled pixels ($\sim$60--240~pc). This is due to the energy balance assumption beginning to break down on scales $\lesssim$1~kpc \citep[e.g.,][]{smith&hayward18}. 

Another reddening diagnostic is the amount of reddening on the ionised gas, \EBVgas, based on the Balmer decrement, $F$(\halpha)/$F$(\hbeta), and available from the VLT/MUSE data,
\begin{equation}
\begin{aligned}\label{eq:EBVgas}
E(B-V)_{\mathrm{gas}} = \frac{\log((F(\mathrm{H}\alpha)/F(\mathrm{H}\beta))/2.86)}{0.4(k(\mathrm{H}\beta)-k(\mathrm{H}\alpha))}  \,,
\end{aligned}
\end{equation}
where 2.86 is the theoretical value expected for the unreddened ratio of $F(\mathrm{H}\alpha)/F(\mathrm{H}\beta)$ undergoing Case B recombination with $T_{\mathrm{e}}=10^4$~K and $n_{\mathrm{e}}=100$~cm$^{-3}$ \citep{osterbrock89,osterbrock&ferland06}, and we assume an average MW extinction curve, $k(\lambda)$, at the wavelengths of \hbeta\ and \halpha, with $k(\mathrm{H}\beta)-k(\mathrm{H}\alpha)=1.160$ \citep{fitzpatrick19}. 

In nearby galaxies it is generally found that the stellar continuum experiences roughly half the amount of reddening relative to the ionised gas on average, \citep[\EBVratioavg$\ \sim0.5$; e.g.,][]{calzetti94, kreckel13, battisti16, emsellem22}. We measure this relationship for galaxies in our sample and found a similar trend (see \ref{app:reddening_compare}), indicating that using either \EBVgas\ or \EBVstar\ is reasonable to trace reddening in a region. 

For our analysis, we will normalise the bump strength using the amount of reddening from the ionised gas, \EBVgas, because it is empirical and (mostly) independent of assumptions (Case B recombination), which helps to avoid circularity issues. While this method relies on an assumed dust extinction curve for the ionised gas reddening, our choice of the MW curve should be reasonable given that the PHANGS sample consists primarily of massive spiral galaxies. Furthermore, the shape of the average MW, LMC, and SMC extinction curves are relatively similar at the optical wavelengths of \halpha\ and \hbeta\ (i.e., average $k(\mathrm{H}\beta)-k(\mathrm{H}\alpha)$ are within 15\%), and show the largest variations at UV wavelengths.
 
For completeness, we also performed the main analysis of this work using stellar continuum reddening ($A_{V,star}$) derived from \magphys\ SED modeling. These results are presented in \ref{app:reddening_compare} and show qualitatively the same results as we find when normalising by the ionised gas reddening. We attribute this to the fact that stellar and gas reddening show a relatively tight correlation at the spatial scales of our regions (see Figure~\ref{fig_all_AV_compare}).

\subsection{PAH Abundance - \rpah }
The abundance of PAHs is inferred using the proxy \rpah\ \citep{chastenet23a, sutter24}, but slightly modified to remove the contribution of emission from the stellar continuum,
\begin{equation}
\begin{aligned}\label{eq:RPAH}
R_\mathrm{PAH} = \frac{f_{\nu,\mathrm{dust}}(\mathrm{F770W})+f_{\nu,\mathrm{dust}}(\mathrm{F1130W})}{f_{\nu,\mathrm{dust}}(\mathrm{F2100W})}  \,,
\end{aligned}
\end{equation}
where $f_{\nu,\mathrm{dust}}(\lambda)$ is the dust-only flux density for a particular filter in units of erg~s$^{-1}$~cm$^{-2}$~Hz$^{-1}$. 
The dust-only emission is determined by subtracting the stellar continuum using the \spitzer/IRAC ch1 data by assuming the stellar emission follows a blackbody,
\begin{equation}
\begin{aligned}\label{eq:f_dust}
f_{\nu,\mathrm{dust}}(\lambda) = f_{\nu}(\lambda) - \frac{B_\nu(\lambda,T_{\star,eff})}{B_\nu(\mathrm{IRAC1},T_{\star,eff})} f_{\nu}(\mathrm{IRAC1}) \,.
\end{aligned}
\end{equation}
where $B_\nu(\lambda,T_{\star,eff})$ is the blackbody function and $T_{\star,eff}$ is the effective temperature of the stellar population. We assume $T_{\star,eff}=5000$~K, which is representative of local star-forming galaxies \citep{draine07}. For reference, the blackbody flux ratio in Eq~(\ref{eq:f_dust}) assuming $T_{\star,eff}=5000$~K is 0.276, 0.133, and 0.042 for F770W, F1130W, and F2100W, respectively. For the regions analysed in this study (selection described in Section~\ref{selection_cuts}), the median fraction of emission removed is $0.056^{+0.037}_{-0.021}$, $0.019^{+0.012}_{-0.007}$, and $0.012^{+0.009}_{-0.005}$ for F770W, F1130, and F2100W, respectively.
This approach is roughly similar to the method used in \cite{sutter24} for PHANGS-JWST, who remove the stellar continuum using SED-fitting of the \jwst/NIRCam+\jwst/MIRI data. We choose to use IRAC instead of NIRCam data because we adopt stellar mass maps derived from IRAC data (see Section~\ref{method:stellar_mass}) and therefore we consider this approach to be more self-consistent. 

We note that the 7.7$\mu$m/11.3$\mu$m ratio changes in galaxies with different ionising spectra \citep[e.g.,][]{draine21}, which may impact the values of \rpah\ in different local environments of galaxies. We explored this by using only single PAH-feature measurements (e.g., F770W/F2100W), but found that the results are qualitatively consistent with those found using \rpah.

\subsection{Star Formation Rate - SFR}\label{method:sfr}
Star formation rates (SFRs) for individual regions are derived from extinction-corrected \halpha, 
\begin{equation}
\begin{aligned}\label{eq:ha_corr}
F(\mathrm{H}\alpha)_\mathrm{corr}=F(\mathrm{H}\alpha) 10^{(0.4 k(\mathrm{H}\alpha)E(B-V)_{\mathrm{gas}})} \,,
\end{aligned}
\end{equation}
where we assume an average MW extinction curve \citep{fitzpatrick19} for $k(\mathrm{H}\alpha)$ and \EBVgas\ is derived following Eq~(\ref{eq:EBVgas}). We convert this to a luminosity based on the luminosity distance (see Table~\ref{tab:properties}) and use the conversion from \cite{calzetti13}, which assumes a \cite{kroupa01} initial mass function (IMF),
\begin{equation}
\begin{aligned}\label{eq:sfr}
\mathrm{SFR}(M_\odot~\mathrm{yr}^{-1}) = 5.5\cdot 10^{-42} L(\mathrm{H}\alpha)_\mathrm{corr} \,,
\end{aligned}
\end{equation}
where the \halpha\ luminosity is measured in erg~s$^{-1}$.  

\subsection{Stellar Mass - $M_\star$}\label{method:stellar_mass}
When available, we use the Independent Component Analysis (ICA) data products produced by the S4G Pipeline 5\footnote{\url{https://irsa.ipac.caltech.edu/data/SPITZER/S4G/overview.html}} \citep[\texttt{P5};][]{querejeta15}. The ICA method separates the emission from old stars and dust that contribute to the observed IRAC ch1 (3.6$\mu$m) flux. \cite{querejeta15} find that as much as 10\%–30\% of the total 3.6$\mu$m flux can be  contributed by dust, with larger fractions occuring for  galaxies with higher specific-SFR (sSFR=SFR/$M_\star$). We use the \texttt{P5} stellar emission maps (e.g., \texttt{NGCXXXX.stellar.fits}) together with eq~(6) in \cite{querejeta15} to estimate the stellar mass contained in a resampled pixel. We adopt a mass-to-light ratio $M/L=0.6$ as recommended by the authors. We note that this mass-to-light ratio assumes a \citet{chabrier03} IMF, which differs slightly from a \citet{kroupa01} IMF (used for SFRs), but that the impact on $M_\star$ and SFR values between these IMFs are very minor \citep[e.g.,][]{speagle14}. 

Three of the S4G galaxies do not have ICA products available (NGC1433, 1512, and IC5332), indicating they have IRAC colours consistent with minimal dust contamination. This is further supported by the weak overall PAH emission in the \jwst/MIRI data for these galaxies. 
For these galaxies, we first estimate the $M/L$ for each region using eq~(7) in \cite{querejeta15}, and then use eq~(6) in \cite{querejeta15} assuming the original IRAC data are representative of the stellar-only emission. 
For NGC2835, we use eq~(8) in \cite{querejeta15} because it was not run through the S4G \texttt{P5}. For the sample with ICA products, we find the ICA-based stellar masses show good agreement (within $\sim$0.1dex) with those derived from eq~(8) in \cite{querejeta15} (by design), such that this should not significantly affect our analysis. 

\subsection{Ionisation Source Classification}
The assumed intrinsic value for the Balmer decrement assumes conditions appropriate for star-forming HII regions of a galaxy. Therefore, we restrict our analysis to regions that are classified as star-forming using the standard Baldwin, Phillips \& Terlevich \citep[BPT;][]{baldwin81} diagram and using the demarcation lines from \cite{kewley01} and \cite{kauffmann03c}. We require that all emission lines ([OIII]/\hbeta\ vs. [NII]/\halpha) have $S/N\ge3$ per region for determining a classification. This requirement does not significantly restrict our sample that satisfies our photometric requirements (conditions (1) and (2) Section~\ref{selection_cuts}), with 73\%-100\% (median 87\%) of regions in each galaxy also satisfying this emission line $S/N$ condition.

\subsection{Gas-phase Metallicity}
We adopt the $Scal$ prescription of \citet{pilyugin&grebel16}, which is the preferred method of the PHANGS team \citep{kreckel19, groves23}, to estimate the gas-phase metallicity. This prescription uses a combination of the following line ratios:
\begin{align}\label{eq:another}
\begin{split}
\mathrm{N}_2 &= (\mathrm{[NII]}\lambda 6548 + \lambda 6584) / \mathrm{H}\beta \,,\\
\mathrm{S}_2 &= (\mathrm{[SII]}\lambda 6717 + \lambda 6731) / \mathrm{H}\beta \,,\\
\mathrm{R}_3 &= (\mathrm{[OIII]}\lambda 4959 + \lambda 5007) / \mathrm{H}\beta \,.
\end{split}
\end{align}
These lines are extinction corrected assuming a MW extinction curve \citep{fitzpatrick19} and \EBVgas\ derived following Eq~(\ref{eq:EBVgas}). We require that all emission lines have $S/N\ge3$ per region to estimate a metallicity for a region. This requirement is very similar to above because it uses many of the same emission lines, with the exception of [SII]. We find 73\%-100\% (median 86\%) of regions in each galaxy that satisfy the photometric requirements also satisfy this condition.

\subsection{Surface Area}
In order to compare resolved regions of galaxies at different distances in a fair manner, we normalize stellar masses and SFRs by the surface area of the resampled regions. The surface area depends on the distance of the galaxy and the inclination according to:
\begin{equation}
\begin{aligned}\label{eq:surface_area}
Area [kpc^2] = \frac{(\theta_\mathrm{region} D_\mathrm{Lum}/206265)^2}{\cos(i)} \,,
\end{aligned}
\end{equation}
where we are assuming the small angle approximation and an infinitely thin disk, $\theta_\mathrm{region}$ is in arcsec (2.5\arcsec\ in our case), $D_\mathrm{Lum}$ is in kpc, and $i$ is the inclination angle. The adopted distance and inclination values for each galaxy are listed in Table~\ref{tab:properties}.

\subsection{Selection Cuts}\label{selection_cuts}
We select a robust sample of star-forming regions by requiring that the following conditions are met for each resampled region:
\begin{enumerate}
\item[(1)] All \swift\ and \jwst\ photometry are $S/N \ge 5$
\item[(2)] Uncontaminated by MW foreground stars
\item[(3)] Classified as `star-forming' on the BPT diagram
\item[(4)] $\sigma(k_{\mathrm{bump}}) < 0.5$
\end{enumerate}

The value of $\sigma(k_{\mathrm{bump}})$ is determined from propagating the uncertainty of the two free parameters in $k_{\mathrm{bump}}$ (i.e., $\sigma(A_{\mathrm{bump}})$ and $\sigma(E(B-V)_{\mathrm{gas}})$).
Condition (4) is imposed to restrict the analysis to regions with reliable bump measurements. The primary factor affecting the bump measurement accuracy is the \swift\ depth, however we note that the sample with the lowest uncertainties on $k_\mathrm{bump}$ do not directly correspond to the deepest \swift\ data (see Table~\ref{tab:swift_exposure}). This is because $k_\mathrm{bump}$ also depends on the uncertainty in reddening (\EBVgas), which depends on the Balmer lines from the VLT/MUSE data. At a fixed $S/N$ value for the emission lines, regions with lower total reddening will have larger uncertainties on $k_\mathrm{bump}$ (i.e., if $A_\mathrm{bump}>0$, you get $k_{\rm{bump}}\rightarrow\infty$ as $E(B-V)_{\mathrm{gas}}\rightarrow0$). As a result, condition (4) limits our analysis to regions with $E(B-V)_{\mathrm{gas}} \gtrsim 0.07$. An example where the reddening term is the dominant source of uncertainty is IC5332, which has a median \EBVgas\ of 0.05 and a majority of its star-forming regions do not satisfy condition (4). Therefore, we do not report on correlation strengths for IC5332 in the subsequent analysis.
An example of the property maps after these selection cuts are shown in Figure~\ref{fig_property_HII_region}.

\begin{figure*}
\centering
\includegraphics[width=0.95\linewidth]{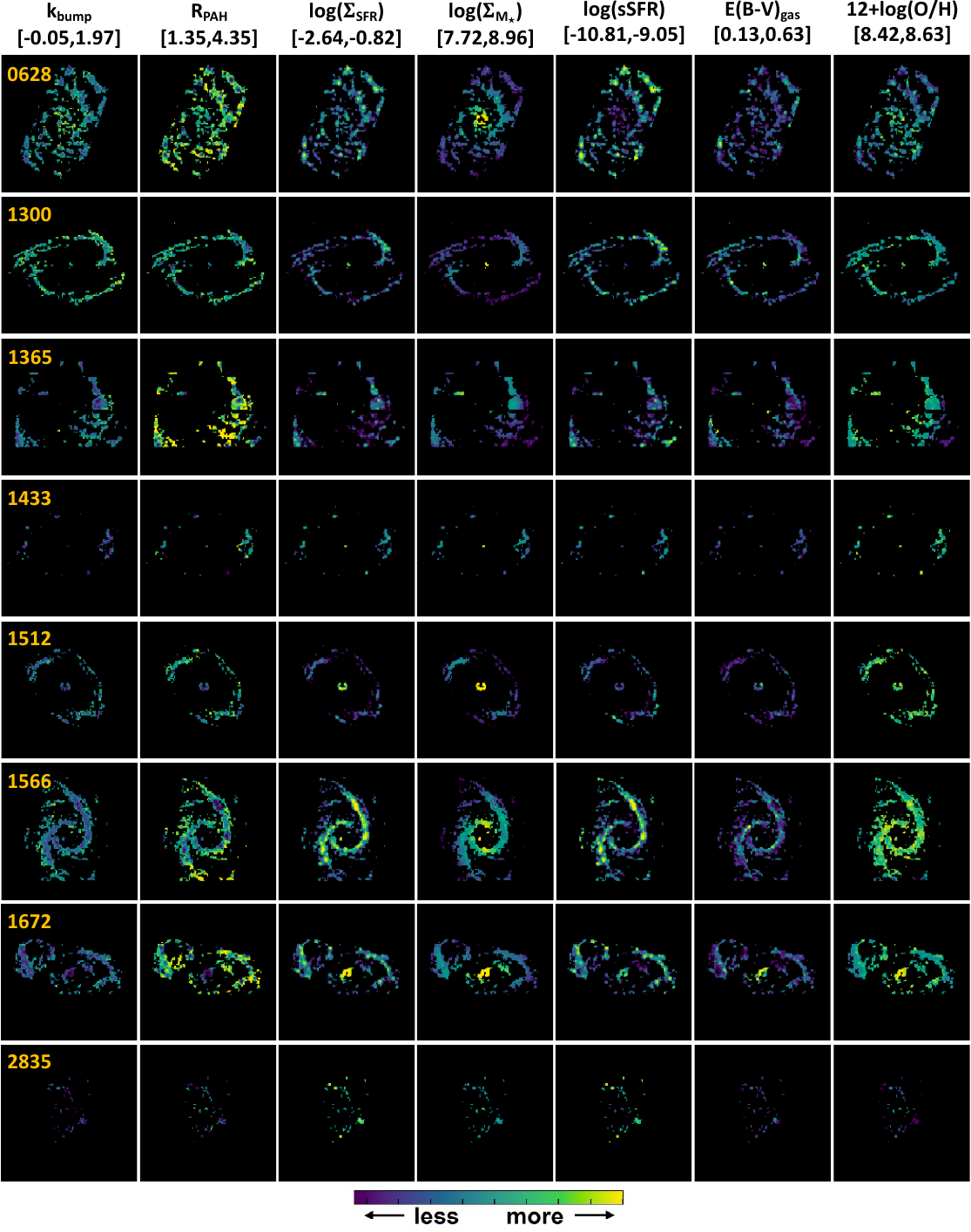}
\vspace{-1.5mm}
\caption{Derived property maps for regions that satisfy the selection cuts described in Section~\ref{selection_cuts}. Each property (i.e., column) uses the same colour-scale range, covering 2.5\%-97.5\% of the full distribution (see brackets at top). From left to right: the intrinsic 2175\AA\ strength (\kbump), PAH abundance (\rpah), SFR surface density (log($\Sigma_\mathrm{SFR}$)), stellar mass surface density (log($\Sigma_{\mathrm{M}\star}$)), specific-SFR ($\mathrm{sSFR}=\log \mathrm{SFR}-\log(M_\star)$), ionised gas reddening (\EBVgas), and gas-phase metallicity ($12+\log(\mathrm{O/H})$; using Scal). A positive correlation between \kbump\ and \rpah\ is evident, which provides support for PAHs as a potential carrier of the bump. Negative correlations between \kbump\ and \rpah\ with log($\Sigma_\mathrm{SFR}$) and sSFR are evident, which may indicate that small dust grains are being destroyed by ionising photons from massive stars. No significant correlations with log($\Sigma_{\mathrm{M}\star}$), \EBVgas, or $12+\log(\mathrm{O/H})$ are evident. There are relatively few regions in NGC~1433, 2835, and IC5332 after our selection cuts, which is due to shallower UV coverage and/or low reddening for these galaxies. (figure continues on the next page)} 
\label{fig_property_HII_region}
\end{figure*}

\begin{figure*}\ContinuedFloat 
\includegraphics[width=0.95\textwidth]{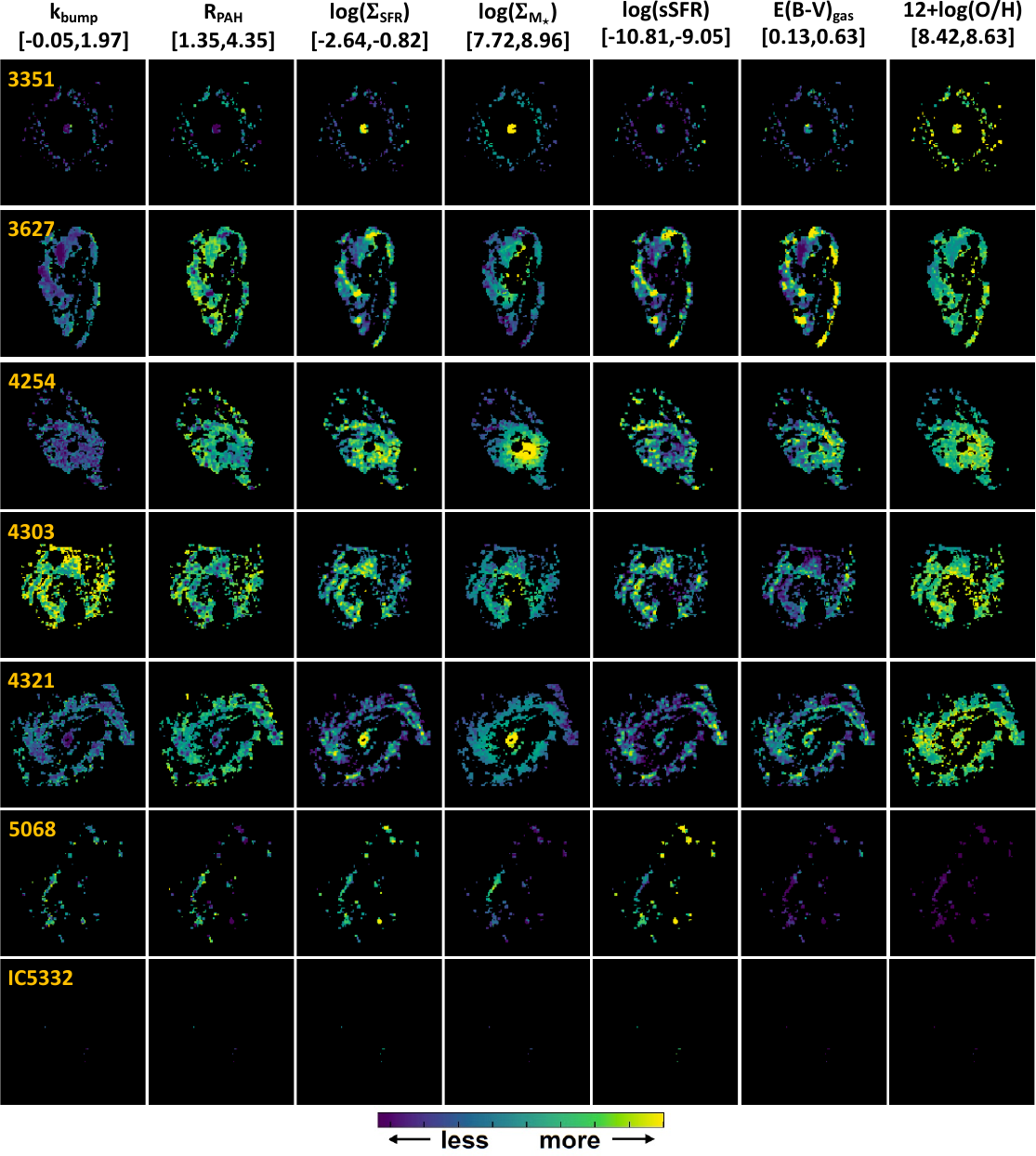}
\caption{(continued figure) 
 \label{fig_property_HII_region2}}
\end{figure*}

\section{Results}\label{results}
\subsection{\kbump\ - \rpah\ Correlation}\label{result:kbump_vs_rpah}
\begin{figure*}[hbt!]
\centering
$\begin{array}{ccc}
\includegraphics[width=0.55\textwidth]{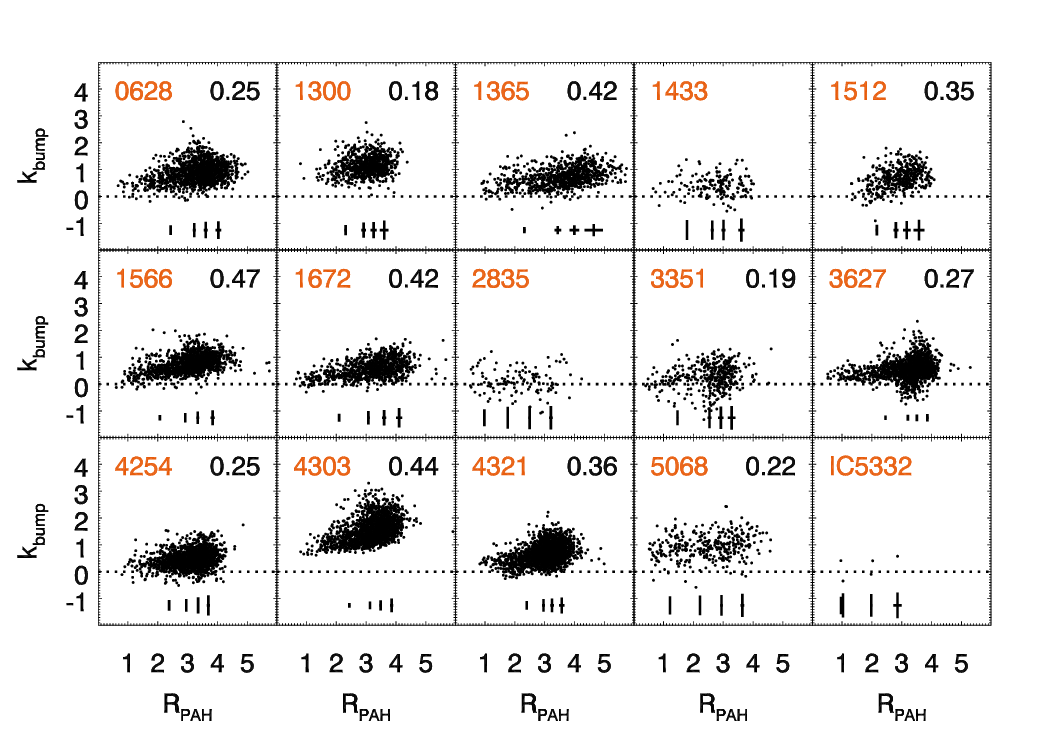} & \hspace{-8mm}
\includegraphics[width=0.5\textwidth]{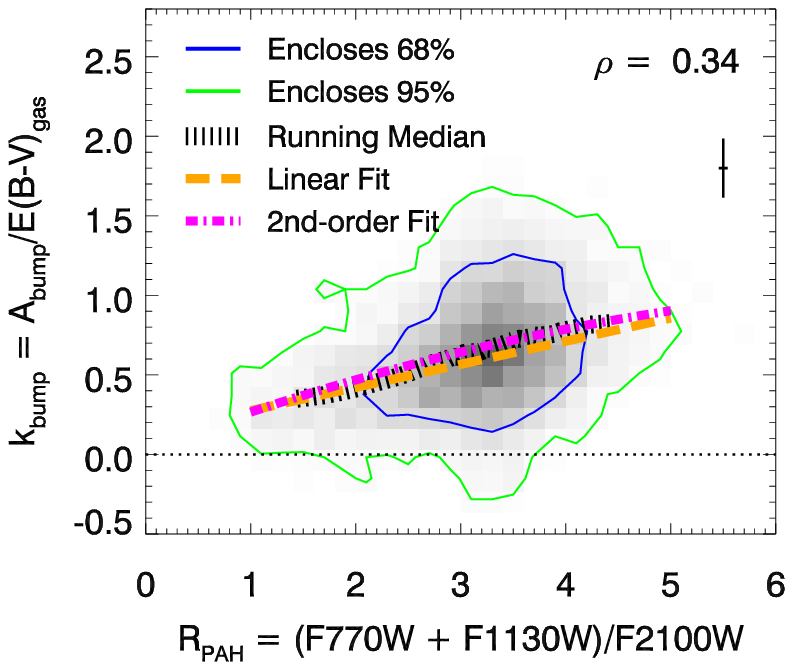} \\
\end{array}$
\vspace{-4mm}
\caption{ (\textit{Left:}) Intrinsic 2175\AA\ feature strength, \kbump=\abumpnorm, vs. PAH abundance, \rpah=(F770W+F1130W)/F2100W, for the 15 galaxies in our sample. The Spearman correlation coefficient for each galaxy is indicated in the upper-right of each panel for cases with a $p$-value<0.01. Representative median error bars  of the regions in quartiles of \rpah\ are shown at the bottom of each panel. 
Most galaxies show a slight correlation between these quantities, with the correlation strength being higher when \kbump\ is better constrained (y-axis error bar). NGC1300 and NGC4303 show a systemic offset toward higher \abumpnorm\ at a given \rpah, the cause of which is unclear.
(\textit{Right:}) 2D histogram of \kbump\ vs. \rpah\ combining five galaxies with median value of $\sigma(k_\mathrm{bump})\lesssim 0.25$ (except NGC~4303): NGC~1365, 1566, 1672, 3627, and 4321. The running median values are indicated by the black hatch lines. Linear and second-order best-fits are shown as a dashed orange line and dash-dot magenta line, respectively, which highlight a positive correlation between these parameters, albeit with large intrinsic scatter. A representative median errorbar is shown in the upper-right.}
\label{fig_kbump_RPAH}
\end{figure*}

We present a comparison between \kbump\ and \rpah\ for individual galaxies in Figure~\ref{fig_kbump_RPAH}, left. We find a slight positive correlation between these parameters, with the Spearman correlation coefficient ranging from $0.3 \lesssim \rho \lesssim 0.5$ for galaxies with a moderate number of regions available after selection cuts. The correlation strength generally increases for galaxies with tighter constraints on \kbump. We note that comparing \abump\ without normalisation by reddening with \rpah\ shows systematically lower correlation strength in our sample (lower by $\rho\sim0.2$). The median values of \kbump\ and \rpah\ for the entire sample are 0.73 and 3.2, respectively. This \kbump\ value corresponds to 22\% the MW strength, however we note that the true strength is likely higher due to the limitations of relying on the \swift/UVOT filters to measure the feature (see \ref{app:swift_reliability}) and the use of ionised gas reddening for the normalisation (see \ref{app:reddening_compare}).

Figure~\ref{fig_kbump_RPAH}, right, shows a 2D histogram combining the five galaxies with the lowest uncertainties on $k_\mathrm{bump}$ (NGC~1365, 1566, 1672, 3627, and 4321; median value of $\sigma(k_\mathrm{bump})\lesssim 0.25$), but excluding NGC~4303 because it shows a noticeable vertical offset toward larger $k_\mathrm{bump}$ from other galaxies. No obvious causes for this offset are apparent from inspecting the \swift\ data, although we note this galaxy has a larger difference between the exposure lengths of UVW2 and UVW1 relative to UVM2 than the rest of the sample (see Table~\ref{tab:swift_exposure}. These five galaxies are among the most massive in the PHANGS sample, ranging from $10.7< \log(M_\star/M_\odot)<11$ with a median of $\log(M_\star/M_\odot)=10.82$ (see Table~\ref{tab:properties}). We note that NGC~1433 ($\log(M_\star/M_\odot)=10.87$) is not in this group because it has shallower \swift\ data relative to the other massive galaxies. We fit a linear relationship to these combined data using the \texttt{MPFITEXY} routine \citep{williams10}, which relies on the \texttt{MPFIT} package \citep{markwardt09}. This routine performs a linear least-squares fit using the error in both variables while also including a term accounting for intrinsic scatter in the data. We also perform a simple second-order polynomial least-square fit and note that the difference relative to the linear fit is very minor. The parameters of both fits are listed in Table~\ref{Tab:bump_rpah_vs_param}.

\begin{table*}
\begin{center}
\caption{Fit Parameters of \kbump\ and \rpah\ as a Function of Galaxy Properties \label{Tab:bump_rpah_vs_param}}
\begin{tabular}{cc|ccccc|ccccc}
\hline\hline 
             &        \hspace{1cm}$\mathbf{y\rightarrow}$     &   \multicolumn{5}{c|}{\kbump} & \multicolumn{5}{c}{\rpah}   \\
      $x$    &     range    &  $p_0$    &  $p_1$    & $p_2$   & $\sigma_{\mathrm{int}}$ & $\rho_S$ & $p_0$    &  $p_1$    & $p_2$ & $\sigma_{\mathrm{int}}$ & $\rho_S$  \\ \hline

\rpah &  [1,5] & 
0.130 $\pm$ 0.015 & 0.145 $\pm$ 0.005 & -- & 0.24 & 0.34 & 
--  & --  & -- & -- & -- \\
 & & 
0.035 $\pm$ 0.030 &  0.246 $\pm$ 0.017 & -0.014 $\pm$ 0.003 & & & 
--  & --  & -- & -- & -- \\ \hline

log($\Sigma_{\mathrm{SFR}}$) & [-2.8,-0.5] & 
0.004 $\pm$ 0.012 & -0.317 $\pm$ 0.007 & -- & 0.21 & -0.55 & 
1.130 $\pm$ 0.027 & -1.054 $\pm$ 0.014 & -- & 0.54 & -0.59 \\ 
 $[M_\odot$yr$^{-1}$kpc$^{-2}]$ & & 
0.124 $\pm$ 0.032 & -0.134 $\pm$ 0.037 &  0.071 $\pm$ 0.011 & & &
0.651 $\pm$ 0.062 & -1.653 $\pm$ 0.072 & -0.167 $\pm$ 0.020 & & \\ \hline

log(sSFR) & [-10.9,-8.8] & 
-2.101 $\pm$ 0.074 & -0.266 $\pm$ 0.007 & -- & 0.23 & -0.45 & 
-6.946 $\pm$ 0.159 & -0.995 $\pm$ 0.016 & -- & 0.61 & -0.52 \\
 $[\mathrm{yr}^{-1}]$ & & 
3.91 $\pm$ 1.20 &  0.99 $\pm$ 0.24 &  0.065 $\pm$ 0.012 & & & 
-30.77 $\pm$ 2.34 & -5.83 $\pm$ 0.47 & -0.244 $\pm$ 0.024 & & \\ \hline

\end{tabular}
\end{center}
\textbf{Notes.} The functional form of these fits is $y = p_0+p_1x+p_2x^2$, where $y$ is \kbump\ or \rpah\ (middle and right columns, respectively). We present both a linear and second-order polynomial fit for each case. We also report the intrinsic dispersion, $\sigma_{\mathrm{int}}$, returned from \texttt{MPFITEXY}, and the Spearman nonparametric correlation coefficient, $\rho_S$. The data used in the fits are a combination of five galaxies (NGC~1365, 1566, 1672, 3627, and 4321).
\end{table*}

\subsection{\kbump\ and \rpah\ Correlation with \sigmaSFR\ and sSFR}
We present a comparison between \kbump\ and \rpah\ with two proxies of the ionisation parameter of a region, the SFR surface density, \sigmaSFR=SFR/Area, and the specific-SFR, sSFR=SFR/$M_\star$. 
We note that MUSE does not cover [OII] to provide a direct tracer of the ionisation parameter through the [OIII]/[OII] ratio \citep{kewley19}.
The \sigmaSFR\ and sSFR parameters are found to have the strongest correlation with \kbump\ and \rpah\ among the parameters that we examined (see Section~\ref{result:kbump_rpah_vs_other}). As described in Section~\ref{method:sfr}, the SFRs (and SFR surface densities) in this work are derived from \halpha\ and will correspond to average star formation on timescales of $\sim$10~Myr. In practice, sSFR can be easier to measure for high-redshift samples than \sigmaSFR\ because the galaxies examined can be unresolved.

\begin{figure*}[hbt!]
\centering
$\begin{array}{ccc}
\includegraphics[width=0.55\textwidth]{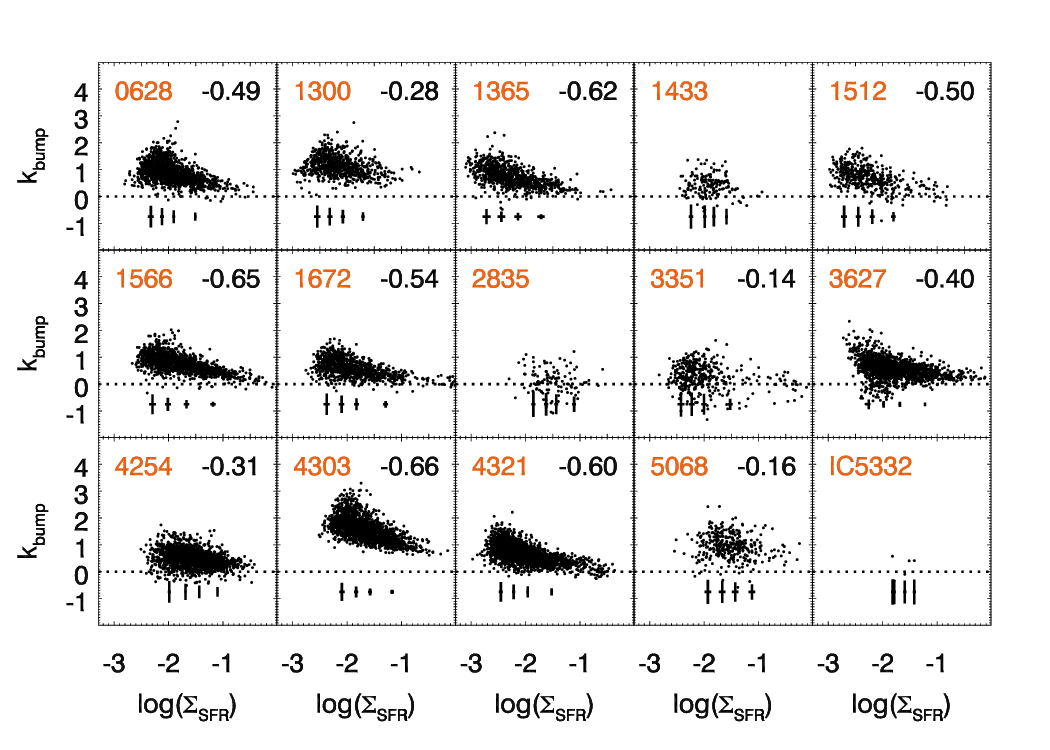} & \hspace{-8mm}
\includegraphics[width=0.5\textwidth]{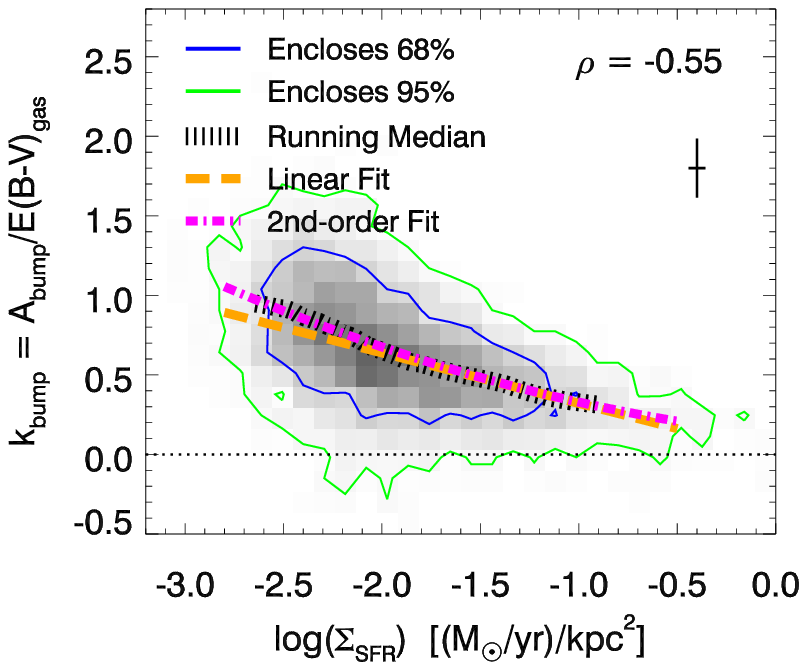} \\
\end{array}$
\vspace{-4mm}
\caption{ Similar to Figure~\ref{fig_kbump_RPAH}, except now showing the intrinsic 2175\AA\ feature strength, \kbump=\abumpnorm, vs. SFR surface density, \sigmaSFR. Representative median error bars are in quartiles of \sigmaSFR. The 2D histogram combines five galaxies with median value of $\sigma(k_\mathrm{bump})\lesssim 0.25$ (except NGC~4303): NGC~1365, 1566, 1672, 3627, and 4321. The best-fits highlight a negative correlation between these parameters, albeit with large intrinsic scatter.}
\label{fig_kbump_SigmaSFR}
\end{figure*}

\begin{figure*}[hbt!]
\centering
$\begin{array}{ccc}
\includegraphics[width=0.55\textwidth]{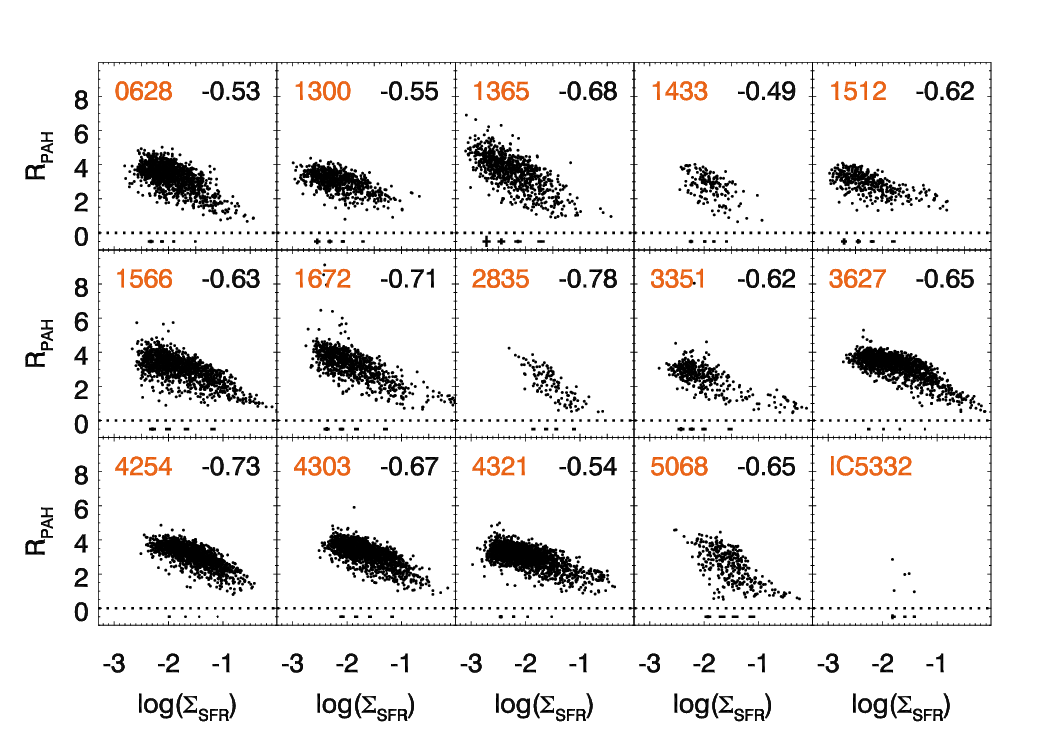} & \hspace{-8mm}
\includegraphics[width=0.5\textwidth]{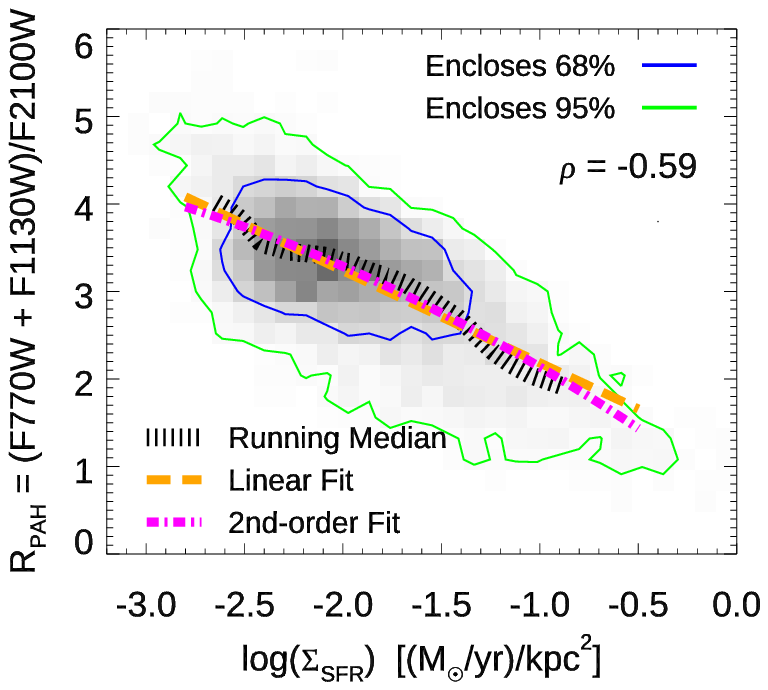} \\
\end{array}$
\vspace{-4mm}
\caption{Similar to Figure~\ref{fig_kbump_RPAH}, except now showing the PAH abundance, \rpah=(F770W+F1130W)/F2100W, vs. SFR surface density, \sigmaSFR. Representative median error bars are in quartiles of \sigmaSFR. For consistency with the previous figures, the 2D histogram combines the same five galaxies with median value of $\sigma(k_\mathrm{bump})\lesssim 0.25$ (except NGC~4303): NGC~1365, 1566, 1672, 3627, and 4321. The best-fits highlight a negative correlation between these parameters, albeit with large intrinsic scatter.}
\label{fig_RPAH_SigmaSFR}
\end{figure*}

A comparison between \kbump\ and \sigmaSFR is shown in Figure~\ref{fig_kbump_SigmaSFR},
and \rpah\ and \sigmaSFR\ in Figure~\ref{fig_RPAH_SigmaSFR}. We find a moderate negative correlation between these parameters, with the Spearman correlation coefficient ranging from $-0.6 \lesssim \rho \lesssim -0.3$ for \kbump\ vs. \sigmaSFR, and $-0.7 \lesssim \rho \lesssim -0.5$ for \rpah\ vs. \sigmaSFR. Linear and second-order polynomial least-square fits parameters to the five galaxies with robust measurements are listed in Table~\ref{Tab:bump_rpah_vs_param}.

\begin{figure*}[hbt!]
\centering
$\begin{array}{ccc}
\includegraphics[width=0.55\textwidth]{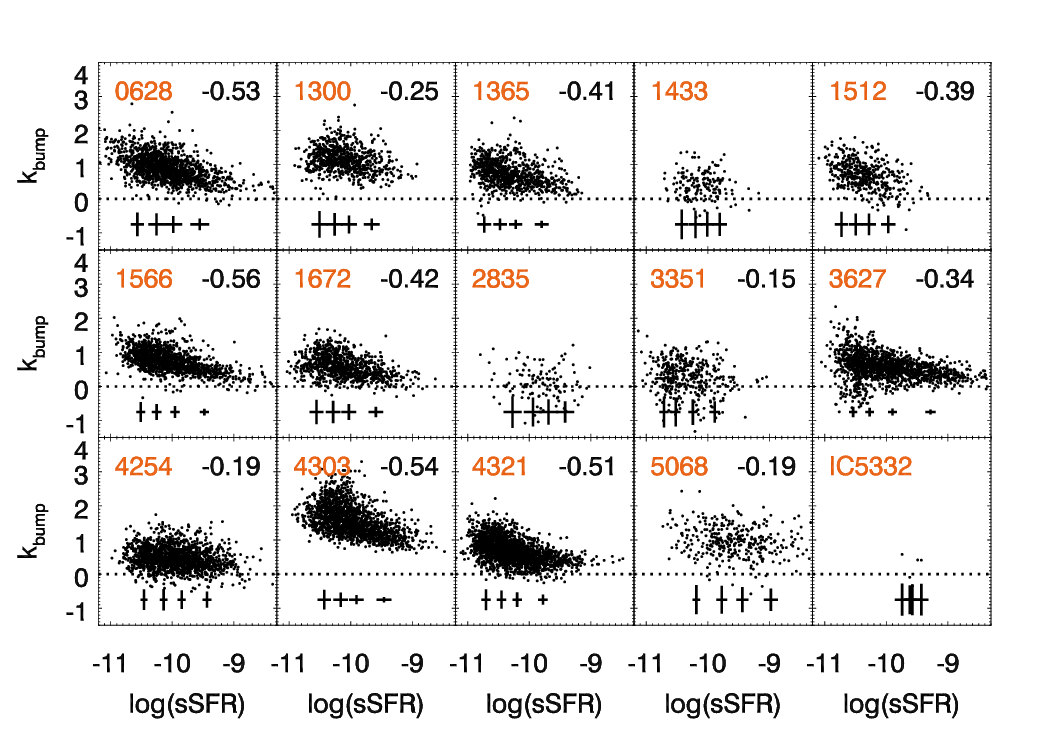} & \hspace{-8mm}
\includegraphics[width=0.5\textwidth]{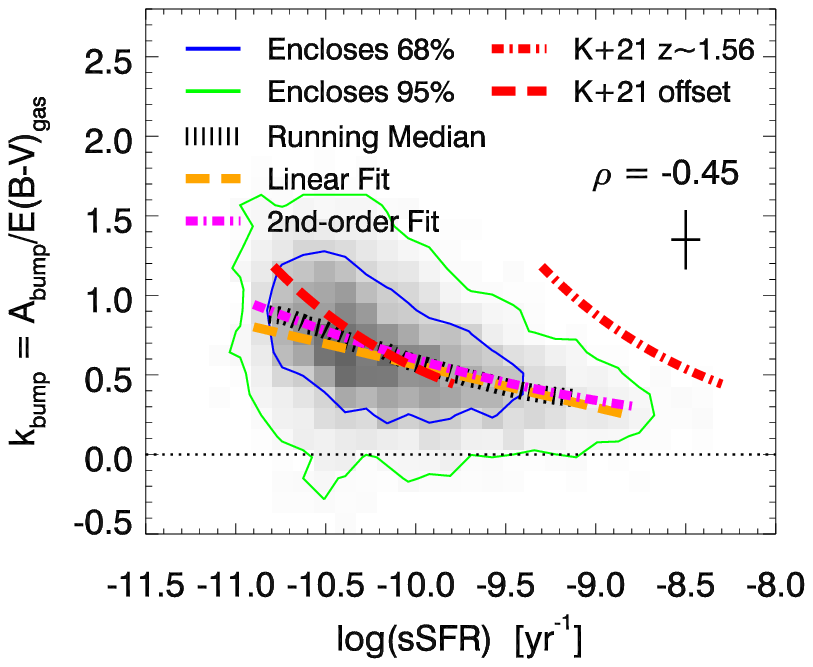} \\
\end{array}$
\vspace{-4mm}
\caption{ Similar to Figure~\ref{fig_kbump_RPAH}, except now showing the intrinsic 2175\AA\ feature strength, \kbump=\abumpnorm, vs. sSFR. Representative median error bars are in quartiles of sSFR. The 2D histogram combines five galaxies with median value of $\sigma(k_\mathrm{bump})\lesssim 0.25$ (except NGC~4303): NGC~1365, 1566, 1672, 3627, and 4321. The best-fits highlight a negative correlation between these parameters, albeit with large intrinsic scatter. The red dash-dot line shows the relation in \cite{kashino21} for galaxies at $z\sim1.56$ (adopting $\log(M_\star/M_\odot)=10.82$; the mean value of the five PHANGS galaxies shown). The red dashed line is the same relation after applying a MS-offset between $z=1.56$ and $z=0$ (see~Section~\ref{lit_compare}) and shows qualitative agreement with the local relation. 
} \label{fig_kbump_sSFR}
\end{figure*}

\begin{figure*}[hbt!]
\centering
$\begin{array}{ccc}
\includegraphics[width=0.55\textwidth]{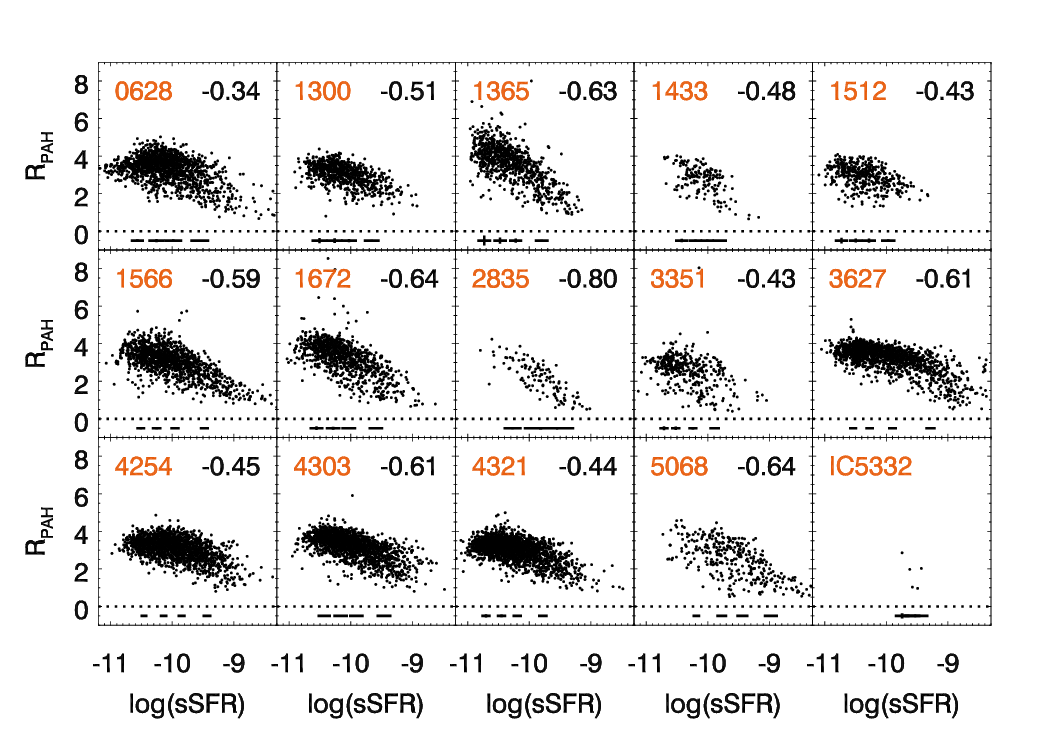} & \hspace{-8mm}
\includegraphics[width=0.5\textwidth]{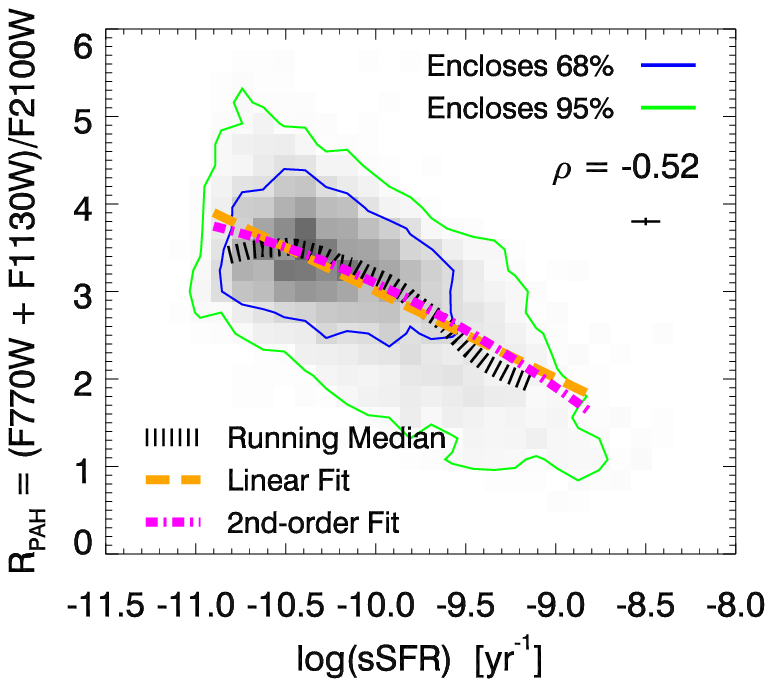} \\
\end{array}$
\vspace{-4mm}
\caption{Similar to Figure~\ref{fig_kbump_RPAH}, except now showing the PAH abundance, \rpah=(F770W+F1130W)/F2100W, vs. sSFR. Representative median error bars are in quartiles of sSFR. For consistency with the previous figures, the 2D histogram combines the same five galaxies with median value of $\sigma(k_\mathrm{bump})\lesssim 0.25$ (except NGC~4303): NGC~1365, 1566, 1672, 3627, and 4321. The best-fits highlight a negative correlation between these parameters, albeit with large intrinsic scatter.}
\label{fig_RPAH_sSFR}
\end{figure*}

A comparison between \kbump\ and sSFR is shown in Figure~\ref{fig_kbump_sSFR},
and \rpah\ and sSFR in Figure~\ref{fig_RPAH_sSFR}. We find similar trends to those with \sigmaSFR, with a moderate negative correlation between these parameters and sSFR, with the Spearman correlation coefficient ranging from $-0.6 \lesssim \rho \lesssim -0.2$ for \kbump\ vs. sSFR, and $-0.6 \lesssim \rho \lesssim -0.3$ for \rpah\ vs. sSFR. Linear and second-order polynomial least-square fits parameters to the five galaxies with robust measurements are listed in Table~\ref{Tab:bump_rpah_vs_param}. The slightly weaker correlations with sSFR than \sigmaSFR\ suggest that these parameters are more closely tied to the presence of young stars (i.e., SFR) than older stars ($M_\star$). This is also supported by the fact that there is no  (or weak) correlation with stellar mass surface density for most galaxies (Section~\ref{result:kbump_rpah_vs_other}).

\subsection{\kbump\ and \rpah\ Correlation with other Parameters}\label{result:kbump_rpah_vs_other}
We explored a range of additional correlations between \kbump\ and \rpah\ with other physical properties. These include gas-phase metallicity, total reddening (\EBVgas), stellar mass surface density, and galactocentric radius. 
We do not attempt a comparison with the UV slope ($\beta$) because the value derived from the \swift/UVOT filters is heavily contaminated by the 2175\AA\ feature (see \ref{app:swift_reliability}).
For nearly all cases the correlation strengths were not found to be significant ($\vert \rho \vert \lesssim 0.3$), and the few exceptions do not share correlation strengths between both \kbump\ and \rpah\ as is found for \sigmaSFR\ and sSFR. 
We note that our sample spans a narrow range in metallicity, with 98\% of the regions after our selection cuts being in the range $8.40 < \log[O/H] < 8.65$ (using Scal) and that previous studies suggest that PAH abundances may only significantly vary at metallicities of $\log[O/H] \lesssim 8.1$ \citep{draine07, marble10}. Ongoing and future measurements of PAH features for nearby dwarf galaxies with \jwst/MIRI will undoubtedly improve our picture of this in the near future due to these objects tending to have lower metallicities. Regarding the galactocentric radius, we do find that the strength of \kbump\ and values of \rpah\ tend to be lower in the very central regions relative to the outskirts, but that this is likely a reflection of the central regions having higher SFRs.

\section{Discussion}\label{discussion}

\subsection{Implications of Trends and Correlation Strengths}\label{implications}
We find a moderate correlation between \kbump\ and \rpah\ for galaxies where the former can be well-constrained. This lends support for PAHs to be a potential carrier of the feature, but this is far from conclusive. For instance, it is almost certainly the case the 2175\AA\ feature is due to small dust grains \citep[e.g.,][]{mathis94, draine03}, and PAHs (which are also small grains) may simply be co-spatial with other small dust grains. The link to small grains is further supported by the anti-correlation of both \kbump\ and \rpah\ with \sigmaSFR\ and sSFR, which are proxies of the ionising radiation strength. Ionising radiation is likely to destroy (via photodissociation) small dust grains responsible for both the 2175\AA\ feature \citep[e.g.,][]{calzetti94, clayton00, fischera&dopita11} and PAHs \citep[e.g.,][]{egorov23, sutter24, chastenet24, chown24}, and it is interesting that the correlation strengths are stronger with these ionisation proxies than the correlation with each other. However, as we discuss below, the trend of \rpah\ with \sigmaSFR\ and sSFR can be also explained with dust heating, rather than PAH destruction. We can test the correlation between \kbump\ and \rpah\ while controlling for \sigmaSFR, by examining the partial correlation coefficient \cite{kendall42},
\begin{equation}
\begin{aligned}\label{eq:partial_corr}
\rho_{AB|C}=\frac{\rho_{AB}-\rho_{AC}\rho_{BC}}{\sqrt{1-\rho_{AC}^2}\sqrt{1-\rho_{BC}^2}} \,,
\end{aligned}
\end{equation}
where $\rho_{AB}$ is the Spearman correlation
coefficient between variables $A$ and $B$, and so on for other combinations. We note this metric assumes a monotonic relation is present between the quantities of interest. In our case, we choose $A$, $B$, and $C$ to be \kbump, \rpah, and \sigmaSFR, respectively. We adopt the correlation coefficients listed in Table~\ref{Tab:bump_rpah_vs_param}, which are based on our five `robust' galaxies. We find a value of $\rho_{AB|C}=0.02$, indicating that the correlation between \kbump\ and \rpah\ may be entirely driven by the correlation of these quantities with \sigmaSFR. If we use sSFR for $C$, we get $\rho_{AB|C}=0.19$.

\begin{figure}[hbt!]
\centering
\includegraphics[width=1\textwidth]{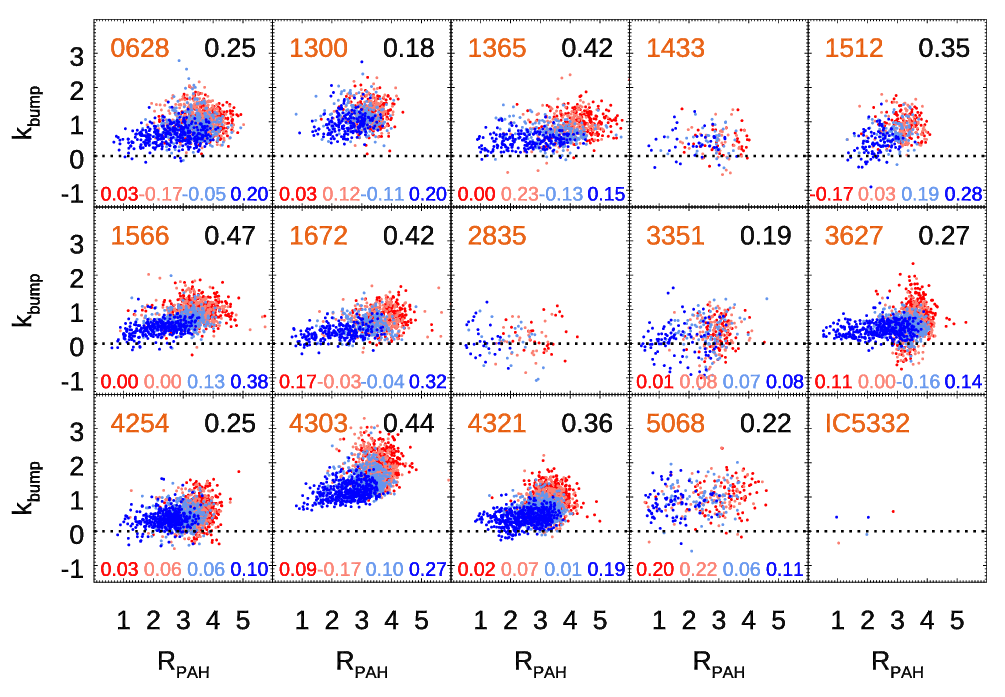}
\vspace{-6mm}
\caption{Similar to Figure~\ref{fig_kbump_RPAH}, \textit{Left}, but subdividing the regions of each galaxy into quartiles of \sigmaSFR. The values of \sigmaSFR\ increase as dark red, light red, light blue, and dark blue, respectively (i.e., dark blue is highest \sigmaSFR). The correlation strengths between \abump\ and \rpah\ for the quartile subsamples are indicated by the numbers at the bottom of each panel and are lower for all galaxies ($\rho \lesssim 0.3$) relative to the value when using all regions (black value, upper-right).
}
\label{fig_kbump_rpah_SigmaSFR_quartiles}
\end{figure}

An independent, and more visually intuitive, test can be done by examining the correlation strengths of these quantities when subdividing the regions of each galaxy into quartiles of \sigmaSFR, which is shown in Figure~\ref{fig_kbump_rpah_SigmaSFR_quartiles}. We find that the correlation strengths between \abump\ and \rpah\ for all galaxies are lower for the quartile subsamples ($\rho \lesssim 0.3$), with only the highest \sigmaSFR\ quartile retaining a moderate trend in a couple cases ($\rho \sim 0.3$; e.g., NGC~1566, 1672). This outcome also suggests that most of the correlation between these quantities may be a consequence of the correlation with \sigmaSFR.

\begin{figure} 
\includegraphics[width=0.95\textwidth]{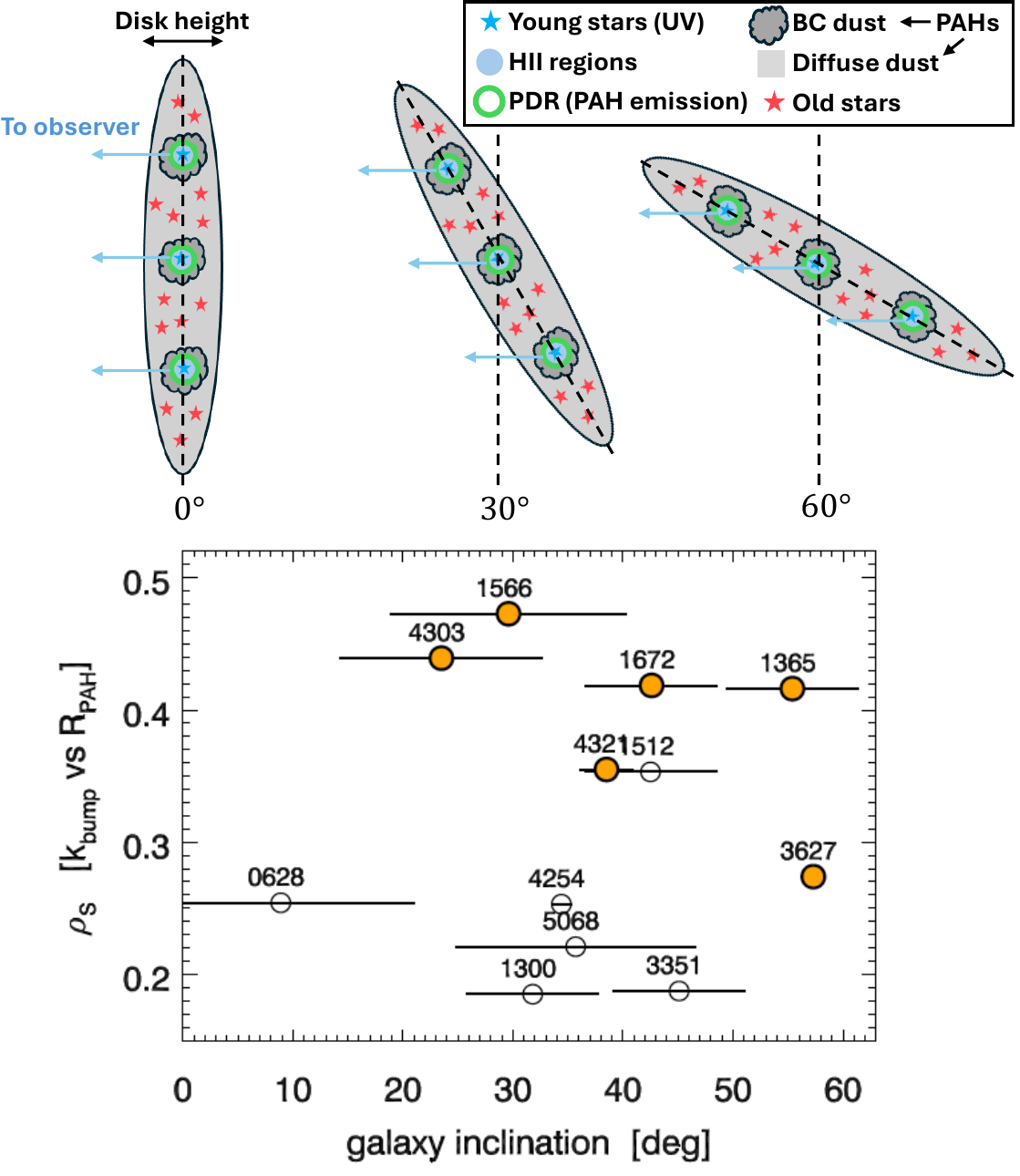}
\caption{\textit{Top:} Cartoon visualisation of how changes in the inclination angle of a disk galaxy could affect the correlation strength between \kbump\ and \rpah . \kbump\ is measured via the effect of dust attenuation toward UV-bright young stars that are affected by both birth-cloud dust and the diffuse dust in the ISM. As the inclination increases from face-on ($0\degree$) to edge-on ($90\degree$), there will be increasing path length of the diffuse ISM between the young stars and the observer. PAHs can be present in both dust mediums, but the majority of PAH emission will come from PDRs surrounding HII regions and will be independent of viewing angle (optically thin). Thus, the association between the 2175\AA\ absorption feature and PAH emission may be weaker as the viewing angle increases since viewing angle only affects the former.  
\textit{Bottom:} correlation strength, $\rho_\mathrm{S}$ between \kbump\ and \rpah\ as a function of the galaxy inclination. Orange circles denote our `robust' sample of galaxies with median value of $\sigma(k_\mathrm{bump})\lesssim 0.25$: NGC~1365, 1566, 1672, 3627, 4303, and 4321. There is a slight preference for stronger correlation strengths between \kbump\ and \rpah\ for galaxies at lower inclination (i.e., closer to face-on). The trends are less apparent among the rest of the sample (open circles), but this may be attributed to the large uncertainty in \kbump\ measurements for these galaxies. 
 \label{fig:inclination_demo}}
\end{figure}

If PAHs are a carrier of the 2175\AA\ feature, one possible factor that could reduce the correlation strength between \abump\ and \rpah\ is that along the line of sight from a star-forming region there can be 2175\AA\ absorption from intervening dust (PAHs) that resides between the source and the observer (e.g., in the diffuse ISM) that is spatially disconnected from the majority of observed PAH emission from the photodissociation region (PDR) that surrounds the star-forming region. This intervening dust would also not be affected by effects of PAH destruction, since it is far away from the strong ionising radiation (i.e., HII region). Depending on the amount of dust around the HII region (birth-cloud dust), the intervening diffuse dust may have a modest effect on the strength of the 2175\AA\ feature. This effect is likely to be more pronounced as the inclination of galaxies go from face-on to edge-on, however it will always be present due to the non-zero thickness of the disk. We show a cartoon visualisation of this scenario in Figure~\ref{fig:inclination_demo} (top). Interestingly, we find a slight preference for higher correlation strength with lower inclination in our `robust' sample ($\sigma(k_\mathrm{bump})\lesssim 0.25$), as shown in Figure~\ref{fig:inclination_demo} (bottom). We include NGC~4303 in our sample here because the correlation strength should be independent of the vertical offset in the \kbump\ vs \rpah\ relation that we see in this galaxy with respect to the other `robust' cases. For example, NGC~1566 and 4303, the least inclined ($i\sim30\degree$) among these sources, show slightly stronger correlation strengths relative to the more inclined galaxies. Performing this comparison in a larger sample, and also across a wider range in inclination, is needed to draw firm conclusions on whether this effect is real.

Assuming a link between PAHs and the UV bump, dust heating could be another factor to help explain a weak correlation between \abump\ and \rpah\ (i.e., if the dust continuum used as the baseline in \rpah\ is varying). \cite{calzetti07} presented calibrations of \spitzer\ 8$\mu$m and 24$\mu$m luminosity density ($S_{8\mu \mathrm{m,dust}}$ and $S_{24\mu \mathrm{m,dust}}$) as \sigmaSFR\ diagnostics (see their eq. 2 and 3; when assuming $\Sigma$(Pa$\alpha_\mathrm{corr}$) is a direct proxy of \sigmaSFR). The slope of their relationship log($S_{8\mu \mathrm{m,dust}}$) with log(\sigmaSFR) is sub-linear ($0.94\pm0.02$), whereas the slope for log($S_{24\mu \mathrm{m,dust}}$) is super-linear ($1.23\pm 0.03$). More recently, \cite{calzetti24} found a shallower super-linear relation ($1.07 \pm 0.01$) between log($S_{24\mu \mathrm{m,dust}}$) (based on F2100W) and log(\sigmaSFR). 
The sub-linear 8$\mu$m relation could be an indication of PAH destruction at higher \sigmaSFR\ values, whereas the super-linear 24$\mu$m could indicate increased dust-heating at higher \sigmaSFR\ values (i.e., a larger fraction of the total IR luminosity will be measured at 24$\mu$m as dust temperature increases). Therefore, both of these factors could affect the ratio of these quantities. We note that the slopes of these relationships are sensitive to the methodology used for local background subtraction and on the size scales being considered \citep[see discussion in][]{calzetti24}, and may also be time-dependent with the emergent state of embedded stellar clusters \citep[e.g.,][]{gregg24}.
Assuming that the 7.7$\mu$m feature-alone relative to a dust continuum proxy (e.g., F2100W or 24$\mu$m) is a good proxy for \rpah, which seems to be reasonable \citep[see Appendix~B in][]{sutter24}, then this would be reflected in similar trends in \rpah\ (ratio of flux densities) vs \sigmaSFR. 

\begin{figure}[hbt!]
\centering
\includegraphics[width=1\textwidth]{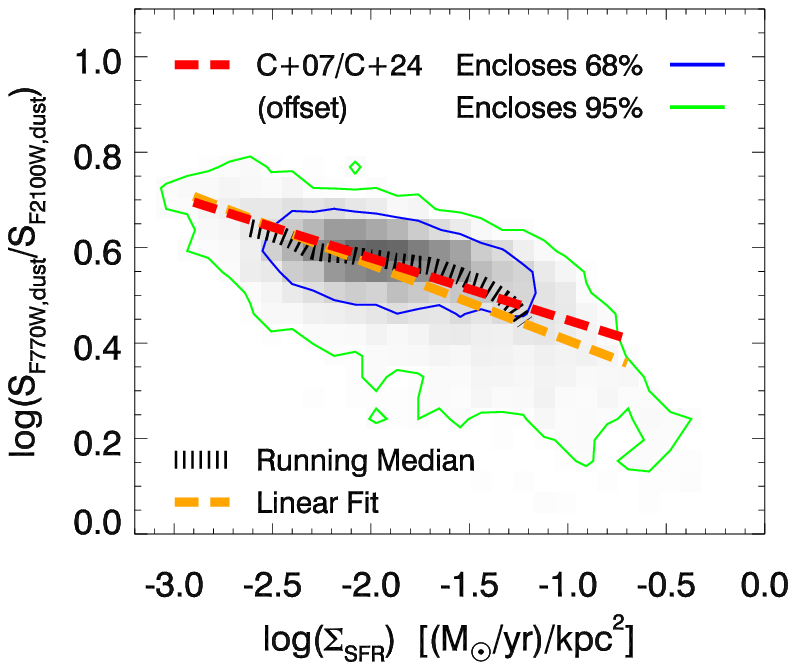}
\caption{2D histogram of the F770W/F2100W surface luminosity ratio as a function of \sigmaSFR\ for the six `robust' galaxies (NGC~1365, 1566, 1672, 3627, 4303, and 4321). The best-fit linear slope (orange dashed line) is consistent with the slope using the \cite{calzetti07} 8$\mu$m and \cite{calzetti24} 24$\mu$m relations (red dashed line), where the former is sub-linear with \sigmaSFR\ and the latter is super-linear. These results imply that both dust destruction and dust heating could be factors in explaining the trend of \rpah\ with \sigmaSFR.
}
\label{fig_F770W2F2100W_vs_SigmaSFR}
\end{figure}

Taking the difference (in log) of the \cite{calzetti07, calzetti24} relations gives $\log(S_{8\mu \mathrm{m,dust}}/S_{24\mu \mathrm{m,dust}}) \propto -0.13 \log(\Sigma_\mathrm{SFR}$). As we show in Figure~\ref{fig_F770W2F2100W_vs_SigmaSFR}, this is similar to the best-fit linear slope that we find in this work if we examine the ratio of F770W and F2100W luminosity densities as a function of \sigmaSFR,
\begin{equation}
\begin{aligned}\label{eq:C07_compare}
\log \left( \frac{S_\mathrm{F770W,dust}}{S_\mathrm{F2100W,dust}} \right) = (-0.16 \pm 0.01)\log(\Sigma_\mathrm{SFR}) + (0.25 \pm 0.01) \,,
\end{aligned}
\end{equation}
which has an intrinsic scatter of 0.09~dex. We note that there is some indication that the trend deviates from a linear form at higher \sigmaSFR, with the trend becoming slightly steeper (more negative) at $\log(\Sigma_\mathrm{SFR})\gtrsim-1.5$. This could correspond to a boundary where a change in dust destruction and/or dust heating is occurring. 
Resolving the degeneracy that these mechanisms have on the \rpah\ value will require mid-IR spectroscopy around the PAH features to properly measure the underlying dust continuum.

Below we highlight a few factors that complicate our ability to provide a more definitive link between the 2175\AA\ feature and PAHs. 
The first is that the spatial scales that we are measuring both quantities at ($\sim 100$~pc) are coarse relative to the scales of HII regions ($\sim$few~pc) where photo-dissociation effects can be studied more precisely \citep[e.g.,][]{berne22, pedrini24}. This situation could be improved through higher spatial resolution UV surveys of nearby galaxies with \hst\ to more closely match the \jwst/MIRI resolution. 
Second, linking dust attenuation to emission is complicated because of the degeneracies that optical depth and geometric effects can have on the observed strength of the feature when dealing with unresolved star/dust distributions (e.g., Figure~\ref{fig:inclination_demo}). This problem is exacerbated by the fact that there are limited UV data available for nearby galaxies, and in most cases the filters available are very broad, to study the feature in detail.
An alternative could be to study UV extinction curves using spectroscopy toward a larger number of MW, LMC, and SMC sightlines where PAH emission measurements are also available \citep[e.g.,][]{massa22, gordon24}. Currently, \hst\ is the only facility capable of characterising  UV extinction curves near the 2175\AA\ region and only $\sim$100 sightlines in the MW, LMC, and SMC have been characterised in this wavelength regime in the past 30 years. 
Finally, studies to date have relied on trying to link measurements of the 2175\AA\ feature in extinction/attenuation to PAHs observed in emission. A much more conclusive link between PAHs as the carrier would come from measuring both features in extinction along the same sightline. 
We highlight avenues that future facilities may provide in addressing these shortcomings in Section~\ref{future_missions}.

\subsection{Literature Comparison}\label{lit_compare}

To our knowledge, there have been relatively few attempts to link the 2175\AA\ feature (measured in attenuation) to PAH emission. \cite{decleir19} performed a similar comparison to NGC~0628 as this work, using \swift/UVOT and \spitzer+\textit{Herschel} data. In that work, they adopted an SED-fitting approach to constrain the intrinsic bump strength and the PAH abundance was inferred from a ratio of IRAC 8~$\mu$m to total far-infrared luminosity. No significant correlation was found, however this could be attributed to the large measurement uncertainties in both quantities. Indeed, we find stronger correlations are present for galaxies with better constraints on \kbump, and that NGC~0628 is not among the best-constrained in our sample, with a median value of $\sigma(k_\mathrm{bump})\sim 0.5$ (for regions satisfying conditions (1)-(3) in Section~\ref{selection_cuts}).
\cite{shivaei22b} examined the relation between the 2175\AA\ feature and PAH emission, constrained using MUSE spectra and \spitzer/MIPS photometry, respectively, in a sample of 86 SFGs at $z\sim2$. Their comparison slightly differs in that it compared \abump\ (i.e., apparent bump strength in magnitude) relative to a PAH abundance (rest-frame 8$\mu$m relative to FIR). They find that galaxies with stronger bumps are positively correlated with higher PAH abundances, which is qualitatively consistent with our findings, although we find a weaker correlation between \abump\ and \rpah\ relative to \kbump\ and \rpah.

Several studies have explored the relationship between the 2175\AA\ feature strength with the SFR and sSFR of galaxies, and which qualitatively follow the trends observed in our study. 
\cite{kriek&conroy13} used composite SEDs from a sample of $\sim$3500 galaxies at $0.5 < z < 2.0$ to characterise the shape of dust attenuation curves and the 2175\AA\ feature. They found that the bump strength decreases with increasing equivalent width of \halpha, which is a proxy for sSFR \citep[e.g.,][]{marmol-queralto16}, with larger EW(\halpha) corresponding to higher sSFR. 
\cite{kashino21} used individual and stacked spectra from a sample of 505 SFGs at $1.3\leq z\leq 1.8$ (median $z=1.56$; $-9.3\lesssim \log\mathrm{sSFR}\lesssim -8.3$) in the zCOSMOS-deep survey, finding that the 2175\AA\ feature amplitude decreases with increasing sSFR, while it increases moderately with stellar mass. 
We show their relationship in Figure~\ref{fig_kbump_sSFR}, which is shifted toward higher sSFR, as well as an offset version that accounts for the difference in the galaxy main-sequence (MS; logSFR vs. log$M_\star$) values with redshift \citep[e.g.,][]{popesso23}. We take this offset as the difference in the MS sSFR between z=0 and z=1.56 for a galaxy with $\log(M_\star/M_\odot)=10.82$ (median of the five PHANGS galaxies we combine) using the galaxy MS relationship from \cite{leroy19} for z=0 and \cite{leslie20} for z=1.56 (latter using the `SF' sample). These give an offset of $\log(\mathrm{sSFR}_{z=1.56})-\log(\mathrm{sSFR}_{z=0})=1.49$. Interestingly, we find that shifting the \cite{kashino21} relation (i.e., subtracting the x-axis values by 1.49) is qualitatively consistent with our local relation. We note that the y-axis agreement is almost certainly coincidental between these studies because the Swift-derived bump strength is likely underestimated and \cite{kashino21} normalise the UV bump using the reddening on the stellar continuum (\EBVstar; which are typically lower than \EBVgas, see \ref{app:reddening_compare}).
Finally, \cite{zhou23} used 134 galaxies in MaNGA \citep{bundy15} that also have \swift\ data \citep{molina20b} to study the 2175\AA\ feature, finding that the 2175\AA\ feature amplitude decreases with increasing sSFR. In that work, they estimate \abump\ in a similar way as this work by taking the difference in attenuation at 2175\AA\ from UVM2 relative to a UV slope fit (i.e., $f_{\lambda}(\mathrm{UVM2},0))$, see Figure~\ref{fig_Abump_example}). However, they normalise their bump strength based on the difference relative to an intrinsic SED estimated with an UV to near-IR SED-fitting approach. Therefore, a direct comparison with their findings is not straightforward.

The relationship between the PAH abundances with the SFR and sSFR of galaxies has been explored by two papers in the PHANGS-JWST team. 
\cite{egorov23} used 4 PHANGS-JWST galaxies to examine the relationship between \rpah\ and various physical properties, finding lower \rpah\ values with larger [SIII]/[SII] and EW(\halpha), which are proxies for the ionisation parameter and sSFR, respectively (correlation: $-0.7 \lesssim \rho \lesssim -0.4$). A larger analysis extending to the full 19 PHANGS-JWST galaxies was recently presented in \cite{sutter24}. In that work, they find that the difference between the average \rpah\ measured in nebular regions and the diffuse gas is correlated with the sSFR. The selection criteria used in this study most closely align with the `nebular regions' in \cite{sutter24}, and they also find a decrease in \rpah\ values with increasing sSFR (see their Figure 11) that is qualitatively consistent with our findings.

\subsection{Future Prospects for Mapping the UV Extinction Curves in the Local Group}\label{future_missions}

Dedicated satellite missions in the UV wavelength range could greatly advance our understanding of UV extinction curves and the 2175\AA\ feature. 
As highlighted previously, studying the feature in extinction provides more direct insight into potential carriers than measuring it in attenuation. Ideally, such missions would provide moderate resolution spectroscopy, instead of photometry, in order to study the detailed shape of the bump (amplitude, central wavelength, and width), which can provide better constraints on the size distribution and chemical composition of interstellar dust grains responsible for the feature. Below, we briefly highlight a few approved or planned future UV spectroscopic missions.

First, the Ultraviolet Explorer (UVEX) is a recently approved NASA Medium-Class Explorer mission \citep[launch $\sim$2030][]{kulkarni21}, that will have FUV (1390–1900\AA) and NUV (2030–2700\AA) imaging (PSF$\sim$2\arcsec) and long-slit spectroscopic capability (1150–2650\AA; $R\gtrsim 1000$). UVEX will perform an all-sky imaging survey, but this will have limited utility for studying the 2175\AA\ feature because it is very difficult to characterise it using only two broad filters (similar issues for \textit{GALEX}). In contrast, its spectroscopic instrument is ideal for studying the feature and the currently envisioned survey design would study $\sim$1000 O-stars in the LMC and SMC \citep{kulkarni21}; increasing the number of studied sightlines in these galaxies by over an order of magnitude.

Second, a small satellite (SmallSat) UV spectroscopic mission concept known as UVESS \citep[Ultra Violet Extinction Sky Survey;][]{mathew24} is being developed by the Australian National University and a collaboration of international partners. UVESS would map the variability in extinction curves and 2175\AA\ feature by acquiring near-UV (1400-2700\AA) spectroscopy ($R\sim 2500$) for thousands of OB-stars in the MW, LMC, and SMC. 
UVESS will explore the adoption of a compact spatial heterodyne spectrograph \citep[SHS;][]{sahoo24}, which employs a modified Michelson interferometer configuration, and offers exceptional performance and cost-effectiveness for UV spectroscopy relative to traditional instruments.
Readers that are interested in helping develop the science case for UVESS are encouraged to contact the first author. 

Finally, the Habitable Worlds Observatory\footnote{\url{https://habitableworldsobservatory.org}}, a proposed NASA flagship mission (launch $\sim$2040), is envisioned to have UV spectroscopic capabilities. It will likely have the potential for characterising UV extinction curves toward single stars in galaxies out to $\sim$10--20~Mpc (dependent on the mirror size). This will be revolutionary in characterising the variation of extinction curves and the 2175\AA\ feature over a wider range of ISM conditions. The incredible sensitivity of such a large facility may also make it feasible to measure \textit{both} the 2175\AA\ feature and PAHs in extinction (latter measured from IR facility), thereby alleviating complications that arise when indirectly linking absorption and emission.  

\section{Conclusion}\label{conclusion}
Using a sample of 15 local galaxies in the PHANGS-\jwst\ survey that have NUV and mid-IR data from \swift/UVOT and \jwst/MIRI, respectively, we examined the correlation between the 2175\AA\ feature and PAH abundances in a spatially-resolved manner. This comparison is made to explore the link of PAHs as a potential carrier of the 2175\AA\ feature.
We find a moderate positive correlation between the 2175\AA\ feature strength and PAH abundance (spearman coefficient, $0.3 \lesssim \rho \lesssim 0.5$), albeit with large intrinsic scatter (Figure~\ref{fig_kbump_RPAH}, Table~\ref{Tab:bump_rpah_vs_param}). If the two are linked, this scatter may arise from limitations in the use of broad- and medium-filters to constrain the amplitude of the feature and also in the fact that the feature is measured in attenuation, whereas the PAHs are measured via emission, such that they may not directly probe the same physical regions (i.e., due to optical depth and geometric effects; e.g., Figure~\ref{fig:inclination_demo}). 

We also compare the strength of the 2175\AA\ feature and PAH abundances with various galaxy properties and the strongest trends are negative correlations with \sigmaSFR and sSFR (proxies of ionising radiation intensity; $\rho\sim-0.6$; Figures~\ref{fig_kbump_SigmaSFR}--\ref{fig_RPAH_sSFR}, Table~\ref{Tab:bump_rpah_vs_param}), which can account for trends between \kbump\ and \rpah\ based on partial correlation analysis (Section~\ref{result:kbump_vs_rpah}). This indicates that the 2175\AA\ feature is weaker and PAH abundances are lower in strongly star-forming regions, and this is consistent with many previous findings (Section~\ref{lit_compare}). 
This implies that both the 2175\AA\ grain carrier and PAHs are linked to small dust grains that can be destroyed by UV photons, but that they are not necessarily linked to each other (i.e., PAHs may not be the 2175\AA\ grain carrier).
Performing similar analyses on a larger sample of galaxies that span a wider range in star-formation activity could provide new insight on these links. We provide prescriptions for \kbump\ and \rpah\ in local massive (metal-rich) galaxies with \sigmaSFR and sSFR, however the \kbump\ relations should be used with caution due to the fact that the bump strengths measured from \swift/UVOT are expected to be underestimated. 
No significant trends in \kbump\ and \rpah\ with gas-phase metallicity or galactocentric radii are found, however the metallicity range of our sample is limited ($8.40 < 12+\log[\mathrm{O/H}] < 8.65$), and a larger sample that extends to dwarf galaxies is needed to explore such trends robustly. 

Finally, this work highlights the limitations in characterising the 2175\AA\ feature and PAH abundances from photometric data. Future UV spectroscopic missions are needed to establish definitive links between the 2175\AA\ feature and its primary grain carrier. An important caveat to measuring PAH abundances from photometry is that the effects of dust heating can also cause a decrease in values of \rpah\ in a manner that is similar to PAH destruction. Breaking this degeneracy will require mid-IR spectroscopy (e.g., with MIRI) to determine dust continuum baselines more accurately.

\begin{acknowledgement}
The authors thank the anonymous referee, whose suggestions helped to clarify and improve the content of this work.
AJB thanks Miguel Querejeta for sharing the reduced \spitzer/IRAC mosaics for NGC2835 (pid 14033, PI J.C. Mu\~noz-Mateos) and for helpful feedback on the methodology for deriving stellar mass from \spitzer\ data. AJB thanks J\'er\'emy Chastenet for providing the \jwst\ saturation mask for NGC1365. 
IS acknowledges support by the Programa de Atracc{\'i}on de Talento of local Government of Madrid grant No. No.2022-T1/TIC-20472.
MD acknowledges support through the ESA fellowship programme.
We acknowledge the invaluable labor of the maintenance and clerical staff at our institutions, whose contributions make our scientific discoveries a reality. 
This research was conducted on Ngunnawal Indigenous land.
We acknowledge the use of public data from the \swift\ data archive. This research has made use of data and/or software provided by
the High Energy Astrophysics Science Archive Research Center
(HEASARC), which is a service of the Astrophysics Science Division
at NASA/GSFC and the High Energy Astrophysics Division
of the Smithsonian Astrophysical Observatory.
Based on observations collected at the European Southern Observatory under ESO programmes 1100.B-0651, 095.C-0473, and 094.C-0623 (PHANGS-MUSE; PI: Schinnerer), as well as 094.B-0321 (MAGNUM; PI: Marconi), 099.B-0242, 0100.B-0116, 098.B-0551 (MAD; PI: Carollo) and 097.B-0640 (TIMER; PI: Gadotti).
This work is based on observations made with the NASA/ ESA/CSA James Webb Space Telescope program 2107. The data were obtained from the Mikulski Archive for Space Telescopes at the Space Telescope Science Institute, which is operated by the Association of Universities for Research in Astronomy, Inc., under NASA contract NAS 5-03127 for JWST. 
This work uses data from SDSS.
Funding for SDSS-III has been provided by the Alfred P. Sloan Foundation, the Participating Institutions, the National Science Foundation, and the U.S. Department of Energy Office of Science. The SDSS-III web site is http://www.sdss3.org/.
SDSS-III is managed by the Astrophysical Research Consortium for the Participating Institutions of the SDSS-III Collaboration including the University of Arizona, the Brazilian Participation Group, Brookhaven National Laboratory, Carnegie Mellon University, University of Florida, the French Participation Group, the German Participation Group, Harvard University, the Instituto de Astrofisica de Canarias, the Michigan State/Notre Dame/JINA Participation Group, Johns Hopkins University, Lawrence Berkeley National Laboratory, Max Planck Institute for Astrophysics, Max Planck Institute for Extraterrestrial Physics, New Mexico State University, New York University, Ohio State University, Pennsylvania State University, University of Portsmouth, Princeton University, the Spanish Participation Group, University of Tokyo, University of Utah, Vanderbilt University, University of Virginia, University of Washington, and Yale University.
This publication makes use of data products from the Two Micron All Sky Survey, which is a joint project of the University of Massachusetts and the Infrared Processing and Analysis Center/California Institute of Technology, funded by the National Aeronautics and Space Administration and the National Science Foundation.

\end{acknowledgement}

\paragraph{Funding Statement}
Parts of this research were supported by the Australian Research Council Centre of Excellence for All Sky Astrophysics in 3 Dimensions (ASTRO 3D), through project No. CE170100013.


\paragraph{Competing Interests}
None

\paragraph{Data Availability Statement}

The galaxies used in this study are part of the Physics at High Angular resolution in Nearby GalaxieS (PHANGS) surveys. The PHANGS team provide public data products for the \jwst\ and VLT/MUSE data on their website\footnote{\url{https://sites.google.com/view/phangs/home/data}}. The raw \swift\ data are publicly available through the NASA High Energy Astrophysics Science Archive Research Center (HEASARC) service\footnote{\url{https://heasarc.gsfc.nasa.gov/docs/archive.html}}. Reduced \swift/UVOT images or other data products can be made available upon reasonable request to the first author.

\printendnotes

\bibliography{AJB_bib}

\begin{thebibliography}{}
\expandafter\ifx\csname natexlab\endcsname\relax\def\natexlab#1{#1}\fi

\bibitem[{{Aniano} {et~al.}(2011){Aniano}, {Draine}, {Gordon}, \&
  {Sandstrom}}]{aniano11}
{Aniano}, G., {Draine}, B.~T., {Gordon}, K.~D., \& {Sandstrom}, K. 2011, \pasp,
  123, 1218

\bibitem[{{Baldwin} {et~al.}(1981){Baldwin}, {Phillips}, \&
  {Terlevich}}]{baldwin81}
{Baldwin}, J.~A., {Phillips}, M.~M., \& {Terlevich}, R. 1981, \pasp, 93, 5

\bibitem[{{Battisti} {et~al.}(2016){Battisti}, {Calzetti}, \&
  {Chary}}]{battisti16}
{Battisti}, A.~J., {Calzetti}, D., \& {Chary}, R.-R. 2016, \apj, 818, 13

\bibitem[{{Battisti} {et~al.}(2017){Battisti}, {Calzetti}, \&
  {Chary}}]{battisti17b}
---. 2017, \apj, 851, 90

\bibitem[{{Battisti} {et~al.}(2020){Battisti}, {da Cunha}, {Shivaei}, \&
  {Calzetti}}]{battisti20}
{Battisti}, A.~J., {da Cunha}, E., {Shivaei}, I., \& {Calzetti}, D. 2020, \apj,
  888, 108

\bibitem[{{Battisti} {et~al.}(2019){Battisti}, {da Cunha}, {Grasha}, {Salvato},
  {Daddi}, {Davies}, {Jin}, {Liu}, {Schinnerer}, {Vaccari}, \& {(COSMOS
  Collaboration}}]{battisti19}
{Battisti}, A.~J., {da Cunha}, E., {Grasha}, K., {et~al.} 2019, \apj, 882, 61

\bibitem[{{Belles} {et~al.}(2023){Belles}, {Decleir}, {Bowman}, {Hagen},
  {Gronwall}, \& {Siegel}}]{belles23}
{Belles}, A., {Decleir}, M., {Bowman}, W.~P., {et~al.} 2023, \apj, 953, 54

\bibitem[{{Bern{\'e}} {et~al.}(2022){Bern{\'e}}, {Habart}, {Peeters},
  {Abergel}, {Bergin}, {Bernard-Salas}, {Bron}, {Cami}, {Dartois}, {Fuente},
  {Goicoechea}, {Gordon}, {Okada}, {Onaka}, {Robberto}, {R{\"o}llig},
  {Tielens}, {Vicente}, {Wolfire}, {Alarc{\'o}n}, {Boersma}, {Canin}, {Chown},
  {Dicken}, {Languignon}, {Le Gal}, {Pound}, {Trahin}, {Simmer}, {Sidhu}, {Van
  De Putte}, {Cuadrado}, {Guilloteau}, {Maragkoudakis}, {Schefter}, {Schirmer},
  {Cazaux}, {Aleman}, {Allamandola}, {Auchettl}, {Baratta}, {Bejaoui}, {Bera},
  {Bilalbegovi{\'c}}, {Black}, {Boulanger}, {Bouwman}, {Brandl}, {Brechignac},
  {Br{\"u}nken}, {Burkhardt}, {Candian}, {Cernicharo}, {Chabot}, {Chakraborty},
  {Champion}, {Colgan}, {Cooke}, {Coutens}, {Cox}, {Demyk}, {Donovan Meyer},
  {Engrand}, {Foschino}, {Garc{\'\i}a-Lario}, {Gavilan}, {Gerin}, {Godard},
  {Gottlieb}, {Guillard}, {Gusdorf}, {Hartigan}, {He}, {Herbst}, {Hornekaer},
  {J{\"a}ger}, {Janot-Pacheco}, {Joblin}, {Kaufman}, {Kemper}, {Kendrew},
  {Kirsanova}, {Klaassen}, {Knight}, {Kwok}, {Labiano}, {Lai}, {Lee},
  {Lefloch}, {Le Petit}, {Li}, {Linz}, {Mackie}, {Madden}, {Mascetti},
  {McGuire}, {Merino}, {Micelotta}, {Misselt}, {Morse}, {Mulas}, {Neelamkodan},
  {Ohsawa}, {Omont}, {Paladini}, {Palumbo}, {Pathak}, {Pendleton},
  {Petrignani}, {Pino}, {Puga}, {Rangwala}, {Rapacioli}, {Ricca},
  {Roman-Duval}, {Roser}, {Roueff}, {Rouill{\'e}}, {Salama}, {Sales},
  {Sandstrom}, {Sarre}, {Sciamma-O'Brien}, {Sellgren}, {Shannon}, {Shenoy},
  {Teyssier}, {Thomas}, {Togi}, {Verstraete}, {Witt}, {Wootten}, {Ysard},
  {Zettergren}, {Zhang}, {Zhang}, \& {Zhen}}]{berne22}
{Bern{\'e}}, O., {Habart}, {\'E}., {Peeters}, E., {et~al.} 2022, \pasp, 134,
  054301

\bibitem[{{Bertin}(2010)}]{bertin10}
{Bertin}, E. 2010, {SWarp: Resampling and Co-adding FITS Images Together},
  ascl:1010.068

\bibitem[{{Bianchi} {et~al.}(1996){Bianchi}, {Clayton}, {Bohlin}, {Hutchings},
  \& {Massey}}]{bianchi96}
{Bianchi}, L., {Clayton}, G.~C., {Bohlin}, R.~C., {Hutchings}, J.~B., \&
  {Massey}, P. 1996, \apj, 471, 203

\bibitem[{{Bradley} {et~al.}(2005){Bradley}, {Dai}, {Erni}, {Browning},
  {Graham}, {Weber}, {Smith}, {Hutcheon}, {Ishii}, {Bajt}, {Floss},
  {Stadermann}, \& {Sandford}}]{bradley05}
{Bradley}, J., {Dai}, Z.~R., {Erni}, R., {et~al.} 2005, Science, 307, 244

\bibitem[{{Buat} {et~al.}(2011){Buat}, {Giovannoli}, {Heinis}, {Charmandaris},
  {Coia}, {Daddi}, {Dickinson}, {Elbaz}, {Hwang}, {Morrison}, {Dasyra},
  {Aussel}, {Altieri}, {Dannerbauer}, {Kartaltepe}, {Leiton}, {Magdis},
  {Magnelli}, \& {Popesso}}]{buat11b}
{Buat}, V., {Giovannoli}, E., {Heinis}, S., {et~al.} 2011, \aap, 533, A93

\bibitem[{{Buat} {et~al.}(2012){Buat}, {Noll}, {Burgarella}, {Giovannoli},
  {Charmandaris}, {Pannella}, {Hwang}, {Elbaz}, {Dickinson}, {Magdis}, {Reddy},
  \& {Murphy}}]{buat12}
{Buat}, V., {Noll}, S., {Burgarella}, D., {et~al.} 2012, \aap, 545, A141

\bibitem[{{Bundy} {et~al.}(2015){Bundy}, {Bershady}, {Law}, {Yan}, {Drory},
  {MacDonald}, {Wake}, {Cherinka}, {S{\'a}nchez-Gallego}, {Weijmans}, {Thomas},
  {Tremonti}, {Masters}, {Coccato}, {Diamond-Stanic}, {Arag{\'o}n-Salamanca},
  {Avila-Reese}, {Badenes}, {Falc{\'o}n-Barroso}, {Belfiore}, {Bizyaev},
  {Blanc}, {Bland-Hawthorn}, {Blanton}, {Brownstein}, {Byler}, {Cappellari},
  {Conroy}, {Dutton}, {Emsellem}, {Etherington}, {Frinchaboy}, {Fu}, {Gunn},
  {Harding}, {Johnston}, {Kauffmann}, {Kinemuchi}, {Klaene}, {Knapen},
  {Leauthaud}, {Li}, {Lin}, {Maiolino}, {Malanushenko}, {Malanushenko}, {Mao},
  {Maraston}, {McDermid}, {Merrifield}, {Nichol}, {Oravetz}, {Pan}, {Parejko},
  {Sanchez}, {Schlegel}, {Simmons}, {Steele}, {Steinmetz}, {Thanjavur},
  {Thompson}, {Tinker}, {van den Bosch}, {Westfall}, {Wilkinson}, {Wright},
  {Xiao}, \& {Zhang}}]{bundy15}
{Bundy}, K., {Bershady}, M.~A., {Law}, D.~R., {et~al.} 2015, \apj, 798, 7

\bibitem[{{Calzetti}(2013)}]{calzetti13}
{Calzetti}, D. 2013, {Star Formation Rate Indicators}, ed.
  J.~{Falc{\'o}n-Barroso} \& J.~H. {Knapen}, 419

\bibitem[{{Calzetti} {et~al.}(2000){Calzetti}, {Armus}, {Bohlin}, {Kinney},
  {Koornneef}, \& {Storchi-Bergmann}}]{calzetti00}
{Calzetti}, D., {Armus}, L., {Bohlin}, R.~C., {et~al.} 2000, \apj, 533, 682

\bibitem[{{Calzetti} {et~al.}(1994){Calzetti}, {Kinney}, \&
  {Storchi-Bergmann}}]{calzetti94}
{Calzetti}, D., {Kinney}, A.~L., \& {Storchi-Bergmann}, T. 1994, \apj, 429, 582

\bibitem[{{Calzetti} {et~al.}(2007){Calzetti}, {Kennicutt}, {Engelbracht},
  {Leitherer}, {Draine}, {Kewley}, {Moustakas}, {Sosey}, {Dale}, {Gordon},
  {Helou}, {Hollenbach}, {Armus}, {Bendo}, {Bot}, {Buckalew}, {Jarrett}, {Li},
  {Meyer}, {Murphy}, {Prescott}, {Regan}, {Rieke}, {Roussel}, {Sheth}, {Smith},
  {Thornley}, \& {Walter}}]{calzetti07}
{Calzetti}, D., {Kennicutt}, R.~C., {Engelbracht}, C.~W., {et~al.} 2007, \apj,
  666, 870

\bibitem[{{Calzetti} {et~al.}(2024){Calzetti}, {Adamo}, {Linden}, {Gregg},
  {Krumholz}, {Bajaj}, {Bik}, {Cignoni}, {Correnti}, {Elmegreen}, {Faustino
  Vieira}, {Gallagher}, {Grasha}, {Gutermuth}, {Johnson}, {Messa}, {Melinder},
  {Ostlin}, {Pedrini}, {Sabbi}, {Smith}, \& {Tosi}}]{calzetti24}
{Calzetti}, D., {Adamo}, A., {Linden}, S.~T., {et~al.} 2024, arXiv e-prints,
  arXiv:2406.01831

\bibitem[{{Camps} \& {Baes}(2015)}]{camps&baes15}
{Camps}, P., \& {Baes}, M. 2015, Astronomy and Computing, 9, 20

\bibitem[{{Camps} \& {Baes}(2020)}]{camps&baes20}
---. 2020, Astronomy and Computing, 31, 100381

\bibitem[{{Cardelli} {et~al.}(1989){Cardelli}, {Clayton}, \&
  {Mathis}}]{cardelli89}
{Cardelli}, J.~A., {Clayton}, G.~C., \& {Mathis}, J.~S. 1989, \apj, 345, 245

\bibitem[{{Chabrier}(2003)}]{chabrier03}
{Chabrier}, G. 2003, \pasp, 115, 763

\bibitem[{{Charlot} \& {Fall}(2000)}]{charlot&fall00}
{Charlot}, S., \& {Fall}, S.~M. 2000, \apj, 539, 718

\bibitem[{{Chastenet} {et~al.}(2023){Chastenet}, {Sutter}, {Sandstrom},
  {Belfiore}, {Egorov}, {Larson}, {Leroy}, {Liu}, {Rosolowsky}, {Thilker},
  {Watkins}, {Williams}, {Barnes}, {Bigiel}, {Boquien}, {Chevance}, {Chiang},
  {Dale}, {Kruijssen}, {Emsellem}, {Grasha}, {Groves}, {Hassani}, {Hughes},
  {Kreckel}, {Meidt}, {Rickards Vaught}, {Sardone}, \&
  {Schinnerer}}]{chastenet23a}
{Chastenet}, J., {Sutter}, J., {Sandstrom}, K., {et~al.} 2023, \apjl, 944, L11

\bibitem[{{Chastenet} {et~al.}(2024){Chastenet}, {Sandstrom}, {Leroy}, {Bot},
  {Chiang}, {Chown}, {Gordon}, {Koch}, {Roussel}, {Sutter}, \&
  {Williams}}]{chastenet24}
{Chastenet}, J., {Sandstrom}, K.~M., {Leroy}, A.~K., {et~al.} 2024, arXiv
  e-prints, arXiv:2410.03835

\bibitem[{{Chown} {et~al.}(2024){Chown}, {Leroy}, {Sandstrom}, {Chastenet},
  {Sutter}, {Koch}, {Koziol}, {Neumann}, {Sun}, {Williams}, {Baron}, {Anand},
  {Barnes}, {Bazzi}, {Belfiore}, {Bolatto}, {Boquien}, {Cao}, {Chevance},
  {Colombo}, {Dale}, {Egorov}, {Eibensteiner}, {Emsellem}, {Hassani},
  {Henshaw}, {He}, {Kim}, {Kreckel}, {Meidt}, {Murphy}, {Oakes}, {Ostriker},
  {Pan}, {Pathak}, {Rosolowsky}, {Sarbadhicary}, {Schinnerer}, \&
  {Teng}}]{chown24}
{Chown}, R., {Leroy}, A.~K., {Sandstrom}, K., {et~al.} 2024, arXiv e-prints,
  arXiv:2410.05397

\bibitem[{{Clark} {et~al.}(2018){Clark}, {Verstocken}, {Bianchi}, {Fritz},
  {Viaene}, {Smith}, {Baes}, {Casasola}, {Cassara}, {Davies}, {De Looze}, {De
  Vis}, {Evans}, {Galametz}, {Jones}, {Lianou}, {Madden}, {Mosenkov}, \&
  {Xilouris}}]{clark18}
{Clark}, C.~J.~R., {Verstocken}, S., {Bianchi}, S., {et~al.} 2018, \aap, 609,
  A37

\bibitem[{{Clayton} {et~al.}(2015){Clayton}, {Gordon}, {Bianchi}, {Massa},
  {Fitzpatrick}, {Bohlin}, \& {Wolff}}]{clayton15}
{Clayton}, G.~C., {Gordon}, K.~D., {Bianchi}, L.~C., {et~al.} 2015, \apj, 815,
  14

\bibitem[{{Clayton} {et~al.}(2000){Clayton}, {Gordon}, \& {Wolff}}]{clayton00}
{Clayton}, G.~C., {Gordon}, K.~D., \& {Wolff}, M.~J. 2000, \apjs, 129, 147

\bibitem[{{Clayton} {et~al.}(2003){Clayton}, {Gordon}, {Salama}, {Allamandola},
  {Martin}, {Snow}, {Whittet}, {Witt}, \& {Wolff}}]{clayton03}
{Clayton}, G.~C., {Gordon}, K.~D., {Salama}, F., {et~al.} 2003, \apj, 592, 947

\bibitem[{{Conroy}(2013)}]{conroy13}
{Conroy}, C. 2013, \araa, 51, 393

\bibitem[{{Conroy} {et~al.}(2010){Conroy}, {Schiminovich}, \&
  {Blanton}}]{conroy10}
{Conroy}, C., {Schiminovich}, D., \& {Blanton}, M.~R. 2010, \apj, 718, 184

\bibitem[{{Cutri} {et~al.}(2003){Cutri}, {Skrutskie}, {van Dyk}, {Beichman},
  {Carpenter}, {Chester}, {Cambresy}, {Evans}, {Fowler}, {Gizis}, {Howard},
  {Huchra}, {Jarrett}, {Kopan}, {Kirkpatrick}, {Light}, {Marsh}, {McCallon},
  {Schneider}, {Stiening}, {Sykes}, {Weinberg}, {Wheaton}, {Wheelock}, \&
  {Zacarias}}]{cutri03}
{Cutri}, R.~M., {Skrutskie}, M.~F., {van Dyk}, S., {et~al.} 2003, {VizieR
  Online Data Catalog: 2MASS All-Sky Catalog of Point Sources (Cutri+ 2003)},
  VizieR On-line Data Catalog: II/246. Originally published in:
  2003yCat.2246....0C

\bibitem[{{da Cunha} {et~al.}(2008){da Cunha}, {Charlot}, \&
  {Elbaz}}]{daCunha08}
{da Cunha}, E., {Charlot}, S., \& {Elbaz}, D. 2008, \mnras, 388, 1595

\bibitem[{{Decleir}(2019)}]{decleir19PhDT}
{Decleir}, M. 2019, PhD thesis, Ghent University, Belgium

\bibitem[{{Decleir} {et~al.}(2019){Decleir}, {De Looze}, {Boquien}, {Baes},
  {Verstocken}, {Calzetti}, {Ciesla}, {Fritz}, {Kennicutt}, {Nersesian}, \&
  {Page}}]{decleir19}
{Decleir}, M., {De Looze}, I., {Boquien}, M., {et~al.} 2019, \mnras, 486, 743

\bibitem[{{Decleir} {et~al.}(2022){Decleir}, {Gordon}, {Andrews}, {Clayton},
  {Cushing}, {Misselt}, {Pendleton}, {Rayner}, {Vacca}, \&
  {Whittet}}]{decleir22}
{Decleir}, M., {Gordon}, K.~D., {Andrews}, J.~E., {et~al.} 2022, \apj, 930, 15

\bibitem[{{Draine}(2003)}]{draine03}
{Draine}, B.~T. 2003, \araa, 41, 241

\bibitem[{{Draine} \& {Li}(2007)}]{draine&li07}
{Draine}, B.~T., \& {Li}, A. 2007, \apj, 657, 810

\bibitem[{{Draine} {et~al.}(2021){Draine}, {Li}, {Hensley}, {Hunt},
  {Sandstrom}, \& {Smith}}]{draine21}
{Draine}, B.~T., {Li}, A., {Hensley}, B.~S., {et~al.} 2021, \apj, 917, 3

\bibitem[{{Draine} {et~al.}(2007){Draine}, {Dale}, {Bendo}, {Gordon}, {Smith},
  {Armus}, {Engelbracht}, {Helou}, {Kennicutt}, {Li}, {Roussel}, {Walter},
  {Calzetti}, {Moustakas}, {Murphy}, {Rieke}, {Bot}, {Hollenbach}, {Sheth}, \&
  {Teplitz}}]{draine07}
{Draine}, B.~T., {Dale}, D.~A., {Bendo}, G., {et~al.} 2007, \apj, 663, 866

\bibitem[{{Egorov} {et~al.}(2023){Egorov}, {Kreckel}, {Sandstrom}, {Leroy},
  {Glover}, {Groves}, {Kruijssen}, {Barnes}, {Belfiore}, {Bigiel}, {Blanc},
  {Boquien}, {Cao}, {Chastenet}, {Chevance}, {Congiu}, {Dale}, {Emsellem},
  {Grasha}, {Klessen}, {Larson}, {Liu}, {Murphy}, {Pan}, {Pessa}, {Pety},
  {Rosolowsky}, {Scheuermann}, {Schinnerer}, {Sutter}, {Thilker}, {Watkins}, \&
  {Williams}}]{egorov23}
{Egorov}, O.~V., {Kreckel}, K., {Sandstrom}, K.~M., {et~al.} 2023, \apjl, 944,
  L16

\bibitem[{{Emsellem} {et~al.}(2022){Emsellem}, {Schinnerer}, {Santoro},
  {Belfiore}, {Pessa}, {McElroy}, {Blanc}, {Congiu}, {Groves}, {Ho}, {Kreckel},
  {Razza}, {Sanchez-Blazquez}, {Egorov}, {Faesi}, {Klessen}, {Leroy}, {Meidt},
  {Querejeta}, {Rosolowsky}, {Scheuermann}, {Anand}, {Barnes},
  {Be{\v{s}}li{\'c}}, {Bigiel}, {Boquien}, {Cao}, {Chevance}, {Dale},
  {Eibensteiner}, {Glover}, {Grasha}, {Henshaw}, {Hughes}, {Koch}, {Kruijssen},
  {Lee}, {Liu}, {Pan}, {Pety}, {Saito}, {Sandstrom}, {Schruba}, {Sun},
  {Thilker}, {Usero}, {Watkins}, \& {Williams}}]{emsellem22}
{Emsellem}, E., {Schinnerer}, E., {Santoro}, F., {et~al.} 2022, \aap, 659, A191

\bibitem[{{Euclid Collaboration} {et~al.}(2024){Euclid Collaboration},
  {Mellier}, {Abdurro'uf}, {Acevedo Barroso}, {Ach{\'u}carro}, {Adamek},
  {Adam}, {Addison}, {Aghanim}, {Aguena}, {Ajani}, {Akrami}, {Al-Bahlawan},
  {Alavi}, {Albuquerque}, {Alestas}, {Alguero}, {Allaoui}, {Allen}, {Allevato},
  {Alonso-Tetilla}, {Altieri}, {Alvarez-Candal}, {Amara}, {Amendola}, {Amiaux},
  {Andika}, {Andreon}, {Andrews}, {Angora}, {Angulo}, {Annibali}, {Anselmi},
  {Anselmi}, {Arcari}, {Archidiacono}, {Aric{\`o}}, {Arnaud}, {Arnouts},
  {Asgari}, {Asorey}, {Atayde}, {Atek}, {Atrio-Barandela}, {Aubert}, {Aubourg},
  {Auphan}, {Auricchio}, {Aussel}, {Aussel}, {Avelino}, {Avgoustidis}, {Avila},
  {Awan}, {Azzollini}, {Baccigalupi}, {Bachelet}, {Bacon}, {Baes}, {Bagley},
  {Bahr-Kalus}, {Balaguera-Antolinez}, {Balbinot}, {Balcells}, {Baldi},
  {Baldry}, {Balestra}, {Ballardini}, {Ballester}, {Balogh}, {Ba{\~n}ados},
  {Barbier}, {Bardelli}, {Barreiro}, {Barriere}, {Barros}, {Barthelemy},
  {Bartolo}, {Basset}, {Battaglia}, {Battisti}, {Baugh}, {Baumont},
  {Bazzanini}, {Beaulieu}, {Beckmann}, {Belikov}, {Bel}, {Bellagamba}, {Bella},
  {Bellini}, {Benabed}, {Bender}, {Benevento}, {Bennett}, {Benson},
  {Bergamini}, {Bermejo-Climent}, {Bernardeau}, {Bertacca}, {Berthe},
  {Berthier}, {Bethermin}, {Beutler}, {Bevillon}, {Bhargava}, {Bhatawdekar},
  {Bisigello}, {Biviano}, {Blake}, {Blanchard}, {Blazek}, {Blot}, {Bosco},
  {Bodendorf}, {Boenke}, {B{\"o}hringer}, {Bolzonella}, {Bonchi}, {Bonici},
  {Bonino}, {Bonino}, {Bonvin}, {Bon}, {Booth}, {Borgani}, {Borlaff},
  {Borsato}, {Bosco}, {Bose}, {Botticella}, {Boucaud}, {Bouche}, {Boucher},
  {Boutigny}, {Bouvard}, {Bouy}, {Bowler}, {Bozza}, {Bozzo}, {Branchini},
  {Brau-Nogue}, {Brekke}, {Bremer}, {Brescia}, {Breton}, {Brinchmann},
  {Brinckmann}, {Brockley-Blatt}, {Brodwin}, {Brouard}, {Brown}, {Bruton},
  {Bucko}, {Buddelmeijer}, {Buenadicha}, {Buitrago}, {Burger}, {Burigana},
  {Busillo}, {Busonero}, {Cabanac}, {Cabayol-Garcia}, {Cagliari}, {Caillat},
  {Caillat}, {Calabrese}, {Calabro}, {Calderone}, {Calura}, {Camacho Quevedo},
  {Camera}, {Campos}, {Canas-Herrera}, {Candini}, {Cantiello}, {Capobianco},
  {Cappellaro}, {Cappelluti}, {Cappi}, {Caputi}, {Cara}, {Carbone}, {Cardone},
  {Carella}, {Carlberg}, {Carle}, {Carminati}, {Caro}, {Carrasco}, {Carretero},
  {Carrilho}, {Carron Duque}, {Carry}, {Carvalho}, {Carvalho}, {Casas},
  {Casas}, {Casenove}, {Casey}, {Cassata}, {Castander}, {Castelao},
  {Castellano}, {Castiblanco}, {Castignani}, {Castro}, {Cavet}, {Cavuoti},
  {Chabaud}, {Chambers}, {Charles}, {Charlot}, {Chartab}, {Chary}, {Chaumeil},
  {Cho}, {Chon}, {Ciancetta}, {Ciliegi}, {Cimatti}, {Cimino}, {Cioni},
  {Claydon}, {Cleland}, {Cl{\'e}ment}, {Clements}, {Clerc}, {Clesse}, {Codis},
  {Cogato}, {Colbert}, {Cole}, {Coles}, {Collett}, {Collins}, {Colodro-Conde},
  {Colombo}, {Combes}, {Conforti}, {Congedo}, {Conseil}, {Conselice},
  {Contarini}, {Contini}, {Conversi}, {Cooray}, {Copin}, {Corasaniti},
  {Corcho-Caballero}, {Corcione}, {Cordes}, {Corpace}, {Correnti}, {Costanzi},
  {Costille}, {Courbin}, {Courcoult Mifsud}, {Courtois}, {Cousinou}, {Covone},
  {Cowell}, {Cragg}, {Cresci}, {Cristiani}, {Crocce}, {Cropper}, {E Crouzet},
  {Csizi}, {Cuby}, {Cucchetti}, {Cucciati}, {Cuillandre}, {Cunha}, {Cuozzo},
  {Daddi}, {D'Addona}, {Dafonte}, {Dagoneau}, {Dalessandro}, {Dalton},
  {D'Amico}, {Dannerbauer}, {Danto}, {Das}, {Da Silva}, {da Silva}, {Daste},
  {Davies}, {Davini}, {de Boer}, {Decarli}, {De Caro}, {Degaudenzi}, {Degni},
  {de Jong}, {de la Bella}, {de la Torre}, {Delhaise}, {Delley}, {Delucchi},
  {De Lucia}, {Denniston}, {De Paolis}, {De Petris}, {Derosa}, {Desai},
  {Desjacques}, {Despali}, {Desprez}, {De Vicente-Albendea}, {Deville}, {Dias},
  {D{\'\i}az-S{\'a}nchez}, {Diaz}, {Di Domizio}, {Diego}, {Di Ferdinando}, {Di
  Giorgio}, {Dimauro}, {Dinis}, {Dolag}, {Dolding}, {Dole}, {Dom{\'\i}nguez
  S{\'a}nchez}, {Dor{\'e}}, {Dournac}, {Douspis}, {Dreihahn}, {Droge}, {Dryer},
  {Dubath}, {Duc}, {Ducret}, {Duffy}, {Dufresne}, {Duncan}, {Dupac}, {Duret},
  {Durrer}, {Durret}, {Dusini}, {Ealet}, {Eggemeier}, {Eisenhardt}, {Elbaz},
  {Elkhashab}, {Ellien}, {Endicott}, {Enia}, {Erben}, {Escartin Vigo},
  {Escoffier}, {Escudero Sanz}, {Essert}, {Ettori}, {Ezziati}, {Fabbian},
  {Fabricius}, {Fang}, {Farina}, {Farina}, {Farinelli}, {Farrens}, {Faustini},
  {Feltre}, {Ferguson}, {Ferrando}, {Ferrari}, {Ferr{\'e}-Mateu}, {Ferreira},
  {Ferreras}, {Ferrero}, {Ferriol}, {Ferruit}, {Filleul}, {Finelli},
  {Finkelstein}, {Finoguenov}, {Fiorini}, {Flentge}, {Focardi}, {Fonseca},
  {Fontana}, {Fontanot}, {Fornari}, {Fosalba}, {Fossati}, {Fotopoulou},
  {Fouchez}, {Fourmanoit}, {Frailis}, {Fraix-Burnet}, {Franceschi}, {Franco},
  {Franzetti}, {Freihoefer}, {Frittoli}, {Frugier}, {Frusciante}, {Fumagalli},
  {Fumagalli}, {Fumana}, {Fu}, {Gabarra}, {Galeotta}, {Galluccio}, {Ganga},
  {Gao}, {Garc{\'\i}a-Bellido}, {Garcia}, {Gardner}, {Garilli},
  {Gaspar-Venancio}, {Gasparetto}, {Gautard}, {Gavazzi}, {Gaztanaga},
  {Genolet}, {Genova Santos}, {Gentile}, {George}, {Ghaffari}, {Giacomini},
  {Gianotti}, {Gibb}, {Gillard}, {Gillis}, {Ginolfi}, {Giocoli}, {Girardi},
  {Giri}, {Goh}, {G{\'o}mez-Alvarez}, {Gonzalez}, {Gonzalez}, {Gonzalez},
  {Gouyou Beauchamps}, {Gozaliasl}, {Gracia-Carpio}, {Grandis}, {Granett},
  {Granvik}, {Grazian}, {Gregorio}, {Grenet}, {Grillo}, {Grupp}, {Gruppioni},
  {Gruppuso}, {Guerbuez}, {Guerrini}, {Guidi}, {Guillard}, {Gutierrez},
  {Guttridge}, {Guzzo}, {Gwyn}, {Haapala}, {Haase}, {Haddow}, {Hailey}, {Hall},
  {Hall}, {Hamaus}, {Haridasu}, {Harnois-D{\'e}raps}, {Harper}, {Hartley},
  {Hasinger}, {Hassani}, {Hatch}, {Haugan}, {H{\"a}u{\ss}ler}, {Heavens},
  {Heisenberg}, {Helmi}, {Helou}, {Hemmati}, {Henares}, {Herent},
  {Hern{\'a}ndez-Monteagudo}, {Heuberger}, {Hewett}, {Heydenreich},
  {Hildebrandt}, {Hirschmann}, {Hjorth}, {Hoar}, {Hoekstra}, {Holland},
  {Holliman}, {Holmes}, {Hook}, {Horeau}, {Hormuth}, {Hornstrup}, {Hosseini},
  {Hu}, {Hudelot}, {Hudson}, {Huertas-Company}, {Huff}, {Hughes}, {Humphrey},
  {Hunt}, {Huynh}, {Ibata}, {Ichikawa}, {Iglesias-Groth}, {Ilbert}, {Ili{\'c}},
  {Ingoglia}, {Iodice}, {Israel}, {Israelsson}, {Izzo}, {Jablonka}, {Jackson},
  {Jacobson}, {Jafariyazani}, {Jahnke}, {Jansen}, {Jarvis}, {Jasche}, {Jauzac},
  {Jeffrey}, {Jhabvala}, {Jimenez-Teja}, {Jimenez Mu{\~n}oz}, {Joachimi},
  {Johansson}, {Joudaki}, {Jullo}, {Kajava}, {Kang}, {Kannawadi}, {Kansal},
  {Karagiannis}, {K{\"a}rcher}, {Kashlinsky}, {Kazandjian}, {Keck},
  {Keih{\"a}nen}, {Kerins}, {Kermiche}, {Khalil}, {Kiessling}, {Kiiveri},
  {Kilbinger}, {Kim}, {King}, {Kirkpatrick}, {Kitching}, {Kluge}, {Knabenhans},
  {Knapen}, {Knebe}, {Kneib}, {Kohley}, {Koopmans}, {Koskinen}, {Koulouridis},
  {Kou}, {Kov{\'a}cs}, {Kova\{{\v{c}}\}i{\'c}}, {Kowalczyk}, {Koyama},
  {Kraljic}, {Krause}, {Kruk}, {Kubik}, {Kuchner}, {Kuijken}, {K{\"u}mmel},
  {Kunz}, {Kurki-Suonio}, {Lacasa}, {Lacey}, {La Franca}, {Lagarde}, {Lahav},
  {Laigle}, {La Marca}, {La Marle}, {Lamine}, {Lam}, {Lan{\c{c}}on}, {Landt},
  {Langer}, {Lapi}, {Larcheveque}, {Larsen}, {Lattanzi}, {Laudisio}, {Laugier},
  {Laureijs}, {Lavaux}, {Lawrenson}, {Lazanu}, {Lazeyras}, {Le Boulc'h}, {Le
  Brun}, {Le Brun}, {Leclercq}, {Lee}, {Le Graet}, {Legrand}, {Leirvik}, {Le
  Jeune}, {Lembo}, {Le Mignant}, {Lepinzan}, {Lepori}, {Lesci}, {Lesgourgues},
  {Leuzzi}, {Levi}, {Liaudat}, {Libet}, {Liebing}, {Ligori}, {Lilje}, {Lin},
  {Linde}, {Linder}, {Lindholm}, {Linke}, {Li}, {Liu}, {Lloro}, {Lobo},
  {Lodieu}, {Lombardi}, {Lombriser}, {Lonare}, {Longo}, {L{\'o}pez-Caniego},
  {Lopez Lopez}, {Alvarez}, {Loureiro}, {Loveday}, {Lusso}, {Macias-Perez},
  {Maciaszek}, {Magliocchetti}, {Magnard}, {Magnier}, {Magro}, {Mahler},
  {Mainetti}, {Maino}, {Maiorano}, {Maiorano}, {Malavasi}, {Mamon}, {Mancini},
  {Mandelbaum}, {Manera}, {Manj{\'o}n-Garc{\'\i}a}, {Mannucci}, {Mansutti},
  {Manteiga Outeiro}, {Maoli}, {Maraston}, {Marcin}, {Marcos-Arenal},
  {Margalef-Bentabol}, {Marggraf}, {Marinucci}, {Marinucci}, {Markovic},
  {Marleau}, {Marpaud}, {Martignac}, {Mart{\'\i}n-Fleitas}, {Martin-Moruno},
  {Martin}, {Martinelli}, {Martinet}, {Martin}, {Martins}, {Marulli},
  {Massari}, {Massey}, {Masters}, {Matarrese}, {Matsuoka}, {Matthew},
  {Maughan}, {Mauri}, {Maurin}, {Maurogordato}, {McCarthy}, {McConnachie},
  {McCracken}, {McDonald}, {McEwen}, {McPartland}, {Medinaceli}, {Mehta},
  {Mei}, {Melchior}, {Melin}, {M{\'e}nard}, {Mendes}, {Mendez-Abreu},
  {Meneghetti}, {Mercurio}, {Merlin}, {Metcalf}, {Meylan}, {Migliaccio},
  {Mignoli}, {Miller}, {Miluzio}, {Milvang-Jensen}, {Mimoso}, {Miquel},
  {Miyatake}, {Mobasher}, {Mohr}, {Monaco}, {Mongui{\'o}}, {Montoro}, {Mora},
  {Moradinezhad Dizgah}, {Moresco}, {Moretti}, {Morgante}, {Morisset},
  {Moriya}, {Morris}, {Mortlock}, {Moscardini}, {Mota}, {Moustakas}, {Moutard},
  {M{\"u}ller}, {Munari}, {Murphree}, {Murray}, {Murray}, {Musi}, {Nadathur},
  {Nagam}, {Nagao}, {Naidoo}, {Nakajima}, {Nally}, {Natoli}, {Navarro-Alsina},
  {Navarro Girones}, {Neissner}, {Nersesian}, {Nesseris}, {Nguyen-Kim},
  {Nicastro}, {Nichol}, {Nielbock}, {Niemi}, {Nieto}, {Nilsson}, {Noller},
  {Norberg}, {Nourizonoz}, {Ntelis}, {Nucita}, {Nugent}, {Nunes}, {Nutma},
  {Ocampo}, {Odier}, {Oesch}, {Oguri}, {Magalhaes Oliveira}, {Onoue},
  {Oosterbroek}, {Oppizzi}, {Ordenovic}, {Osato}, {Pacaud}, {Pace}, {Padilla},
  {Paech}, {Pagano}, {Page}, {Palazzi}, {Paltani}, {Pamuk}, {Pandolfi},
  {Paoletti}, {Paolillo}, {Papaderos}, {Pardede}, {Parimbelli}, {Parmar},
  {Partmann}, {Pasian}, {Passalacqua}, {Paterson}, {Patrizii}, {Pattison},
  {Paulino-Afonso}, {Paviot}, {Peacock}, {Pearce}, {Pedersen}, {Peel},
  {Peletier}, {Pellejero Ibanez}, {Pello}, {Penny}, {Percival},
  {Perez-Garrido}, {Perotto}, {Pettorino}, {Pezzotta}, {Pezzuto}, {Philippon},
  {Piersanti}, {Pietroni}, {Piga}, {Pilo}, {Pires}, {Pisani}, {Pizzella},
  {Pizzuti}, {Plana}, {Polenta}, {Pollack}, {Poncet}, {P{\"o}ntinen}, {Pool},
  {Popa}, {Popa}, {Popp}, {Porciani}, {Porth}, {Potter}, {Poulain},
  {Pourtsidou}, {Pozzetti}, {Prandoni}, {Pratt}, {Prezelus}, {Prieto}, {Pugno},
  {Quai}, {Quilley}, {Racca}, {Raccanelli}, {R{\'a}cz}, {Radinovi{\'c}},
  {Radovich}, {Ragagnin}, {Ragnit}, {Raison}, {Ramos-Chernenko}, {Ranc},
  {Raylet}, {Rebolo}, {Refregier}, {Reimberg}, {Reiprich}, {Renk}, {Renzi},
  {Retre}, {Revaz}, {Reyl{\'e}}, {Reynolds}, {Rhodes}, {Ricci}, {Ricci},
  {Riccio}, {Ricken}, {Rissanen}, {Risso}, {Rix}, {Robin}, {Rocca-Volmerange},
  {Rocci}, {Rodenhuis}, {Rodighiero}, {Rodriguez Monroy}, {Rollins},
  {Romanello}, {Roman}, {Romelli}, {Romero-Gomez}, {Roncarelli}, {Rosati},
  {Rosset}, {Rossetti}, {Roster}, {Rottgering}, {Rozas-Fern{\'a}ndez}, {Ruane},
  {Rubino-Martin}, {Rudolph}, {Ruppin}, {Rusholme}, {Sacquegna},
  {S{\'a}ez-Casares}, {Saga}, {Saglia}, {Sahl{\'e}n}, {Saifollahi}, {Sakr},
  {Salvalaggio}, {Salvaterra}, {Salvati}, {Salvato}, {Salvignol},
  {S{\'a}nchez}, {Sanchez}, {Sanders}, {Sapone}, {Saponara}, {Sarpa}, {Sarron},
  {Sartori}, {Sassolas}, {Sauniere}, {Sauvage}, {Sawicki}, {Scaramella},
  {Scarlata}, {Scharr{\'e}}, {Schaye}, {Schewtschenko}, {Schindler},
  {Schinnerer}, {Schirmer}, {Schmidt}, {Schmidt}, {Schmidt}, {Schneider},
  {Schneider}, {Schneider}, {Sch{\"o}neberg}, {Schrabback}, {Schultheis},
  {Schulz}, {Schwartz}, {Sciotti}, {Scodeggio}, {Scognamiglio}, {Scott},
  {Scottez}, {Secroun}, {Sefusatti}, {Seidel}, {Seiffert}, {Sellentin},
  {Selwood}, {Semboloni}, {Sereno}, {Serjeant}, {Serrano}, {Shankar},
  {Sharples}, {Short}, {Shulevski}, {Shuntov}, {Sias}, {Sikkema}, {Silvestri},
  {Simon}, {Sirignano}, {Sirri}, {Skottfelt}, {Slezak}, {Sluse}, {Smith},
  {Smith}, {Smith}, {Smit}, {Soldano}, {Solheim}, {Sorce}, {Sorrenti},
  {Soubrie}, {Spinoglio}, {Spurio Mancini}, {Stadel}, {Stagnaro}, {Stanco},
  {Stanford}, {Starck}, {Stassi}, {Steinwagner}, {Stern}, {Stone}, {Strada},
  {Strafella}, {Stramaccioni}, {Surace}, {Sureau}, {Suyu}, {Swindells},
  {Szafraniec}, {Szapudi}, {Taamoli}, {Talia}, {Tallada-Cresp{\'\i}},
  {Tanidis}, {Tao}, {Tarr{\'\i}o}, {Tavagnacco}, {Taylor}, {Taylor}, {Taylor},
  {Teixeira}, {Tenti}, {Teodoro Idiago}, {Teplitz}, {Tereno}, {Tessore},
  {Testa}, {Testera}, {Tewes}, {Teyssier}, {Theret}, {Thizy}, {Thomas}, {Toba},
  {Toft}, {Toledo-Moreo}, {Tolstoy}, {Tommasi}, {Torbaniuk}, {Torradeflot},
  {Tortora}, {Tosi}, {Tosti}, {Trifoglio}, {Troja}, {Trombetti}, {Tronconi},
  {Tsedrik}, {Tsyganov}, {Tucci}, {Tutusaus}, {Uhlemann}, {Ulivi}, {Urbano},
  {Vacher}, {Vaillon}, {Valdes}, {Valentijn}, {Valenziano}, {Valieri},
  {Valiviita}, {Van den Broeck}, {Vassallo}, {Vavrek}, {Venemans}, {Venhola},
  {Ventura}, {Verdoes Kleijn}, {Vergani}, {Verma}, {Vernizzi}, {Veropalumbo},
  {Verza}, {Vescovi}, {Vibert}, {Viel}, {Vielzeuf}, {Viglione}, {Viitanen},
  {Villaescusa-Navarro}, {Vinciguerra}, {Visticot}, {Voggel}, {von
  Wietersheim-Kramsta}, {Vriend}, {Wachter}, {Walmsley}, {Walth}, {Walton},
  {Walton}, {Wander}, {Wang}, {Wang}, {Weaver}, {Weller}, {Whalen}, {Wiesmann},
  {Wilde}, {Williams}, {Winther}, {Wittje}, {Wong}, {Wright}, {Yankelevich},
  {Yeung}, {Youles}, {Yung}, {Zacchei}, {Zalesky}, {Zamorani}, {Zamorano
  Vitorelli}, {Zanoni Marc}, {Zennaro}, {Zerbi}, {Zinchenko}, {Zoubian},
  {Zucca}, \& {Zumalacarregui}}]{euclid24}
{Euclid Collaboration}, {Mellier}, Y., {Abdurro'uf}, {et~al.} 2024, arXiv
  e-prints, arXiv:2405.13491

\bibitem[{{Ferreras} {et~al.}(2021){Ferreras}, {Tress}, {Bruzual}, {Charlot},
  {Page}, {Yershov}, {Kuin}, {Kawata}, \& {Cropper}}]{ferreras21}
{Ferreras}, I., {Tress}, M., {Bruzual}, G., {et~al.} 2021, \mnras, 505, 283

\bibitem[{{Fischera} \& {Dopita}(2011)}]{fischera&dopita11}
{Fischera}, J., \& {Dopita}, M. 2011, \aap, 533, A117

\bibitem[{{Fitzpatrick}(1999)}]{fitzpatrick99}
{Fitzpatrick}, E.~L. 1999, \pasp, 111, 63

\bibitem[{{Fitzpatrick} \& {Massa}(1986)}]{fitzpatrick&massa86}
{Fitzpatrick}, E.~L., \& {Massa}, D. 1986, \apj, 307, 286

\bibitem[{{Fitzpatrick} \& {Massa}(1990)}]{fitzpatrick&massa90}
---. 1990, \apjs, 72, 163

\bibitem[{{Fitzpatrick} {et~al.}(2019){Fitzpatrick}, {Massa}, {Gordon},
  {Bohlin}, \& {Clayton}}]{fitzpatrick19}
{Fitzpatrick}, E.~L., {Massa}, D., {Gordon}, K.~D., {Bohlin}, R., \& {Clayton},
  G.~C. 2019, \apj, 886, 108

\bibitem[{{Gallerani} {et~al.}(2010){Gallerani}, {Maiolino}, {Juarez}, {Nagao},
  {Marconi}, {Bianchi}, {Schneider}, {Mannucci}, {Oliva}, {Willott}, {Jiang},
  \& {Fan}}]{gallerani10}
{Gallerani}, S., {Maiolino}, R., {Juarez}, Y., {et~al.} 2010, \aap, 523, A85

\bibitem[{{Galliano} {et~al.}(2018){Galliano}, {Galametz}, \&
  {Jones}}]{galliano18}
{Galliano}, F., {Galametz}, M., \& {Jones}, A.~P. 2018, \araa, 56, 673

\bibitem[{{Gehrels} {et~al.}(2004){Gehrels}, {Chincarini}, {Giommi}, {Mason},
  {Nousek}, {Wells}, {White}, {Barthelmy}, {Burrows}, {Cominsky}, {Hurley},
  {Marshall}, {M{\'e}sz{\'a}ros}, {Roming}, {Angelini}, {Barbier}, {Belloni},
  {Campana}, {Caraveo}, {Chester}, {Citterio}, {Cline}, {Cropper}, {Cummings},
  {Dean}, {Feigelson}, {Fenimore}, {Frail}, {Fruchter}, {Garmire}, {Gendreau},
  {Ghisellini}, {Greiner}, {Hill}, {Hunsberger}, {Krimm}, {Kulkarni}, {Kumar},
  {Lebrun}, {Lloyd-Ronning}, {Markwardt}, {Mattson}, {Mushotzky}, {Norris},
  {Osborne}, {Paczynski}, {Palmer}, {Park}, {Parsons}, {Paul}, {Rees},
  {Reynolds}, {Rhoads}, {Sasseen}, {Schaefer}, {Short}, {Smale}, {Smith},
  {Stella}, {Tagliaferri}, {Takahashi}, {Tashiro}, {Townsley}, {Tueller},
  {Turner}, {Vietri}, {Voges}, {Ward}, {Willingale}, {Zerbi}, \&
  {Zhang}}]{gehrels04}
{Gehrels}, N., {Chincarini}, G., {Giommi}, P., {et~al.} 2004, \apj, 611, 1005

\bibitem[{{Gordon} {et~al.}(1997){Gordon}, {Calzetti}, \& {Witt}}]{gordon97}
{Gordon}, K.~D., {Calzetti}, D., \& {Witt}, A.~N. 1997, \apj, 487, 625

\bibitem[{{Gordon} {et~al.}(2009){Gordon}, {Cartledge}, \&
  {Clayton}}]{gordon09}
{Gordon}, K.~D., {Cartledge}, S., \& {Clayton}, G.~C. 2009, \apj, 705, 1320

\bibitem[{{Gordon} {et~al.}(2003){Gordon}, {Clayton}, {Misselt}, {Landolt}, \&
  {Wolff}}]{gordon03}
{Gordon}, K.~D., {Clayton}, G.~C., {Misselt}, K.~A., {Landolt}, A.~U., \&
  {Wolff}, M.~J. 2003, \apj, 594, 279

\bibitem[{{Gordon} {et~al.}(2021){Gordon}, {Misselt}, {Bouwman}, {Clayton},
  {Decleir}, {Hines}, {Pendleton}, {Rieke}, {Smith}, \& {Whittet}}]{gordon21}
{Gordon}, K.~D., {Misselt}, K.~A., {Bouwman}, J., {et~al.} 2021, \apj, 916, 33

\bibitem[{{Gordon} {et~al.}(2024){Gordon}, {Fitzpatrick}, {Massa}, {Bohlin},
  {Chastenet}, {Murray}, {Clayton}, {Lennon}, {Misselt}, \&
  {Sandstrom}}]{gordon24}
{Gordon}, K.~D., {Fitzpatrick}, E.~L., {Massa}, D., {et~al.} 2024, arXiv
  e-prints, arXiv:2405.12792

\bibitem[{{Gregg} {et~al.}(2024){Gregg}, {Calzetti}, {Adamo}, {Bajaj}, {Ryon},
  {Linden}, {Correnti}, {Cignoni}, {Messa}, {Sabbi}, {Gallagher}, {Grasha},
  {Pedrini}, {Gutermuth}, {Melinder}, {Kotulla}, {P{\'e}rez}, {Krumholz},
  {Bik}, {{\"O}stlin}, {Johnson}, {Bortolini}, {Smith}, {Tosi}, {Maji}, \&
  {Faustino Vieira}}]{gregg24}
{Gregg}, B., {Calzetti}, D., {Adamo}, A., {et~al.} 2024, arXiv e-prints,
  arXiv:2405.09667

\bibitem[{{Groves} {et~al.}(2023){Groves}, {Kreckel}, {Santoro}, {Belfiore},
  {Zavodnik}, {Congiu}, {Egorov}, {Emsellem}, {Grasha}, {Leroy}, {Scheuermann},
  {Schinnerer}, {Watkins}, {Barnes}, {Bigiel}, {Dale}, {Glover}, {Pessa},
  {Sanchez-Blazquez}, \& {Williams}}]{groves23}
{Groves}, B., {Kreckel}, K., {Santoro}, F., {et~al.} 2023, \mnras, 520, 4902

\bibitem[{{Hagen} {et~al.}(2017){Hagen}, {Siegel}, {Hoversten}, {Gronwall},
  {Immler}, \& {Hagen}}]{hagen17}
{Hagen}, L.~M.~Z., {Siegel}, M.~H., {Hoversten}, E.~A., {et~al.} 2017, \mnras,
  466, 4540

\bibitem[{{Hensley} \& {Draine}(2020)}]{hensley&draine20}
{Hensley}, B.~S., \& {Draine}, B.~T. 2020, \apj, 895, 38

\bibitem[{{Hensley} \& {Draine}(2021)}]{hensley&draine21}
---. 2021, \apj, 906, 73

\bibitem[{{Hensley} \& {Draine}(2023)}]{hensley&draine23}
---. 2023, \apj, 948, 55

\bibitem[{{Hopkins} {et~al.}(2004){Hopkins}, {Strauss}, {Hall}, {Richards},
  {Cooper}, {Schneider}, {Vanden Berk}, {Jester}, {Brinkmann}, \&
  {Szokoly}}]{hopkins04}
{Hopkins}, P.~F., {Strauss}, M.~A., {Hall}, P.~B., {et~al.} 2004, \aj, 128,
  1112

\bibitem[{{Hoversten} {et~al.}(2011){Hoversten}, {Gronwall}, {Vanden Berk},
  {Basu-Zych}, {Breeveld}, {Brown}, {Kuin}, {Page}, {Roming}, \&
  {Siegel}}]{hoversten11}
{Hoversten}, E.~A., {Gronwall}, C., {Vanden Berk}, D.~E., {et~al.} 2011, \aj,
  141, 205

\bibitem[{{Kashino} {et~al.}(2021){Kashino}, {Lilly}, {Silverman}, {Renzini},
  {Daddi}, {Bardelli}, {Cucciati}, {Kartaltepe}, {Mainieri}, {Pell{\'o}},
  {Peng}, {Sanders}, \& {Zucca}}]{kashino21}
{Kashino}, D., {Lilly}, S.~J., {Silverman}, J.~D., {et~al.} 2021, \apj, 909,
  213

\bibitem[{{Kauffmann} {et~al.}(2003){Kauffmann}, {Heckman}, {Tremonti},
  {Brinchmann}, {Charlot}, {White}, {Ridgway}, {Brinkmann}, {Fukugita}, {Hall},
  {Ivezi{\'c}}, {Richards}, \& {Schneider}}]{kauffmann03c}
{Kauffmann}, G., {Heckman}, T.~M., {Tremonti}, C., {et~al.} 2003, \mnras, 346,
  1055

\bibitem[{{Kendall}(1942)}]{kendall42}
{Kendall}, M.~G. 1942, Biometrika, 32, 277

\bibitem[{{Kennicutt} \& {Evans}(2012)}]{kennicutt&evans12}
{Kennicutt}, R.~C., \& {Evans}, N.~J. 2012, \araa, 50, 531

\bibitem[{{Kewley} {et~al.}(2001){Kewley}, {Dopita}, {Sutherland}, {Heisler},
  \& {Trevena}}]{kewley01}
{Kewley}, L.~J., {Dopita}, M.~A., {Sutherland}, R.~S., {Heisler}, C.~A., \&
  {Trevena}, J. 2001, \apj, 556, 121

\bibitem[{{Kewley} {et~al.}(2019){Kewley}, {Nicholls}, \&
  {Sutherland}}]{kewley19}
{Kewley}, L.~J., {Nicholls}, D.~C., \& {Sutherland}, R.~S. 2019, \araa, 57, 511

\bibitem[{{Kreckel} {et~al.}(2013){Kreckel}, {Groves}, {Schinnerer}, {Johnson},
  {Aniano}, {Calzetti}, {Croxall}, {Draine}, {Gordon}, {Crocker}, {Dale},
  {Hunt}, {Kennicutt}, {Meidt}, {Smith}, \& {Tabatabaei}}]{kreckel13}
{Kreckel}, K., {Groves}, B., {Schinnerer}, E., {et~al.} 2013, \apj, 771, 62

\bibitem[{{Kreckel} {et~al.}(2019){Kreckel}, {Ho}, {Blanc}, {Groves},
  {Santoro}, {Schinnerer}, {Bigiel}, {Chevance}, {Congiu}, {Emsellem}, {Faesi},
  {Glover}, {Grasha}, {Kruijssen}, {Lang}, {Leroy}, {Meidt}, {McElroy}, {Pety},
  {Rosolowsky}, {Saito}, {Sandstrom}, {Sanchez-Blazquez}, \&
  {Schruba}}]{kreckel19}
{Kreckel}, K., {Ho}, I.~T., {Blanc}, G.~A., {et~al.} 2019, \apj, 887, 80

\bibitem[{{Kriek} \& {Conroy}(2013)}]{kriek&conroy13}
{Kriek}, M., \& {Conroy}, C. 2013, \apjl, 775, L16

\bibitem[{{Kroupa}(2001)}]{kroupa01}
{Kroupa}, P. 2001, \mnras, 322, 231

\bibitem[{{Kulkarni} {et~al.}(2021){Kulkarni}, {Harrison}, {Grefenstette},
  {Earnshaw}, {Andreoni}, {Berg}, {Bloom}, {Cenko}, {Chornock}, {Christiansen},
  {Coughlin}, {Wuollet Criswell}, {Darvish}, {Das}, {De}, {Dessart}, {Dixon},
  {Dorsman}, {El-Badry}, {Evans}, {Ford}, {Fremling}, {Gansicke}, {Gezari},
  {Goetberg}, {Green}, {Graham}, {Heida}, {Ho}, {Jaodand}, {Johns-Krull},
  {Kasliwal}, {Lazzarini}, {Lu}, {Margutti}, {Martin}, {Masters}, {McKernan},
  {Naze}, {Nissanke}, {Parazin}, {Perley}, {Phinney}, {Piro}, {Raaijmakers},
  {Rauw}, {Rodriguez}, {Sana}, {Senchyna}, {Singer}, {Spake}, {Stassun},
  {Stern}, {Teplitz}, {Weisz}, \& {Yao}}]{kulkarni21}
{Kulkarni}, S.~R., {Harrison}, F.~A., {Grefenstette}, B.~W., {et~al.} 2021,
  arXiv e-prints, arXiv:2111.15608

\bibitem[{{Lang} {et~al.}(2020){Lang}, {Meidt}, {Rosolowsky}, {Nofech},
  {Schinnerer}, {Leroy}, {Emsellem}, {Pessa}, {Glover}, {Groves}, {Hughes},
  {Kruijssen}, {Querejeta}, {Schruba}, {Bigiel}, {Blanc}, {Chevance},
  {Colombo}, {Faesi}, {Henshaw}, {Herrera}, {Liu}, {Pety}, {Puschnig}, {Saito},
  {Sun}, \& {Usero}}]{lang20}
{Lang}, P., {Meidt}, S.~E., {Rosolowsky}, E., {et~al.} 2020, \apj, 897, 122

\bibitem[{{Lee} {et~al.}(2023){Lee}, {Sandstrom}, {Leroy}, {Thilker},
  {Schinnerer}, {Rosolowsky}, {Larson}, {Egorov}, {Williams}, {Schmidt},
  {Emsellem}, {Anand}, {Barnes}, {Belfiore}, {Be{\v{s}}li{\'c}}, {Bigiel},
  {Blanc}, {Bolatto}, {Boquien}, {den Brok}, {Cao}, {Chandar}, {Chastenet},
  {Chevance}, {Chiang}, {Congiu}, {Dale}, {Deger}, {Eibensteiner}, {Faesi},
  {Glover}, {Grasha}, {Groves}, {Hassani}, {Henny}, {Henshaw}, {Hoyer},
  {Hughes}, {Jeffreson}, {Jim{\'e}nez-Donaire}, {Kim}, {Kim}, {Klessen},
  {Koch}, {Kreckel}, {Kruijssen}, {Li}, {Liu}, {Lopez}, {Maschmann}, {Chen},
  {Meidt}, {Murphy}, {Neumann}, {Neumayer}, {Pan}, {Pessa}, {Pety},
  {Querejeta}, {Pinna}, {Rodr{\'\i}guez}, {Saito}, {S{\'a}nchez-Bl{\'a}zquez},
  {Santoro}, {Sardone}, {Smith}, {Sormani}, {Scheuermann}, {Stuber}, {Sutter},
  {Sun}, {Teng}, {Tre{\ss}}, {Usero}, {Watkins}, {Whitmore}, \&
  {Razza}}]{lee23}
{Lee}, J.~C., {Sandstrom}, K.~M., {Leroy}, A.~K., {et~al.} 2023, \apjl, 944,
  L17

\bibitem[{{Leitherer} {et~al.}(1999){Leitherer}, {Schaerer}, {Goldader},
  {Delgado}, {Robert}, {Kune}, {de Mello}, {Devost}, \&
  {Heckman}}]{leitherer99}
{Leitherer}, C., {Schaerer}, D., {Goldader}, J.~D., {et~al.} 1999, \apjs, 123,
  3

\bibitem[{{Leroy} {et~al.}(2019){Leroy}, {Sandstrom}, {Lang}, {Lewis}, {Salim},
  {Behrens}, {Chastenet}, {Chiang}, {Gallagher}, {Kessler}, \&
  {Utomo}}]{leroy19}
{Leroy}, A.~K., {Sandstrom}, K.~M., {Lang}, D., {et~al.} 2019, \apjs, 244, 24

\bibitem[{{Leroy} {et~al.}(2021){Leroy}, {Schinnerer}, {Hughes}, {Rosolowsky},
  {Pety}, {Schruba}, {Usero}, {Blanc}, {Chevance}, {Emsellem}, {Faesi},
  {Herrera}, {Liu}, {Meidt}, {Querejeta}, {Saito}, {Sandstrom}, {Sun},
  {Williams}, {Anand}, {Barnes}, {Behrens}, {Belfiore}, {Benincasa},
  {Be{\v{s}}li{\'c}}, {Bigiel}, {Bolatto}, {den Brok}, {Cao}, {Chandar},
  {Chastenet}, {Chiang}, {Congiu}, {Dale}, {Deger}, {Eibensteiner}, {Egorov},
  {Garc{\'\i}a-Rodr{\'\i}guez}, {Glover}, {Grasha}, {Henshaw}, {Ho}, {Kepley},
  {Kim}, {Klessen}, {Kreckel}, {Koch}, {Kruijssen}, {Larson}, {Lee}, {Lopez},
  {Machado}, {Mayker}, {McElroy}, {Murphy}, {Ostriker}, {Pan}, {Pessa},
  {Puschnig}, {Razza}, {S{\'a}nchez-Bl{\'a}zquez}, {Santoro}, {Sardone},
  {Scheuermann}, {Sliwa}, {Sormani}, {Stuber}, {Thilker}, {Turner}, {Utomo},
  {Watkins}, \& {Whitmore}}]{leroy21}
{Leroy}, A.~K., {Schinnerer}, E., {Hughes}, A., {et~al.} 2021, \apjs, 257, 43

\bibitem[{{Leslie} {et~al.}(2020){Leslie}, {Schinnerer}, {Liu}, {Magnelli},
  {Algera}, {Karim}, {Davidzon}, {Gozaliasl}, {Jim{\'e}nez-Andrade}, {Lang},
  {Sargent}, {Novak}, {Groves}, {Smol{\v{c}}i{\'c}}, {Zamorani}, {Vaccari},
  {Battisti}, {Vardoulaki}, {Peng}, \& {Kartaltepe}}]{leslie20}
{Leslie}, S.~K., {Schinnerer}, E., {Liu}, D., {et~al.} 2020, \apj, 899, 58

\bibitem[{{Li}(2020)}]{li20}
{Li}, A. 2020, Nature Astronomy, 4, 339

\bibitem[{{Li} \& {Draine}(2001)}]{li&draine01}
{Li}, A., \& {Draine}, B.~T. 2001, \apj, 554, 778

\bibitem[{{Lin} {et~al.}(2023){Lin}, {Yang}, \& {Li}}]{lin23}
{Lin}, Q., {Yang}, X.~J., \& {Li}, A. 2023, \mnras, 525, 2380

\bibitem[{{Ma} {et~al.}(2017){Ma}, {Ge}, {Zhao}, {Prochaska}, {Zhang}, {Ji}, \&
  {Schneider}}]{ma17}
{Ma}, J., {Ge}, J., {Zhao}, Y., {et~al.} 2017, \mnras, 472, 2196

\bibitem[{{Marble} {et~al.}(2010){Marble}, {Engelbracht}, {van Zee}, {Dale},
  {Smith}, {Gordon}, {Wu}, {Lee}, {Kennicutt}, {Skillman}, {Johnson}, {Block},
  {Calzetti}, {Cohen}, {Lee}, \& {Schuster}}]{marble10}
{Marble}, A.~R., {Engelbracht}, C.~W., {van Zee}, L., {et~al.} 2010, \apj, 715,
  506

\bibitem[{{Markov} {et~al.}(2024){Markov}, {Gallerani}, {Ferrara},
  {Pallottini}, {Parlanti}, {Di Mascia}, {Sommovigo}, \& {Kohandel}}]{markov24}
{Markov}, V., {Gallerani}, S., {Ferrara}, A., {et~al.} 2024, arXiv e-prints,
  arXiv:2402.05996

\bibitem[{{Markwardt}(2009)}]{markwardt09}
{Markwardt}, C.~B. 2009, in Astronomical Society of the Pacific Conference
  Series, Vol. 411, Astronomical Data Analysis Software and Systems XVIII, ed.
  D.~A. {Bohlender}, D.~{Durand}, \& P.~{Dowler}, 251

\bibitem[{{M{\'a}rmol-Queralt{\'o}} {et~al.}(2016){M{\'a}rmol-Queralt{\'o}},
  {McLure}, {Cullen}, {Dunlop}, {Fontana}, \& {McLeod}}]{marmol-queralto16}
{M{\'a}rmol-Queralt{\'o}}, E., {McLure}, R.~J., {Cullen}, F., {et~al.} 2016,
  \mnras, 460, 3587

\bibitem[{{Massa} {et~al.}(2022){Massa}, {Gordon}, \& {Fitzpatrick}}]{massa22}
{Massa}, D., {Gordon}, K.~D., \& {Fitzpatrick}, E.~L. 2022, \apj, 925, 19

\bibitem[{{Mathew} {et~al.}(2024){Mathew}, {Battisti}, {Vaughn}, {Jain},
  {Mohan}, \& {Murthy}}]{mathew24}
{Mathew}, J., {Battisti}, A., {Vaughn}, I., {et~al.} 2024, arXiv e-prints,
  arXiv:2407.15093

\bibitem[{{Mathis}(1994)}]{mathis94}
{Mathis}, J.~S. 1994, \apj, 422, 176

\bibitem[{{Molina} {et~al.}(2020){Molina}, {Ajgaonkar}, {Yan}, {Ciardullo},
  {Gronwall}, {Eracleous}, {Ji}, \& {Blanton}}]{molina20b}
{Molina}, M., {Ajgaonkar}, N., {Yan}, R., {et~al.} 2020, \apjs, 251, 11

\bibitem[{{Noll} {et~al.}(2009){Noll}, {Pierini}, {Cimatti}, {Daddi}, {Kurk},
  {Bolzonella}, {Cassata}, {Halliday}, {Mignoli}, {Pozzetti}, {Renzini},
  {Berta}, {Dickinson}, {Franceschini}, {Rodighiero}, {Rosati}, \&
  {Zamorani}}]{noll09a}
{Noll}, S., {Pierini}, D., {Cimatti}, A., {et~al.} 2009, \aap, 499, 69

\bibitem[{{Osterbrock}(1989)}]{osterbrock89}
{Osterbrock}, D.~E. 1989, {Astrophysics of gaseous nebulae and active galactic
  nuclei} (University Science Books)

\bibitem[{{Osterbrock} \& {Ferland}(2006)}]{osterbrock&ferland06}
{Osterbrock}, D.~E., \& {Ferland}, G.~J. 2006, {Astrophysics of gaseous nebulae
  and active galactic nuclei} (University Science Books)

\bibitem[{{Papoular} \& {Papoular}(2009)}]{papoular&papoular09}
{Papoular}, R.~J., \& {Papoular}, R. 2009, \mnras, 394, 2175

\bibitem[{{Pedrini} {et~al.}(2024){Pedrini}, {Adamo}, {Calzetti}, {Bik},
  {Gregg}, {Linden}, {Bajaj}, {Ryon}, {Ali}, {Bortolini}, {Correnti},
  {Elmegreen}, {Elmegreen}, {Gallagher}, {Grasha}, {Gutermuth}, {Johnson},
  {Melinder}, {Messa}, {{\"O}stlin}, {Sabbi}, {Smith}, {Tosi}, \& {Faustino
  Vieira}}]{pedrini24}
{Pedrini}, A., {Adamo}, A., {Calzetti}, D., {et~al.} 2024, arXiv e-prints,
  arXiv:2406.01666

\bibitem[{{Pilyugin} \& {Grebel}(2016)}]{pilyugin&grebel16}
{Pilyugin}, L.~S., \& {Grebel}, E.~K. 2016, \mnras, 457, 3678

\bibitem[{{Poole} {et~al.}(2008){Poole}, {Breeveld}, {Page}, {Landsman},
  {Holland}, {Roming}, {Kuin}, {Brown}, {Gronwall}, {Hunsberger}, {Koch},
  {Mason}, {Schady}, {vanden Berk}, {Blustin}, {Boyd}, {Broos}, {Carter},
  {Chester}, {Cucchiara}, {Hancock}, {Huckle}, {Immler}, {Ivanushkina},
  {Kennedy}, {Marshall}, {Morgan}, {Pandey}, {de Pasquale}, {Smith}, \&
  {Still}}]{poole08}
{Poole}, T.~S., {Breeveld}, A.~A., {Page}, M.~J., {et~al.} 2008, \mnras, 383,
  627

\bibitem[{{Popesso} {et~al.}(2023){Popesso}, {Concas}, {Cresci}, {Belli},
  {Rodighiero}, {Inami}, {Dickinson}, {Ilbert}, {Pannella}, \&
  {Elbaz}}]{popesso23}
{Popesso}, P., {Concas}, A., {Cresci}, G., {et~al.} 2023, \mnras, 519, 1526

\bibitem[{{Querejeta} {et~al.}(2015){Querejeta}, {Meidt}, {Schinnerer},
  {Cisternas}, {Mu{\~n}oz-Mateos}, {Sheth}, {Knapen}, {van de Ven}, {Norris},
  {Peletier}, {Laurikainen}, {Salo}, {Holwerda}, {Athanassoula}, {Bosma},
  {Groves}, {Ho}, {Gadotti}, {Zaritsky}, {Regan}, {Hinz}, {Gil de Paz},
  {Menendez-Delmestre}, {Seibert}, {Mizusawa}, {Kim}, {Erroz-Ferrer}, {Laine},
  \& {Comer{\'o}n}}]{querejeta15}
{Querejeta}, M., {Meidt}, S.~E., {Schinnerer}, E., {et~al.} 2015, \apjs, 219, 5

\bibitem[{{Querejeta} {et~al.}(2021){Querejeta}, {Schinnerer}, {Meidt}, {Sun},
  {Leroy}, {Emsellem}, {Klessen}, {Mu{\~n}oz-Mateos}, {Salo}, {Laurikainen},
  {Be{\v{s}}li{\'c}}, {Blanc}, {Chevance}, {Dale}, {Eibensteiner}, {Faesi},
  {Garc{\'\i}a-Rodr{\'\i}guez}, {Glover}, {Grasha}, {Henshaw}, {Herrera},
  {Hughes}, {Kreckel}, {Kruijssen}, {Liu}, {Murphy}, {Pan}, {Pety}, {Razza},
  {Rosolowsky}, {Saito}, {Schruba}, {Usero}, {Watkins}, \&
  {Williams}}]{querejeta21}
{Querejeta}, M., {Schinnerer}, E., {Meidt}, S., {et~al.} 2021, \aap, 656, A133

\bibitem[{{Reddy} {et~al.}(2015){Reddy}, {Kriek}, {Shapley}, {Freeman},
  {Siana}, {Coil}, {Mobasher}, {Price}, {Sanders}, \& {Shivaei}}]{reddy15}
{Reddy}, N.~A., {Kriek}, M., {Shapley}, A.~E., {et~al.} 2015, \apj, 806, 259

\bibitem[{{Roming} {et~al.}(2005){Roming}, {Kennedy}, {Mason}, {Nousek}, {Ahr},
  {Bingham}, {Broos}, {Carter}, {Hancock}, {Huckle}, {Hunsberger}, {Kawakami},
  {Killough}, {Koch}, {McLelland}, {Smith}, {Smith}, {Soto}, {Boyd},
  {Breeveld}, {Holland}, {Ivanushkina}, {Pryzby}, {Still}, \&
  {Stock}}]{roming05}
{Roming}, P. W.~A., {Kennedy}, T.~E., {Mason}, K.~O., {et~al.} 2005, \ssr, 120,
  95

\bibitem[{{Sahoo} {et~al.}(2024){Sahoo}, {Mathew}, {Battisti}, \&
  {Tucker}}]{sahoo24}
{Sahoo}, A., {Mathew}, J., {Battisti}, A., \& {Tucker}, B. 2024, Sensors, 24,
  4709

\bibitem[{{Salim} {et~al.}(2018){Salim}, {Boquien}, \& {Lee}}]{salim18}
{Salim}, S., {Boquien}, M., \& {Lee}, J.~C. 2018, \apj, 859, 11

\bibitem[{{Salim} \& {Narayanan}(2020)}]{salim&narayanan20}
{Salim}, S., \& {Narayanan}, D. 2020, \araa, 58, 529

\bibitem[{{Salmon} {et~al.}(2016){Salmon}, {Papovich}, {Long}, {Willner},
  {Finkelstein}, {Ferguson}, {Dickinson}, {Duncan}, {Faber}, {Hathi},
  {Koekemoer}, {Kurczynski}, {Newman}, {Pacifici}, {P{\'e}rez-Gonz{\'a}lez}, \&
  {Pforr}}]{salmon16}
{Salmon}, B., {Papovich}, C., {Long}, J., {et~al.} 2016, \apj, 827, 20

\bibitem[{{Salvato} {et~al.}(2019){Salvato}, {Ilbert}, \& {Hoyle}}]{salvato19}
{Salvato}, M., {Ilbert}, O., \& {Hoyle}, B. 2019, Nature Astronomy, 3, 212

\bibitem[{{Schady} {et~al.}(2012){Schady}, {Dwelly}, {Page}, {Kr{\"u}hler},
  {Greiner}, {Oates}, {de Pasquale}, {Nardini}, {Roming}, {Rossi}, \&
  {Still}}]{schady12}
{Schady}, P., {Dwelly}, T., {Page}, M.~J., {et~al.} 2012, \aap, 537, A15

\bibitem[{{Scoville} {et~al.}(2015){Scoville}, {Faisst}, {Capak}, {Kakazu},
  {Li}, \& {Steinhardt}}]{scoville15}
{Scoville}, N., {Faisst}, A., {Capak}, P., {et~al.} 2015, \apj, 800, 108

\bibitem[{{Seon} \& {Draine}(2016)}]{seon&draine16}
{Seon}, K.-I., \& {Draine}, B.~T. 2016, \apj, 833, 201

\bibitem[{{Sheth} {et~al.}(2010){Sheth}, {Regan}, {Hinz}, {Gil de Paz},
  {Men{\'e}ndez-Delmestre}, {Mu{\~n}oz-Mateos}, {Seibert}, {Kim},
  {Laurikainen}, {Salo}, {Gadotti}, {Laine}, {Mizusawa}, {Armus},
  {Athanassoula}, {Bosma}, {Buta}, {Capak}, {Jarrett}, {Elmegreen},
  {Elmegreen}, {Knapen}, {Koda}, {Helou}, {Ho}, {Madore}, {Masters},
  {Mobasher}, {Ogle}, {Peng}, {Schinnerer}, {Surace}, {Zaritsky},
  {Comer{\'o}n}, {de Swardt}, {Meidt}, {Kasliwal}, \& {Aravena}}]{sheth10}
{Sheth}, K., {Regan}, M., {Hinz}, J.~L., {et~al.} 2010, \pasp, 122, 1397

\bibitem[{{Shivaei} {et~al.}(2020){Shivaei}, {Reddy}, {Rieke}, {Shapley},
  {Kriek}, {Battisti}, {Mobasher}, {Sanders}, {Fetherolf}, {Azadi}, {Coil},
  {Freeman}, {de Groot}, {Leung}, {Price}, {Siana}, \& {Zick}}]{shivaei20a}
{Shivaei}, I., {Reddy}, N., {Rieke}, G., {et~al.} 2020, \apj, 899, 117

\bibitem[{{Shivaei} {et~al.}(2022){Shivaei}, {Boogaard}, {D{\'\i}az-Santos},
  {Battisti}, {da Cunha}, {Brinchmann}, {Maseda}, {Matthee}, {Monreal-Ibero},
  {Nanayakkara}, {Popping}, {Vidal-Garc{\'\i}a}, \& {Weilbacher}}]{shivaei22b}
{Shivaei}, I., {Boogaard}, L., {D{\'\i}az-Santos}, T., {et~al.} 2022, \mnras,
  514, 1886

\bibitem[{{Smith} \& {Hayward}(2018)}]{smith&hayward18}
{Smith}, D.~J.~B., \& {Hayward}, C.~C. 2018, \mnras, 476, 1705

\bibitem[{{Speagle} {et~al.}(2014){Speagle}, {Steinhardt}, {Capak}, \&
  {Silverman}}]{speagle14}
{Speagle}, J.~S., {Steinhardt}, C.~L., {Capak}, P.~L., \& {Silverman}, J.~D.
  2014, \apjs, 214, 15

\bibitem[{{Sutter} {et~al.}(2024){Sutter}, {Sandstrom}, {Chastenet}, {Leroy},
  {Koch}, {Williams}, {Chown}, {Belfiore}, {Bigiel}, {Boquien}, {Cao},
  {Chevance}, {Dale}, {Egorov}, {Glover}, {Groves}, {Klessen}, {Kreckel},
  {Larson}, {Oakes}, {Pathak}, {Ramambason}, {Rosolowsky}, \&
  {Watkins}}]{sutter24}
{Sutter}, J., {Sandstrom}, K., {Chastenet}, J., {et~al.} 2024, arXiv e-prints,
  arXiv:2405.15102

\bibitem[{{Tielens}(2008)}]{tielens08}
{Tielens}, A.~G.~G.~M. 2008, \araa, 46, 289

\bibitem[{{Valencic} {et~al.}(2004){Valencic}, {Clayton}, \&
  {Gordon}}]{valencic04}
{Valencic}, L.~A., {Clayton}, G.~C., \& {Gordon}, K.~D. 2004, \apj, 616, 912

\bibitem[{{Verstocken} {et~al.}(2020){Verstocken}, {Nersesian}, {Baes},
  {Viaene}, {Bianchi}, {Casasola}, {Clark}, {Davies}, {De Looze}, {De Vis},
  {Dobbels}, {Galliano}, {Jones}, {Madden}, {Mosenkov}, {Tr{\v{c}}ka}, \&
  {Xilouris}}]{verstocken20}
{Verstocken}, S., {Nersesian}, A., {Baes}, M., {et~al.} 2020, \aap, 637, A24

\bibitem[{{Wang} {et~al.}(2022{\natexlab{a}}){Wang}, {Gao}, {Ren}, \&
  {Chen}}]{wangY22}
{Wang}, Y., {Gao}, J., {Ren}, Y., \& {Chen}, B. 2022{\natexlab{a}}, \apjs, 260,
  41

\bibitem[{{Wang} {et~al.}(2022{\natexlab{b}}){Wang}, {Zhai}, {Alavi},
  {Massara}, {Pisani}, {Benson}, {Hirata}, {Samushia}, {Weinberg}, {Colbert},
  {Dor{\'e}}, {Eifler}, {Heinrich}, {Ho}, {Krause}, {Padmanabhan}, {Spergel},
  \& {Teplitz}}]{wangYun22}
{Wang}, Y., {Zhai}, Z., {Alavi}, A., {et~al.} 2022{\natexlab{b}}, \apj, 928, 1

\bibitem[{{Weingartner} \& {Draine}(2001)}]{weingartner&draine01}
{Weingartner}, J.~C., \& {Draine}, B.~T. 2001, \apj, 548, 296

\bibitem[{{Wild} {et~al.}(2011){Wild}, {Charlot}, {Brinchmann}, {Heckman},
  {Vince}, {Pacifici}, \& {Chevallard}}]{wild11}
{Wild}, V., {Charlot}, S., {Brinchmann}, J., {et~al.} 2011, \mnras, 417, 1760

\bibitem[{{Williams} {et~al.}(2010){Williams}, {Bureau}, \&
  {Cappellari}}]{williams10}
{Williams}, M.~J., {Bureau}, M., \& {Cappellari}, M. 2010, \mnras, 409, 1330

\bibitem[{{Witstok} {et~al.}(2023){Witstok}, {Shivaei}, {Smit}, {Maiolino},
  {Carniani}, {Curtis-Lake}, {Ferruit}, {Arribas}, {Bunker}, {Cameron},
  {Charlot}, {Chevallard}, {Curti}, {de Graaff}, {D'Eugenio}, {Giardino},
  {Looser}, {Rawle}, {Rodr{\'\i}guez del Pino}, {Willott}, {Alberts}, {Baker},
  {Boyett}, {Egami}, {Eisenstein}, {Endsley}, {Hainline}, {Ji}, {Johnson},
  {Kumari}, {Lyu}, {Nelson}, {Perna}, {Rieke}, {Robertson}, {Sandles},
  {Saxena}, {Scholtz}, {Sun}, {Tacchella}, {Williams}, \&
  {Willmer}}]{witstok23}
{Witstok}, J., {Shivaei}, I., {Smit}, R., {et~al.} 2023, \nat, 621, 267

\bibitem[{{Witt} \& {Gordon}(2000)}]{witt&gordon00}
{Witt}, A.~N., \& {Gordon}, K.~D. 2000, \apj, 528, 799

\bibitem[{{York} {et~al.}(2006){York}, {Khare}, {Vanden Berk}, {Kulkarni},
  {Crotts}, {Lauroesch}, {Richards}, {Schneider}, {Welty}, {Alsayyad}, {Kumar},
  {Lundgren}, {Shanidze}, {Smith}, {Vanlandingham}, {Baugher}, {Hall},
  {Jenkins}, {Menard}, {Rao}, {Tumlinson}, {Turnshek}, {Yip}, \&
  {Brinkmann}}]{york06}
{York}, D.~G., {Khare}, P., {Vanden Berk}, D., {et~al.} 2006, \mnras, 367, 945

\bibitem[{{Zafar} {et~al.}(2011){Zafar}, {Watson}, {Fynbo}, {Malesani},
  {Jakobsson}, \& {de Ugarte Postigo}}]{zafar11}
{Zafar}, T., {Watson}, D., {Fynbo}, J.~P.~U., {et~al.} 2011, \aap, 532, A143

\bibitem[{{Zafar} {et~al.}(2012){Zafar}, {Watson}, {El{\'\i}asd{\'o}ttir},
  {Fynbo}, {Kr{\"u}hler}, {Schady}, {Leloudas}, {Jakobsson}, {Th{\"o}ne},
  {Perley}, {Morgan}, {Bloom}, \& {Greiner}}]{zafar12}
{Zafar}, T., {Watson}, D., {El{\'\i}asd{\'o}ttir}, {\'A}., {et~al.} 2012, \apj,
  753, 82

\bibitem[{{Zafar} {et~al.}(2018){Zafar}, {Watson}, {M{\o}ller}, {Selsing},
  {Fynbo}, {Schady}, {Wiersema}, {Levan}, {Heintz}, {de Ugarte Postigo},
  {D'Elia}, {Jakobsson}, {Bolmer}, {Japelj}, {Covino}, {Gomboc}, \&
  {Cano}}]{zafar18b}
{Zafar}, T., {Watson}, D., {M{\o}ller}, P., {et~al.} 2018, \mnras, 479, 1542

\bibitem[{{Zeimann} {et~al.}(2015){Zeimann}, {Ciardullo}, {Gronwall}, {Bridge},
  {Brooks}, {Fox}, {Gawiser}, {Gebhardt}, {Hagen}, {Schneider}, \&
  {Trump}}]{zeimann15}
{Zeimann}, G.~R., {Ciardullo}, R., {Gronwall}, C., {et~al.} 2015, \apj, 814,
  162

\bibitem[{{Zhou} {et~al.}(2023){Zhou}, {Li}, {Li}, {Mo}, {Yan}, {Eracleous},
  {Molina}, {Gronwall}, {Ajgaonkar}, {Cheng}, \& {Guo}}]{zhou23}
{Zhou}, S., {Li}, C., {Li}, N., {et~al.} 2023, \apj, 957, 75

\end{thebibliography}

\appendix

\section{Reliability of UV slope and 2175\AA\ Feature Strength from the \swift/UVOT Filters}\label{app:swift_reliability}
In this study, we rely on the three \swift/UVOT filters to constrain the 2175\AA\ feature. This is based on first measuring the UV slope from the UVW2 and UVW1 wide filters and measuring the difference of the observed UVM2 medium filter relative to this continuum. Due to the width of the filters, it is difficult to translate these measurements to the intrinsic bump strength. 

\begin{figure}[hbt!]
\centering
\includegraphics[width=0.99\linewidth]{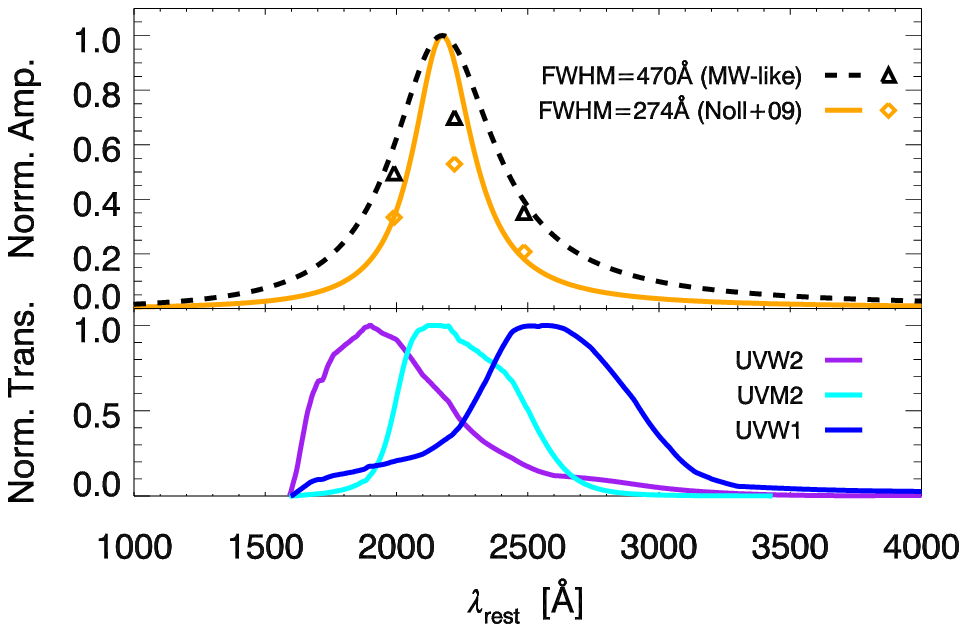}
\caption{The upper panel demonstrates two Drude profiles with fixed central wavelengths at 2175\AA\ but different widths (FWHM). The dashed black line shows the average MW feature \citep{fitzpatrick&massa90} and the orange solid line shows the average from spectroscopic data of $z\sim2$ galaxies \citep{noll09a}. The values of these features after convolving them with the \swift/UVOT filters (shown in lower panel) is indicated by the symbols. All filters extend into the 2175\AA\ feature, which complicates their use to measure the feature. However, the bump values inferred from the UVOT filters (see Section~\ref{method:abump}) are well-behaved over a wide range of UV slopes and under different assumptions for the FWHM of the feature (see Figure~\ref{fig_beta_bump_compare}).
}
\label{fig_filters_vs_bump}
\end{figure}

We demonstrate the position of the UVOT filters relative to a Drude profile, which is commonly used to model the shape of the 2175\AA\ dust absorption feature \citep{fitzpatrick&massa86}, in Figure~\ref{fig_filters_vs_bump}. As can be seen, all filters are affected by the feature to some extent, regardless of the feature width (FWHM value), where we illustrate the cases of the average MW value \citep[FWHM$\sim$470\AA;][]{fitzpatrick&massa90} and the average value from spectroscopy of $z\sim2$ galaxies \citep[FWHM$\sim$274\AA;][]{noll09a}. This range should be fairly representative of the possible FWHM ranges we might expect to see in our resolved regions. As we will demonstrate later in this section, this has a larger impact on the reliability of UV slopes that are measured, but only results in fixed offsets in 2175\AA\ bump amplitude estimates over a wide range of UV slopes. 
Figure~\ref{fig_filters_vs_bump} implies that for the FWHM$\sim$470\AA\ profile, the amplitude of the UVW2, UVM2, and UVW1 filters will reflect 50\%, 70\%, and 35\% of the peak bump amplitude, respectively. 
For the FWHM$\sim$274\AA\ profile, the amplitude of the UVW2, UVM2, and UVW1 filters will reflect 33\%, 53\%, and 21\% of the peak bump amplitude, respectively.
Therefore, the rough expectation is that the observed bump strength from UVM2, using UVW2 and UVW1 as a continuum baseline, would be $\sim$25-30\% of the true value (i.e., difference between UVM2 and the average of the two wide filters). We test this more rigorously below.

\begin{figure*}[hbt!]
\centering
$\begin{array}{ccc}
\hspace{-8mm} \includegraphics[width=0.5\textwidth]{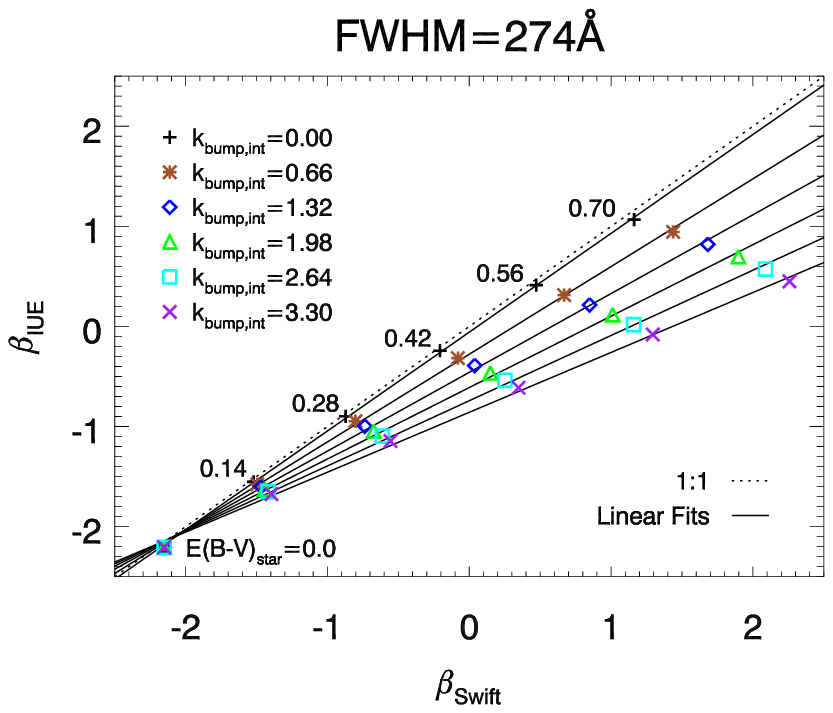} & \hspace{-8mm}
\includegraphics[width=0.5\textwidth]{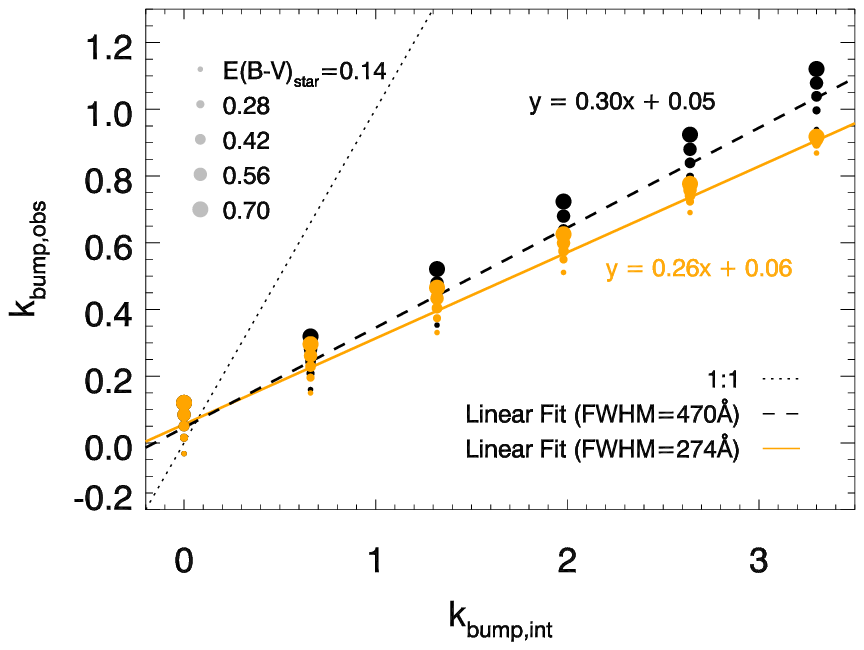} \\
\end{array}$
\vspace{-4mm}
\caption{\textit{Left:} Comparison between UV slopes $\beta_\mathrm{IUE}$, derived using the 10 spectral windows from \cite{calzetti94}, and $\beta_\mathrm{swift}$ (see eq~\ref{eq:beta_swift}). The symbols indicate 6 different values of the intrinsic bump strength $k_\mathrm{bump,int}$ and are shown for 6 different values of reddening (\EBVstar), which start at 0 at the bottom-left and increase toward the top-right with the values indicated in the plot. The slopes agree in the absence of a bump, but deviate with increasing bump strength in a roughly linear manner. This panel shows results for the FWHM=274\AA\ bump and the FWHM=470\AA\ case is qualitatively similar but with slightly larger deviations for each point (not shown for clarity). \textit{Right:} Comparison between the bump measured using UVOT filters $k_\mathrm{bump,obs}$ (see eq~\ref{eq:Abump} and \ref{eq:kbump}) relative to the intrinsic bump strength $k_\mathrm{bump,int}$. It can be seen that they follow roughly linear relations (black dashed and orange solid lines) with minimal impact from the UV slope variation (i.e., \EBVstar\ variation; see \textit{left} panel). The symbol size relates to \EBVstar\ as indicated in the legend and the colours match the Drude profiles used (see Figure~\ref{fig_filters_vs_bump}). This implies that the correlations observed with $k_\mathrm{bump,obs}$ in this work should reflect true correlations that would be seen with $k_\mathrm{bump,int}$, albeit with different scaling factors. 
}
\label{fig_beta_bump_compare}
\end{figure*}

To examine the behaviour of the \swift/UVOT filters on measurements of the UV slope and the 2175\AA\ feature, we will adopt a 10~Gyr stellar population experiencing constant star formation and Solar metallicity from the \textit{Starburst99} models \citep{leitherer99} as our reference galaxy SED. These parameters are a reasonable approximation for the massive star-forming galaxies in our sample and are sufficient for the purpose of the qualitative discussion in this section. We adopt a \cite{calzetti00} dust attenuation curve with a Drude profile superimposed with varying intrinsic amplitude, \kbump, values and FWHM widths. We then impose this attenuation curve with varying the amount of reddening, parameterised through \EBVstar, onto the stellar population model. We examine 6 values of bump amplitudes, ranging from \kbump=0 to 3.3 (MW average), and 6 values of reddening, ranging from \EBVstar=0 to 0.70, where the latter corresponds to the largest values we see in regions for our sample. The results are shown in Figure~\ref{fig_beta_bump_compare} and summarised below.

First, we compare the UV slope estimate from \swift\ ($\beta_\mathrm{swift}$). For our reference UV slope, we perform a fit (in log-space) on the reddened SED using the windows defined in \cite{calzetti94} for spectroscopic observations from the \textit{International Ultraviolet Explorer} (\textit{IUE}) telescope, ($\beta_\mathrm{IUE}$). The 10 UV windows used are between $1200<\lambda <2600$\AA\ and are designed to avoid strong stellar absorption features, including the 2175\AA\ feature. However, we note that the window red-ward of the 2175\AA\ feature used in $\beta_\mathrm{IUE}$ (2400-2580\AA) can become affected by the feature in the case of large values of \kbump\ ($k_\mathrm{bump}\gtrsim1$) when there is moderate reddening, which were not seen in the starburst galaxy sample being studied by \cite{calzetti94}. In cases with large values of \kbump, $\beta_\mathrm{IUE}$ will have bluer values (lower $\beta$) than the true UV slope. Adopting an even redder window (e.g., 2600-3000\AA) could provide a more reliable measure of the true continuum, but this is complicated by the fact that older stars can have non-negligible contributions at $\lambda>2600$\AA\ and cause deviations from a power-law form. Therefore, we simply adopt the standard $\beta_\mathrm{IUE}$ for the purpose of this comparison. Figure~\ref{fig_beta_bump_compare}, \textit{left} compares $\beta_\mathrm{IUE}$ vs $\beta_\mathrm{Swift}$ for the case of a FWHM=274\AA\ Drude profile with different \EBVstar\ and \kbump\ values. In the case of no bump ($k_\mathrm{bump,int}=0$), the value of $\beta_\mathrm{swift}$ shows excellent agreement with $\beta_\mathrm{IUE}$. These parameters deviate from 1:1 as the value of $k_\mathrm{bump,int}$ increases, with larger disagreement at larger \EBVstar\ for a fixed $k_\mathrm{bump,int}$ value.

Second, we compare the bump measured using UVOT filters $k_\mathrm{bump,obs}$ (see eq~\ref{eq:Abump} and \ref{eq:kbump}) relative to the intrinsic bump strength $k_\mathrm{bump,int}$, which corresponds to the true peak amplitude value. We find that for the two FWHM tested, they follow roughly linear relations with minimal impact from the UV slope variation (i.e., \EBVstar\ variation). This implies that the correlations observed with $k_\mathrm{bump,obs}$ in this work should reflect true correlations with the intrinsic bump, albeit with different scaling factors. The slopes of the two relations are 0.30 and 0.26 for FWHM=470\AA\ and 274\AA\, respectively, which agrees with our simple assessment in the beginning of this section.

Finally, we stress that this comparison assumed the galaxy redshift is $z\sim0$ and that these relationships are subject to change as the UVOT filters change position in terms of their rest-frame wavelength.

\section{Comparison between Stellar Continuum and Ionised Gas Reddening and Impact of Normalisation Choice}\label{app:reddening_compare}

We performed a comparison of reddening on the stellar continuum and the ionised gas for all galaxies in our sample.
To estimate the stellar continuum reddening we perform SED modeling using the \magphys\ code \citep{daCunha08, battisti20}. For details on the assumptions and priors used by \magphys, we refer readers to the papers above. The most relevant aspect for this section is that \magphys\ uses the two-component dust attenuation prescription of \cite{charlot&fall00}, where young stars ($t\leq10$~Myr) experience a steeper, SMC-like attenuation curve (birth cloud dust) and older stars ($t>10$~Myr) experience a shallower, starburst-like attenuation curve (diffuse ISM dust), with a flexible 2175\AA\ feature on the diffuse ISM dust curve \citep[see Figure~1 in][]{battisti20}. 

\begin{figure*}[hbt!]
\centering
\includegraphics[width=0.99\textwidth]{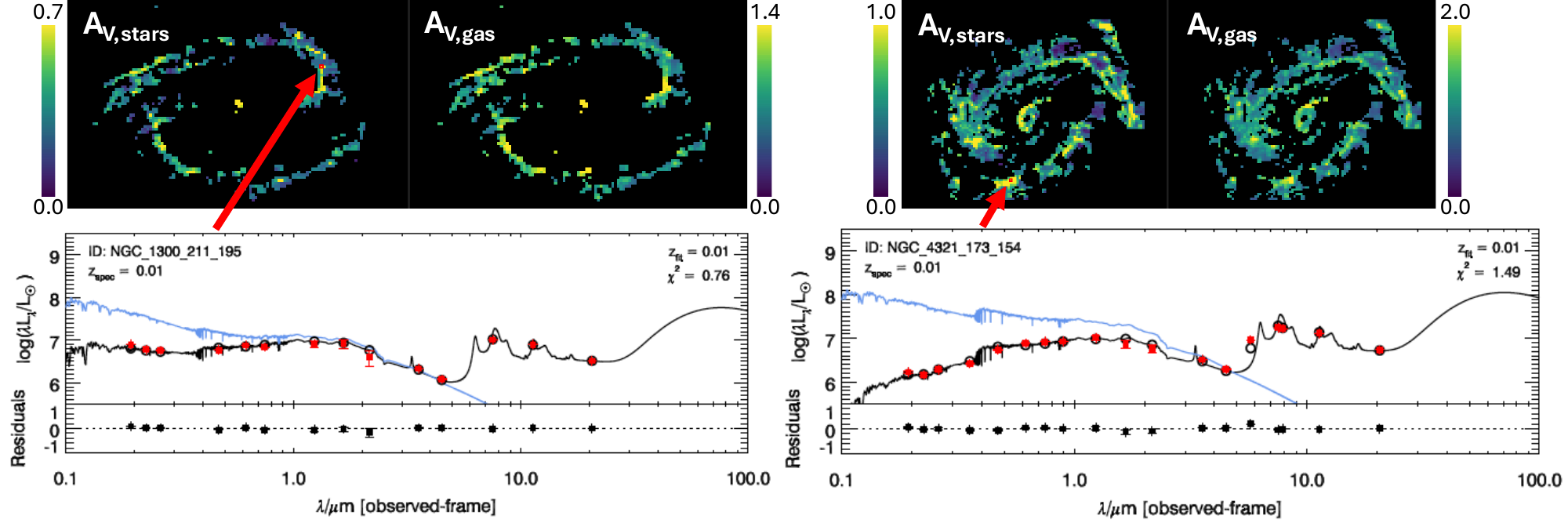}
\vspace{-1mm}
\caption{\textbf{\textit{Top panels:}} Maps of $A_{V,\mathrm{stars}}$, derived from \magphys\ SED modelling of the photometry of individual regions, and $A_{V,\mathrm{gas}}$, derived from the Balmer decrement using VLT/MUSE IFS data and assuming a MW-extinction curve for the nebular reddening for NGC~1300 (left) and NGC~4321 (right). The regions shown are restricted to those satisfying the criteria described in Section~\ref{selection_cuts}. \textbf{\textit{Bottom panels:}} Example region fits with \magphys\ for the regions indicated by the red arrows and boxes.
}
\label{fig_NGC1300_NGC4321_AV_compare}
\end{figure*}

The photometric data used in the SED fits include \swift/UVOT (UVW2, UVM2, UVW1), SDSS ($ugriz$) when available (PHANGS-MUSE ($gri$) otherwise), 2MASS ($JHK$), \spitzer/IRAC (3.6, 4.5; when available, includes 5.8, 8.0$\mu$m), and \jwst/MIRI (F770W, F1130W, F2100W). The SDSS, 2MASS, and \spitzer\ data were retrieved from the NASA/IPAC Extragalactic Database\footnote{\url{https://ned.ipac.caltech.edu/}}.
We emphasize that the outcomes of the SED modelling are dependent on the assumption of energy balance being reliable at the size scales of our regions (out to 21$\mu$m). Given that the mid-IR probes warmer dust primarily heated by young stars, this assumption may not be completely unreasonable.
Prior to SED fitting, all data were convolved to match the \swift\ resolution (2.5\arcsec) and resampled to a common grid, following the procedure outlined in Section~\ref{image_resampling}. Example SED fits to individual regions in NGC1300 and NGC4321 are shown in the Bottom Panels of Figure~\ref{fig_NGC1300_NGC4321_AV_compare}.
As the primary output parameter relating to attenuation in \magphys\ is $A_{V,\mathrm{stars}}$, we compare this to the value $A_{V,\mathrm{gas}}$ derived from the Balmer decrement using VLT/MUSE IFS data and assuming an average MW-extinction curve for the nebular reddening. The relationships between these quantities and \EBVratio\ will depend on the shape of the attenuation/extinction curves (i.e., $k_V=A_V/E(B-V)$), which is flexible in \magphys\ and therefore complicated to recover. However, this would only introduce a scaling factor assuming that the $k_V$ values (i.e., $R_V$) for the stellar attenuation curve do not vary dramatically across each galaxy. 

\begin{figure*}[hbt!]
\centering
\includegraphics[width=0.99\textwidth]{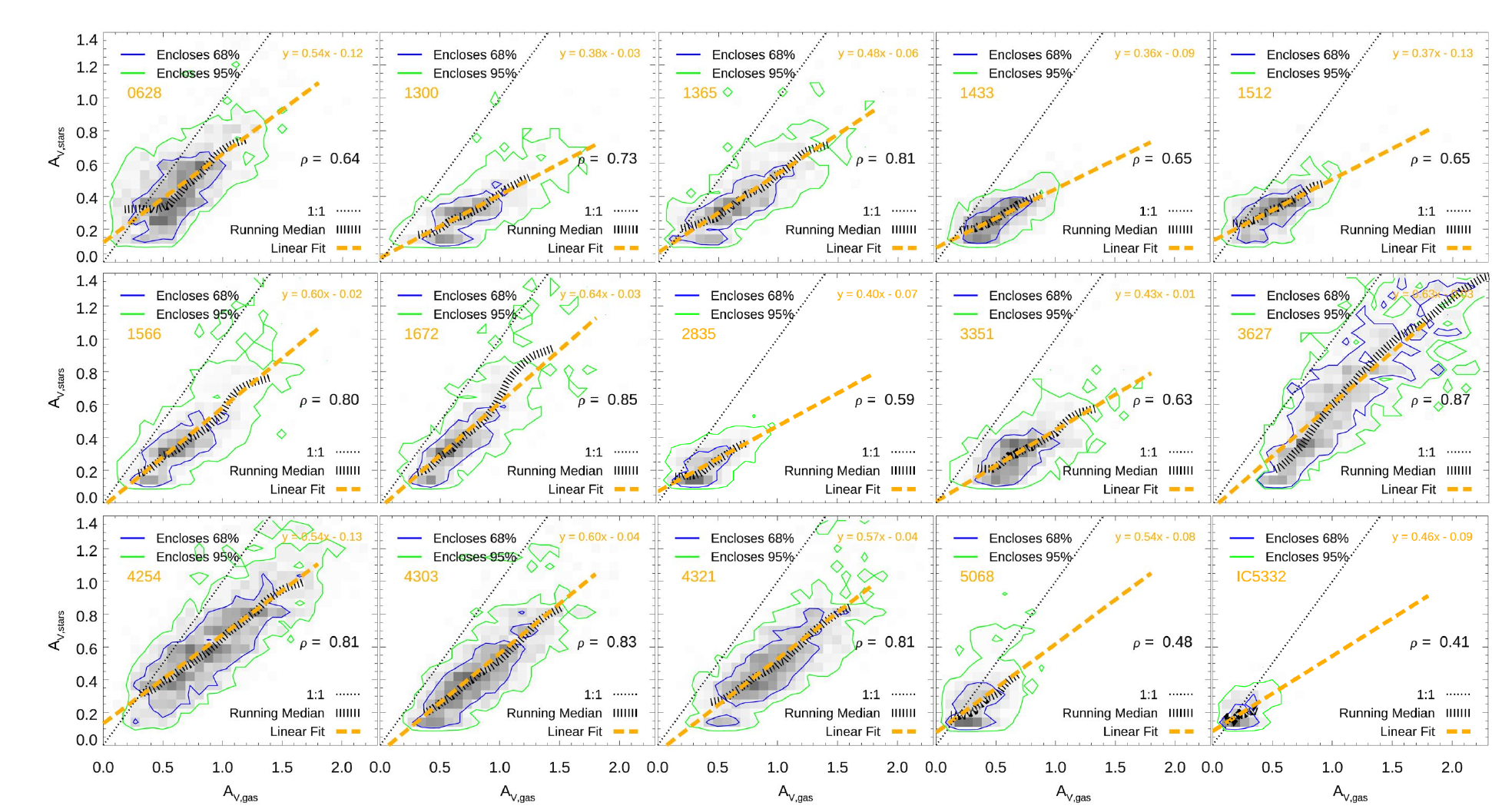}
\vspace{-1mm}
\caption{2D histograms of $A_{V,\mathrm{stars}}$ vs $A_{V,\mathrm{gas}}$ for all galaxies, which show moderately tight ($0.4\lesssim \rho \lesssim 0.9$), linear correlations. The regions shown are restricted to those satisfying criteria (1)-(3) in Section~\ref{selection_cuts} and also $A_V>0.1$ and $\sigma(A_V)<0.3$ (for both cases). The stellar continuum experiences less reddening than the nebular emission in all cases (linear fit slopes are typically between 0.4 and 0.6), similar to previous findings. This implies that using ionised gas reddening is a reasonable proxy for stellar continuum reddening for normalising the 2175\AA\ bump. Dustier galaxies (those with larger $A_V$) tend to have larger slopes (smaller difference in reddening) and tighter correlations than less dusty galaxies.
}
\label{fig_all_AV_compare}
\end{figure*}

Comparisons between $A_{V,\mathrm{stars}}$ and $A_{V,\mathrm{gas}}$ are shown visually in the Top Panels of Figure~\ref{fig_NGC1300_NGC4321_AV_compare} and as 2D histograms for all galaxies in Figure~\ref{fig_all_AV_compare}.
All cases show a moderately tight, linear correlation with the stellar continuum experiencing less reddening than the nebular emission. The average between the two cases is that the stellar continuum experiences roughly half as much reddening as the nebular emission (the average slope is $\sim$0.5), which agrees with previous studies of local galaxies \citep[\EBVratioavg$\ \sim0.5$; e.g.,][]{calzetti94, kreckel13, battisti16, emsellem22}. These results imply that using ionised gas reddening is a reasonable proxy for stellar continuum reddening for normalising the 2175\AA\ bump, as adopted in the main analysis. 

\begin{figure*}[hbt!]
\centering
$\begin{array}{ccc}
\includegraphics[width=0.55\textwidth]{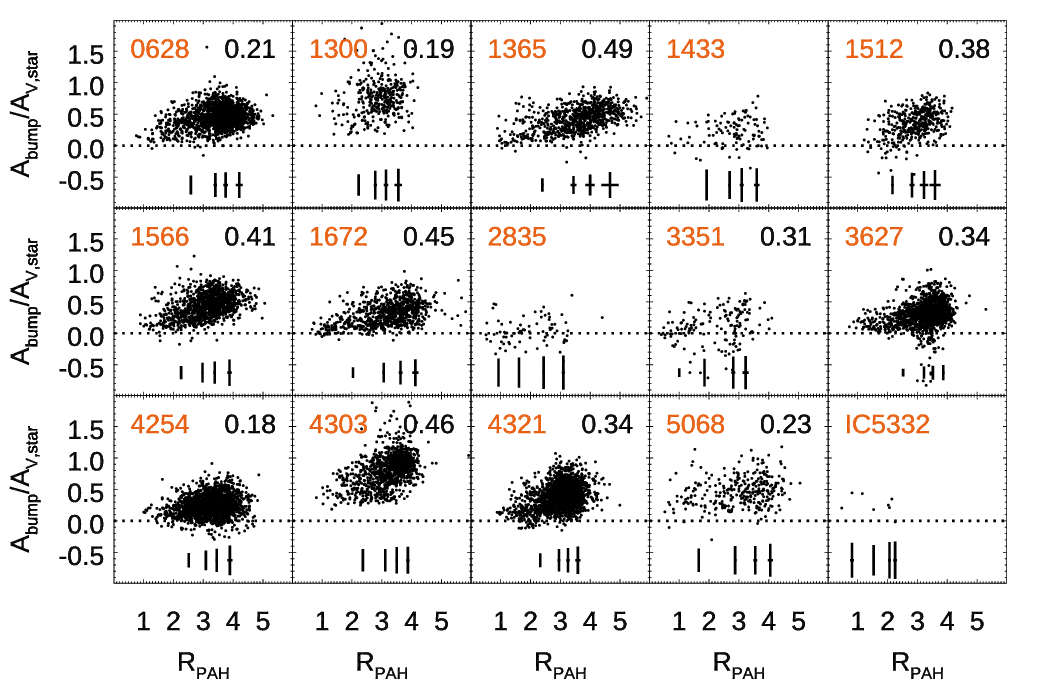} & \hspace{-8mm}
\includegraphics[width=0.5\textwidth]{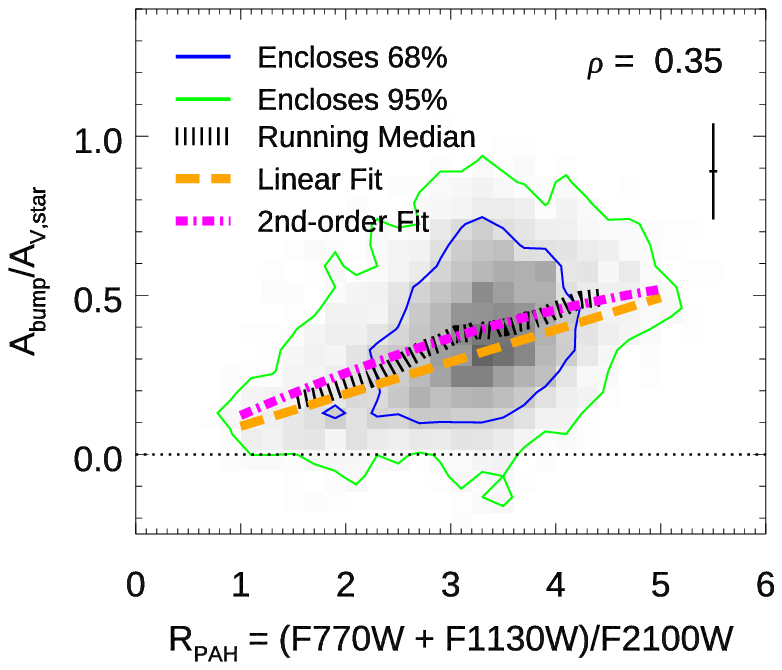} \\
\end{array}$
\vspace{-4mm}
\caption{Similar to Figure~\ref{fig_kbump_RPAH} but now normalising the 2175\AA\ bump by $A_{V,\mathrm{stars}}$. (\textit{Left:}) Intrinsic 2175\AA\ feature strength in terms of $A_\mathrm{bump}/A_{V,\mathrm{stars}}$ vs. PAH abundance, \rpah=(F770W+F1130W)/F2100W, for the 15 galaxies in our sample. 
(\textit{Right:}) 2D histogram of $A_\mathrm{bump}/A_{V,\mathrm{stars}}$ vs. \rpah\ combining the five `robust' galaxies: NGC~1365, 1566, 1672, 3627, and 4321. A representative median errorbar is shown in the upper-right, which is slightly larger than when normalising by the ionised gas reddening. Trends are qualitatively similar to those found when normalising by the ionised gas reddening, \EBVgas.
}
\label{fig_AbumpAVstar_RPAH}
\end{figure*}

\begin{figure*}[hbt!]
\centering
$\begin{array}{ccc}
\includegraphics[width=0.55\textwidth]{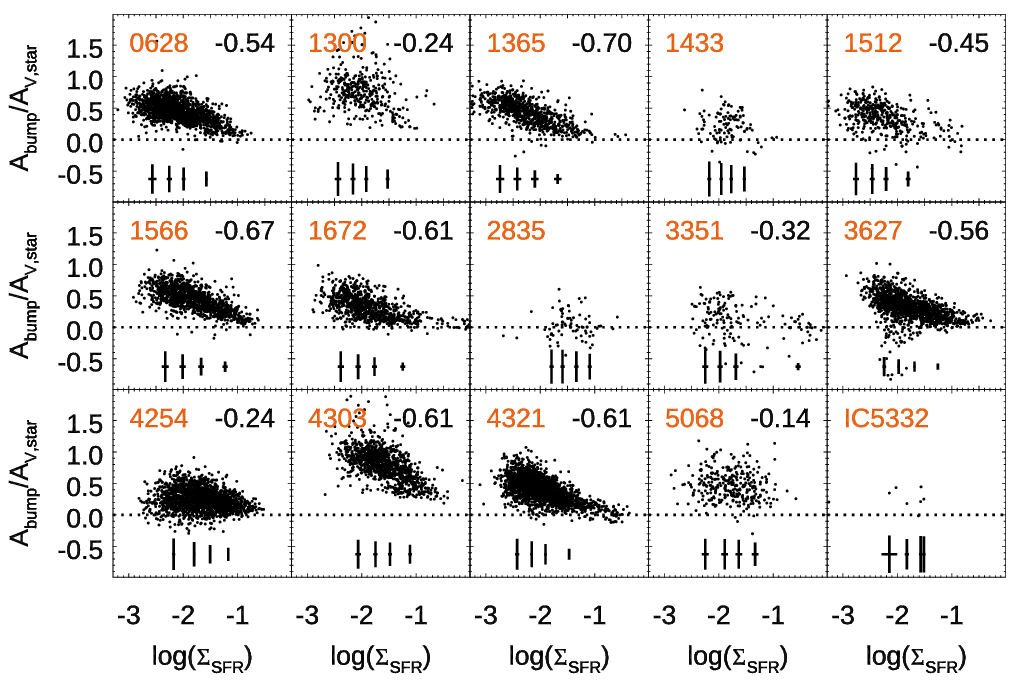} & \hspace{-8mm}
\includegraphics[width=0.5\textwidth]{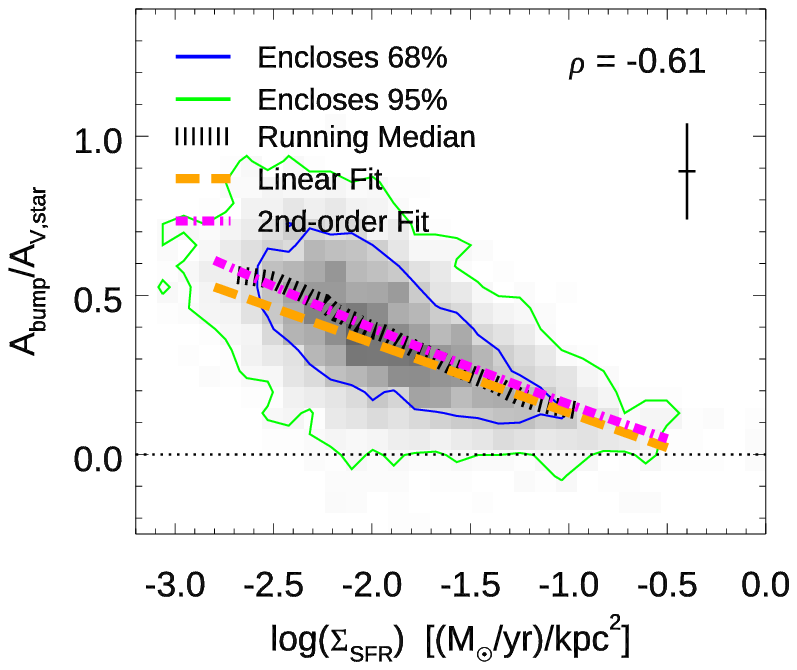} \\
\end{array}$
\vspace{-4mm}
\caption{Similar to Figure~\ref{fig_kbump_SigmaSFR} but now showing the intrinsic 2175\AA\ feature strength in terms of $A_\mathrm{bump}/A_{V,\mathrm{stars}}$ vs. SFR surface density, \sigmaSFR. The 2D histogram combines NGC~1365, 1566, 1672, 3627, and 4321. Trends are qualitatively similar to those found when normalising by the ionised gas reddening, \EBVgas.}
\label{fig_AbumpAVstar_SigmaSFR}
\end{figure*}

For completeness, we also test our main results when normalising by $A_{V,\mathrm{stars}}$ instead of \EBVgas. We adopt the same primary selection criteria in Section~\ref{selection_cuts} ((1)-(3)), because we also compared to quantities derived from the emission lines (e.g., SFR), with a variation to Condition (4) as $\sigma(A_\mathrm{bump}/A_{V,\mathrm{stars}})<0.3$. The $A_\mathrm{bump}/A_{V,\mathrm{stars}}$-\rpah\ relationship is shown in Figure~\ref{fig_AbumpAVstar_RPAH}. The uncertainties on $A_{V,\mathrm{stars}}$ are slightly larger relative to $A_{V,\mathrm{gas}}$ (i.e., \EBVgas) for nearly all galaxies, which we attribute to the age-dust degeneracy in the SED modelling. In \magphys, both the amplitude and shape of the attenuation curve are free parameters, with the final `effective' attenuation curve being a combination of the ISM and birth-cloud dust curves in the \cite{charlot&fall00} model \citep[e.g.,][]{battisti20}. As a reminder, for the nebular reddening we assume a fixed average MW extinction curve. We also find similar outcomes between $A_\mathrm{bump}/A_{V,\mathrm{stars}}$ and other parameters. We show the comparison with \sigmaSFR\ in Figure~\ref{fig_AbumpAVstar_SigmaSFR}.
The parameters of both the linear and second-order fits (including as a function of sSFR, not shown) are listed in Table~\ref{Tab:AbumpAVstar_vs_param}.

\begin{table*}
\begin{center}
\caption{Fit Parameters of $A_\mathrm{bump}/A_{V,\mathrm{stars}}$ as a Function of Galaxy Properties \label{Tab:AbumpAVstar_vs_param}}
\begin{tabular}{cc|ccccc}
\hline\hline 
             &        \hspace{1cm}$y\rightarrow$     &   \multicolumn{5}{c}{$A_\mathrm{bump}/A_{V,\mathrm{stars}}$}   \\
      $x$    &     range    &  $p_0$    &  $p_1$    & $p_2$   & $\sigma_{\mathrm{int}}$ & $\rho_S$  \\ \hline

\rpah &  [1,5] & 
-0.013 $\pm$ 0.008 & 0.101 $\pm$ 0.003 & -- & 0.10 & 0.35 \\
 & & 
-0.034 $\pm$ 0.022 &  0.167 $\pm$ 0.013 & -0.011 $\pm$ 0.002 & & \\ \hline

log($\Sigma_{\mathrm{SFR}}$) & [-2.8,-0.5] & 
-0.088 $\pm$ 0.005 & -0.220 $\pm$ 0.003 & -- & 0.06 & -0.61 \\ 
 $[M_\odot$yr$^{-1}$kpc$^{-2}]$ & & 
-0.052 $\pm$ 0.020 & -0.196 $\pm$ 0.022 &  0.015 $\pm$ 0.006 & & \\ \hline

log(sSFR) & [-10.9,-8.8] & 
-1.778 $\pm$ 0.039 & -0.208 $\pm$ 0.004 & -- & 0.08 & -0.56 \\
 $[\mathrm{yr}^{-1}]$ & & 
2.697 $\pm$ 0.743 &  0.713 $\pm$ 0.149 &  0.048 $\pm$ 0.007 & & \\ \hline

\end{tabular}
\end{center}
\textbf{Notes.} The functional form of these fits is $y = p_0+p_1x+p_2x^2$, where $y$ is $A_\mathrm{bump}/A_{V,\mathrm{stars}}$. We present both a linear and second-order polynomial fit for each case. We also report the intrinsic dispersion, $\sigma_{\mathrm{int}}$, returned from \texttt{MPFITEXY}, and the Spearman nonparametric correlation coefficient, $\rho_S$. The data used in the fits are a combination of five galaxies (NGC~1365, 1566, 1672, 3627, and 4321).
\end{table*}

\end{document}